\documentclass[11pt,a4paper,notitlepage]{article}
\usepackage[T1]{fontenc}
\usepackage{amsmath}
\usepackage{amsfonts}
\usepackage{amssymb}
\usepackage{graphicx}
\usepackage[margin=1in]{geometry}
\usepackage[varg]{txfonts}
\usepackage{hyperref,natbib}
\usepackage{bm,caption,subcaption}
\usepackage{authblk}
\usepackage{xcolor}
\allowdisplaybreaks

\title{The effect of AC electric field on the dynamics of a vesicle under shear flow in the small deformation regime}
\author{Kumari Priti Sinha, Rochish M Thaokar \\Department of Chemical Engineering, Indian Institute of Technology Bombay, Powai, Mumbai-400076, India. \\ E-mail: pritisinha.026@gmail.com, rochish@che.iitb.ac.in}

\begin{document}

\maketitle

\begin{center} \textbf{\Large Abstract} \end{center} 
Vesicles or biological cells under simultaneous shear and electric field can be encountered in dielectrophoretic devices or devices used for continuous flow electrofusion or electroporation. In this work, the dynamics of a vesicle subjected to simultaneous shear and uniform AC electric field is investigated in the small deformation limit. The coupled  equations for vesicle orientation and shape evolution are derived theoretically and the resulting nonlinear equations are handled numerically to generate relevant phase diagrams that demonstrate the effect of electrical parameters on the different dynamical regimes such as tank-treading (TT), trembling (TR), and tumbling (TU). It is found that while the electric Mason number ($Mn$) which represents the relative strength of the electrical forces to the shear forces, promotes TT regime, the response itself is found to be sensitive to the applied frequency as well as the conductivity ratio. While higher outer conductivity promotes orientation along the flow axis, orientation along the electric field is favored when the inner conductivity is higher. Similarly a switch of orientation from the direction of the electric field to the direction of flow is possible by mere change of frequency when the outer conductivity is higher. Interestingly, in some cases, a coupling between electric field induced deformation and shear can result in the system admitting an intermediate TU regime while attaining the TT regime at high $Mn$. The results could enable designing better dielectrophoretic devices wherein the residence time as well as the dynamical states of the vesicular suspension can be controlled as per the application. 	

\section{Introduction}
Vesicles, that are bounded by a  lipid bilayer, can acquire a variety of equilibrium shapes such as prolate and oblate spheroids, discocytes, stomatocytes, etc, a result of  minimizing the bending energy for a prescribed reduced volume. On the other hand,  non-equilibrium, dynamical states are observed under externally applied forces such as  hydrodynamic flow or electric field, and have received extensive attention because of their relevance in bio-microfluidics. Moreover, flow of bio-fluids in organisms involve flow of vesicle-like objects such as Red blood cells (RBC) amongst others, suspended in an ambient fluid. An interplay of hydrodynamic and membrane forces as well as forces due to applied  electric or shear fields, determine the shape of these vesicles and influence the surrounding flow field.\\

A vesicle resists deformation due to external forces, such as viscous stresses due to shear flow, on account of its membrane fluidity, bending resistivity, and area-incompressibility. A linear shear flow can be decomposed into two parts: an extensional (symmetric part) and a rotational component (antisymmetric part).  In the case of a vesicle in shear flow, the extensional component causes extension and deformation of a vesicle into an  ellipsoidal shape, and orients it along the extensional axis of the flow (making an angle of $\pi/4$ with the direction of  flow). On the other hand, the rotational component (antisymmetric part) provides a  rigid body like rotation to the vesicle by applying a torque to it \cite{Mader2007ShInt}. Unlike in rigid objects, shear flows can induce tank-treading motion in a membrane  bound vesicle, wherein lipid molecules are transported by the local velocity because of the fluidic nature of the membrane. \\

A vesicle under shear flow exhibits various dynamic modes such as tank-treading (TT), trembling (TR), and tumbling (TU). The existence of each of these regimes depends upon several geometric as well as flow parameters (such as excess area stored in the  membrane of the vesicle, viscosity contrast across the membrane, and flow capillary number). Influenced by these parameters, a vesicle can acquire a non-axisymmetric ellipsoidal shape which is inclined at a stationary angle to the flow direction. The membrane rotates around its fixed ellipsoidal shape and the resulting motion is called as TT. Beyond a certain viscosity contrast across the membrane, the TT regime is inhibited and a vesicle shows full periodic rotations (flipping motion) with respect to the direction of shear flow. This is referred to as the TU mode. An intermediate regime between TT and TU modes, called the TR mode, is observed  wherein a vesicle's long axis oscillates around the shear direction and undergoes asymmetric shape deformations in the vorticity direction (also called, vacillating breathing or swinging). \\
 
The pioneering theoretical work by ~\cite{skalak1982ShInt} on an undeforming, ellipsoidal shaped RBC  in shear flow admitting a non-zero, position dependent surface velocity, showed that a vesicle can have tank-treading or tumbling motion. The first experimental study on vesicle under shear flow was conducted by ~\cite{Haas1997ShInt} who observed the TT and the TU regimes. While the early theoretical studies on vesicles under shear flow, without viscosity contrast, showed a TT regime, ~\cite{Fujitani1995ShInt,Kraus1996ShInt}, rigorous numerical studies ~\cite{Misbah2003ShInt, Misbah2004ShInt} indicated a TT-TB transition. Results of systematic experiments  
~\cite{Steinberg2005ShInt} in the TT regime were found to be in agreement with theory~\cite{Seifert1999ShInt}. The numerical predictions of TT-TU transition by ~\cite{Misbah2003ShInt, Misbah2004ShInt} were later experimentally observed by ~\cite{Mader2006ShInt}.  ~\cite{Steinberg2006ShInt} reported a new experimental observation  in which a vesicle in shear flow trembles around the flow direction with strong shape fluctuations, which they called as Trembling. This was called as the "swinging state" in another experimental study ~\cite{Noguchi2007ShInt}. Later, a theoretical study \cite{Misbah2006ShInt} explained this mode, which was termed as the "vacillating breathing" mode.  This coupling of shape and orientation angle was further studied  ~\cite{Vla2006ShInt} using a mechanical force balance approach. These different dynamical states around the TU regime could be  qualitatively described by a 2D analytical model  ~\cite{Mader2007ShInt}. \\

In the earlier theoretical studies \cite{Misbah2006ShInt} the deviation of the vesicle shape from a sphere, as well as the membrane forces are considered only upto  leading order in the capillary number, a ratio of deforming shear forces to the restoring bending forces. The resulting evolution equations are independent of the bending rigidity as well as the capillary number, but depend upon the viscosity contrast and the excess area. An extension of the leading order theory to higher order corrections in the shape deviation as well as the membrane forces ~\cite{Lebedev2007ShInt,Danker2007ShInt,Noguchi2007ShInt, Lebedev2008ShInt} led to an evolution equation that exhibited explicit dependence on the capillary number. The expression is more complicated when higher order hydrodynamic terms are also added ~\cite{Misbah2009ShInt,Farutin2010ShInt}. The resulting shape evolution equations admit several new dynamical states. These were also investigated for a vesicle with reduced volume same as RBCs using 3D numerical simulations ~\cite{Farutin2012ShInt}. In another numerical study ~\cite{Barrett2015ShInt}, vesicle shapes under shear flow were obtained by using the area difference elasticity (ADE model) and the spontaneous curvature model. \\

The detailed phase diagram for vesicle dynamics in shear flow ~\cite{Lebedev2007ShInt} was experimentally verified  ~\cite{Steinberg2009PRLShInt, Steinberg2009PNASShInt} by  conducting a series of experiments in pure shear flow when there is viscosity contrast across the vesicle $\eta_{in}/\eta_{ex}>1$ ~\cite{Steinberg2009PRLShInt}. This was later extended to a general shear flow with no viscosity contrast across the vesicle ~\cite{Steinberg2009PNASShInt}. It was found that the experimental results ~\cite{Steinberg2009PRLShInt, Steinberg2009PNASShInt} were in good agreement with the theory ~\cite{Lebedev2007ShInt, Lebedev2008ShInt}. \\ 

The work on a vesicle under simultaneous shear flow and an applied uniform DC electric field ~\cite{Vla2011ShInt} exhibited significant effect on the TT and TB dynamics. The electric field was found to suppress the tumbling dynamics. Since then, there have been at least three works on understanding the  effect of simultaneous shear and DC electric fields using numerical methods, such as the level set method \cite{kolahdouz2015EHD,kolahdouz2015Dyn}, the immersed boundary method \cite{hu2016vesicle} and the boundary integral method \cite{mcconnell2013vesicle} which extend the studies to large deformation of the vesicle. All these numerical studies found that the TU regime is transformed into a TT regime by application of strong DC electric fields.  \\
	
The motivation of the present work is to extend these studies on vesicles under simultaneous shear and DC electric fields to AC electric fields. DC fields are seldom used in experiments in vesicles, unless used for electroporation studies. AC dielectrophoresis is commonly used in biotechnological applications, and in microfulidic devices. It is therefore important to understand the effect of frequency and magnitude of AC fields on the TT, TR and TU regimes observed in the absence of electric fields. Moreover, the conductivity ratio is known to be critical in shape deformations of vesicles in axisymmetric AC electric fields. It is therefore expected to non-trivially alter the dynamics of vesicles in combined shear and AC electric fields as well. Additionally the theoretical analysis of ~\cite{Vla2011ShInt} needs to be modified to include higher order membrane deformation forces. 
 
\section{Mathematical formulation}

\subsection{Model description}
Consider a vesicle of radius $R_0$ such that $V=\frac{4}{3} \pi R_o^3$ with a  non-conducting, bilayer  membrane, subjected to linear shear flow ($\boldsymbol{u_x}=\dot{\gamma}y \boldsymbol{\hat{e}_x}, \boldsymbol{u_y}=\boldsymbol{u_z}=0$) induced by moving two walls in opposite directions along the X-axis, here $\dot{\gamma}$ is the applied shear rate. The $r, \theta, \Phi$ directions in the corresponding spherical coordinates system are the radial position, the azimuthal angle measured from the $Z$ axis towards the $X-Y$ plane and the polar angle measured anticlockwise from the $X$ axis in the $X-Y$ plane. The centroid of the vesicle remains fixed at a position where the velocity of the flowing fluid is zero. The inner and outer regions of the vesicle have different physical properties such as conductivities ($\sigma_{ex},\sigma_{in}$), permittivities ($\epsilon_{ex},\epsilon_{in}$), and viscosities ($\eta_{ex}, \eta_{in}$). The membrane has a finite thickness of $h=5nm$, it is non-conducting (zero conductance) and has capacitance $C_{mem}=\epsilon_{mem}/h$. The dimensionless ratios of interest are  $\sigma_r=\sigma_{in}/\sigma_{ex}, \epsilon_r=\epsilon_{in}/\epsilon_{ex}, \eta=\eta_{in}/\eta_{ex}$. Here subscripts "ex" and "in" represents quantities outside and inside the vesicle respectively. \\

\begin{figure}[tp]
 \centering
 \includegraphics[width=0.6\textwidth]{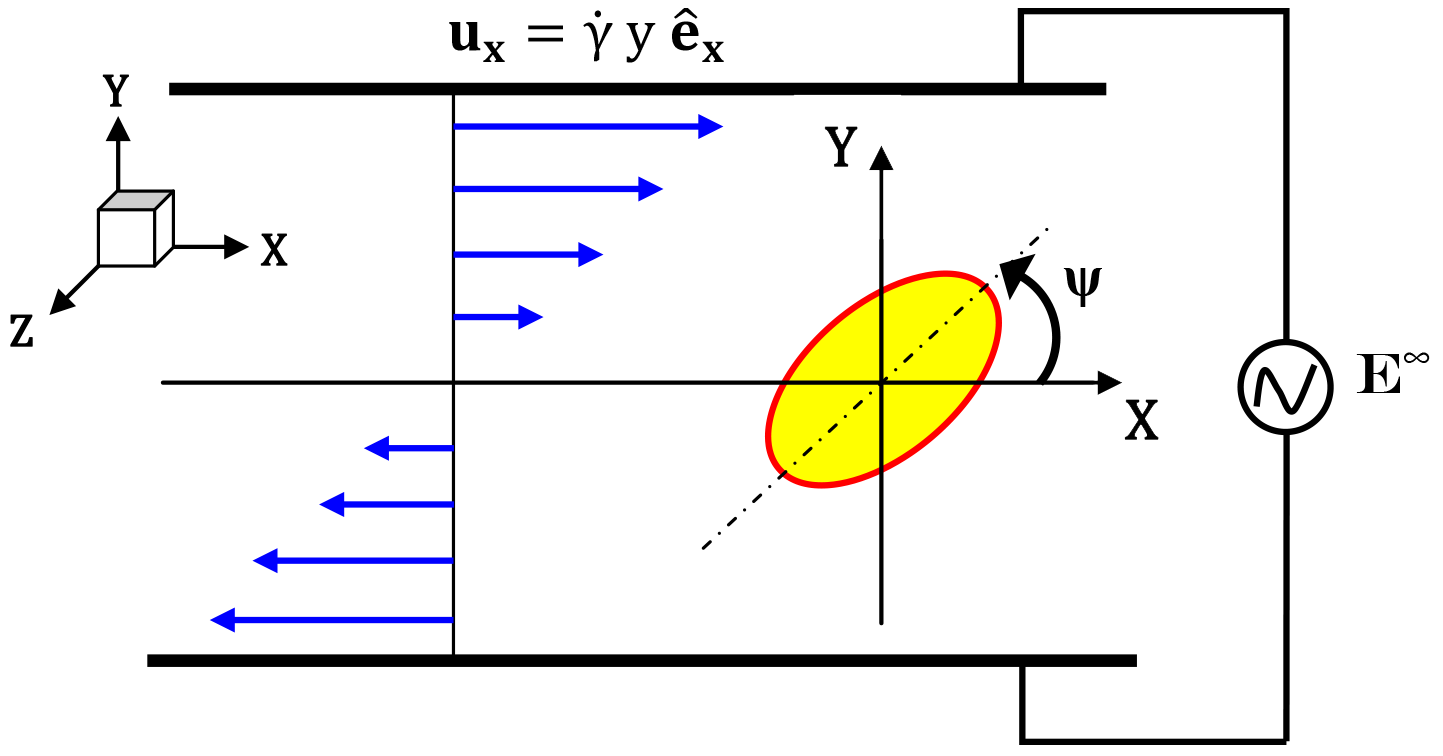}
 \caption{Schematic of vesicle under AC electric field and shear flow}
 \label{schematic}
 \end{figure}

An externally applied AC electric field $\boldsymbol{E^\infty}$ in the Y direction, with frequency $\omega$ acts, on a vesicle suspended in the fluid. Such a vesicle, under the action of an applied shear flow, can get inclined at an angle $\psi$ to the direction of shear flow (fig.\ref{schematic}), measured anticlockwise, and can show both steady and unsteady dynamics depending upon the system parameters. This dynamical state is a result of a balance between shape stabilizing membrane stresses (bending as well as tension stress) and the destabilizing electric and hydrodynamic stresses.     \\

There are various scales associated with different quantities in the model to non-dimensionalize the governing equations. All lengths are nondimensionalized by $R_0$, time by inverse shear rate $\dot{\gamma}^{-1}$, velocity by $R_0 \dot{\gamma}$, electric field by $E_0$, frequency by $\dot{\gamma}$, electric potential by $E_0 R_0$ and stresses by $\dot{\gamma} \eta_{ex}$. The electric stresses are of the order $\epsilon_{ex} E_0^2$,  and membrane stresses scale as $\kappa_b/R_0^3$. Under an electric field, relaxation of bulk charge takes place on a time scale of $t_c^{ex}=\epsilon_{ex}/\sigma_{ex} t_c^{in}=\epsilon_{in}/\sigma_{in}$, , interfacial polarization on time scales $t_{MW}=(\epsilon_{in}+2 \epsilon_{ex})/(\sigma_{in}+2 \sigma_{ex})$,  the charging of a capacitive membrane takes place on a time scale of $t_{mem} = R_0 C_{mem} (\frac{1}{2}+\frac{1}{\sigma_r})/\sigma_{ex}$,  flow induced vesicle deformation takes place on a time scale of $t_{el}=\eta_{ex}/(\epsilon_{ex} E_0^2)$ while shape distortion due to extensional part of applied shear take place on a time scale of $t_{\dot{\gamma}}=\dot{\gamma}^{-1}$. The shape deformations  relax on a time scale of $t_{k}=\eta_{ex} R_0^3/\kappa_b$. \\

Several non-dimensional numbers that compare the relative strength of applied stresses thereby emerge. Flow capillary number, $Ca=\dot{\gamma}\eta_{ex} R_0^3/\kappa_b$ is the ratio of shape-relaxation time scale ($t_{k}$) to shape the shear rate ($t_{\dot{\gamma}}$) and the Mason number $Mn=\epsilon_{ex} E_0^2/(\eta_{ex} \dot{\gamma})$ compares the relative strength of electric and viscous stresses. All equations hereby are presented in non-dimensional form only and no superscript is used for simplicity and brevity. \\

\subsection{Shear-AC coupled model}
\textbf{Hydrodynamics}\\
The flow is described in the low Reynolds number limit using the vectors spherical harmonics while the pressure and the electric potentials are expanded in spherical harmonics. As is the practice in modeling these systems,  ~\cite{Seifert1995ShInt, Seifert1999ShInt, Seifert2008ShInt, Seifert2012ShInt, Steinberg2011ShInt, Steinberg2012ShInt, Seifert2013ShInt} the membrane thermal fluctuations are ignored.       
The velocity fields (${\bf u_{in}, u_{ex}}$) as well as the pressure fields ($p_{in}, p_{ex}$)) in the inner and outer regions of the vesicle are given by: 
\begin{align}
&\nabla p_{in}-\nabla^2 \boldsymbol{u_{in}}=0 &\\
&\nabla.\boldsymbol{u_{in}}=0&
\end{align}
and in the inner region and 
\begin{align}
&\nabla p_{ex}-\nabla^2 \boldsymbol{u_{ex}}=0 &\\
&\nabla.\boldsymbol{u_{ex}}=0&
\end{align}
in the outer region. Note that the viscosity ratio is appropriately taken into account in the stress balance.  Hydrodynamic stresses in the inner and outer ($j=in, ex$) regions of a vesicle are given by
\begin{equation}
\ \boldsymbol{\tau_{j}^h}=-p_{j}\boldsymbol{I}+\mu_{j} \left[\nabla \boldsymbol{u_{j}}+(\nabla \boldsymbol{u_{j}})^T\right]
\end{equation}
where $\boldsymbol{I}$ is the identity matrix and superscript $T$ represents the transpose of the matrix. The traction vectors on the vesicle surface (at $r=1$) are given by
\begin{align}
&\ \tau_{in}^h.\bm{\hat{e}_r}=-p_{in} \bm{\hat{e}_r}+Z_{in} \qquad \mathrm{at\ r=1}& \label{Hin}\\ 
&\ \tau_{ex}^h.\bm{\hat{e}_r}=-p_{ex} \bm{\hat{e}_r}+Z_{ex} \qquad \mathrm{at\ r=1}  \label{Hex}&
\end{align}
with
\begin{equation}
\  Z_{j}=\bm{\hat{e}_r}.\left(\boldsymbol{\nabla} \boldsymbol{u_{j}}+(\boldsymbol{\nabla} \boldsymbol{u_{j}})^T\right)=r \frac{d}{dr}\left(\frac{\boldsymbol{u_{j}}}{r}\right)+\frac{1}{r}\nabla((\boldsymbol{u_{j}}.\bm{\hat{e}_r})r)
\end{equation}
where $j=in, ex$. Expressions for  $Z_{in}|_{r=1}, Z_{ex}|_{r=1}$ along with the final hydrodynamic traction matrices are provided in appendix-A \\

\textbf{Electrostatics}\\

For a vesicle subjected to AC electric field, the expressions for the electric potential in the inner and the outer regions are given by
\begin{align}
\phi_{in}=&P_{in}r\sum_{n=-1}^{n=1}e_{1n}^\infty Y_{1n} \label{phiin}&\\
\phi_{ex}=&\phi^\infty +\frac{ P_{ex}}{r^2} \sum_{n=-1}^{n=1}e_{1n}^\infty Y_{1n} \label{phiex}&
\end{align}
where $\phi^\infty=-r \sum_{n=-1}^{n=1}e_{1n}^\infty Y_{1n}$ is the externally applied unperturbed electric potential. Here $Y_{1n}$ is the associated Legendre Polynomial of degree $1$ and order $n$. If the applied electric acts in the velocity gradient and the vorticity direction then $\boldsymbol{E^\infty}=E_0 (\alpha \boldsymbol{\hat{e}_y}+\beta \boldsymbol{\hat{e}_z})$, where $\alpha, \beta$ are related to the spherical harmonic coefficients by 
$ e_{10}^\infty=\beta \sqrt{\frac{4\pi}{3}}$, 
$ e_{\pm 1}^\infty=\alpha i \sqrt{\frac{2\pi}{3}}$. Unknown coefficients ($P_{in}$, $P_{ex}$) in eq. \eqref{phiin} and \eqref{phiex} are determined by applying electrostatic boundary conditions at the vesicle surface (at $r=1$)
\begin{align}
\phi_{in}-\phi_{ex}=&V_{mem} \sum_{n=-1}^{n=1}e_{1n}^\infty Y_{1n} &\\
\ (1+i \omega \zeta)\frac{d\phi_{ex}}{dr}=&(\sigma_r+i \omega \epsilon_r \zeta)\frac{d\phi_{in}}{dr}&\\
\ -(1+i \omega \zeta)\frac{d\phi_{ex}}{dr}=&i C_{mem} \omega \zeta V_{mem} \sum_{n=-1}^{n=1}e_{1n}^\infty Y_{1n}&
\end{align}
here $\zeta=\dot{\gamma} t_c^{ex}$ is a dimensionless parameter introduced since  in the electrostatics equations the time as well as the frequency non-dimensionalization are done using the  shear rate and $t_c^{ex}$ is of the order of $10^{-6}$ sec.\\

This yields,
\begin{align}
\ P_{in}=&-\frac{3C_{mem} \zeta \omega (-1+i \omega \zeta)}{-2 \sigma_r - 
 I (C_{mem} (2 + \sigma_r) + 
    2 (\epsilon_r + \sigma_r)) \omega \zeta+ (2 \epsilon_r + 
    C_{mem} (2 + \epsilon_r)) \omega^2 \zeta^2}&\\
\ P_{ex}=&\frac{(i \sigma_r - (C_{mem} + \epsilon_r + \sigma_r - 
      C_{mem} \sigma_r) \omega \zeta+ 
   i (C_{mem} (-1 + \epsilon_r) - \epsilon_r) \omega^2 \zeta^2)}{\sigma_r (-2 i + (2 + C_{mem}) \omega \zeta) + \omega \zeta (2 (C_{mem} + \epsilon_r) + i (2 \epsilon_r + C_{mem} (2 + \epsilon_r)) \omega \zeta)}&\\
\ V_{mem}=&\frac{3 (-i + \omega \zeta) (-i \sigma_r + \epsilon_r \omega \zeta)}{-2 \sigma_r - i (C_{mem} (2 + \sigma_r) + 2 (\epsilon_r + \sigma_r)) \omega \zeta+ (2 \epsilon_r + C_{mem} (2 + \epsilon_r)) \omega^2 \zeta^2} &   
\end{align}
Here, the normal and tangential electric field components in the inner region are: $\boldsymbol{E_{in, \ ex, r}}=-\frac{d\phi_{in, \ ex}}{dr} \ \boldsymbol{\hat{e}_r}$, $\boldsymbol{E_{in, \ ex, \theta}}=-\frac{1}{r}\frac{d\phi_{in, \ ex}}{d\theta}\boldsymbol{\hat{e}_\theta}$, and $\boldsymbol{E_{in, \ ex, \Phi}}=-\frac{1}{r\sin\theta}\frac{d\phi_{in, \ ex}}{d\Phi}\boldsymbol{\hat{e}_\Phi}$. The induced surface change on the membrane interface is given by $Q_c=Re[\boldsymbol{E_{ex,r}}]-\epsilon_rRe[\boldsymbol{E_{in,r}}]$, where Re[] represents the real part of the quantity enclosed in the square bracket, to give \\
\begin{align}
\ Q_c=-\frac{3 C_{mem} (\epsilon_r - \sigma_r) (C_{mem} (2 + \sigma_r) + 2 (\epsilon_r + \sigma_r)) \omega^2 \zeta^2 (\beta \cos\theta + \alpha \sin\theta \sin\Phi)}{4 \sigma_r^2 + (4 (C_{mem} + \epsilon_r)^2 + 4 C_{mem}^2 \sigma_r + (2 + 
C_{mem})^2 \sigma_r^2) \omega^2 \zeta^2+ (2 \epsilon_r + C_{mem} (2 + \epsilon_r))^2 \omega^4 \zeta^4}
\end{align}

The time-independent electric stress in the $r, \theta, \Phi$ directions inside as well as outside the vesicle are
\begin{align}
\tau_{in,ex,r}=&\frac{1}{4}(E_{in,ex,r} E_{in,ex,r}^* - E_{in,ex,\theta} E_{in,ex,\theta}^* - E_{in,ex,\phi} E_{in,ex,\Phi}^*)&\\
\tau_{in,ex,\theta}=&\frac{1}{4}(E_{in,ex,r}^* E_{in,ex,\theta} + E_{in,ex,r} E_{in,ex,\theta}^*)&\\
\tau_{in,ex, \Phi}=&\frac{1}{4}(E_{in,ex,r}^* E_{in,ex,\Phi} + E_{in,ex,r} E_{in,ex,\Phi}^*)&
\end{align}
Here $*$ represents the complex conjugate. We consider only time independent part in this work since the electric time scales and the frequency of the applied field are considered to be much faster than the shear rate (typically $\omega=500$ Hz onwards).
\\

Since the electric stresses are calculated on an undeformed sphere, the net normal and tangential components of electric stresses at the vesicle surface are
\begin{align}
\ \tau_r=&\tau_{ex, r}-\epsilon_r \tau_{in, r}&\\
\ \tau_\theta=&\tau_{ex, \theta}-\epsilon_r \tau_{in, \theta}&\\
\ \tau_\Phi=&\tau_{ex, \Phi}-\epsilon_r \tau_{in, \Phi}&
\end{align} 
The final expressions for the above resultant stress components are provided in appendix-B in terms of $\alpha, \beta$. Deformation causing normal electric force is obtained by subtracting the isotropic part of the normal stress ($\tau_0$, provided in appendix-B) from $\tau_r$. Thus $\tau_n=\tau_r-\tau_0 Y_{00}(\theta, \Phi)$, where $Y_{00}(\theta, \Phi)=1/\sqrt{4\pi}$ and the isotropic part is given by
\begin{align}
\tau_0=\frac{\int_{\Phi=0}^{\Phi=2\pi}\int_{\theta=0}^{\theta=\pi}\tau_r \sin\theta Y_{00}(\theta, \Phi)}{\int_{\Phi=0}^{\Phi=2\pi}\int_{\theta=0}^{\theta=\pi} \sin\theta Y_{00}(\theta, \Phi)Y_{00}(\theta, \Phi)}
\end{align}
The resulting normal ($\bm{f_{n}^E}=\boldsymbol{\hat{e}_r}.\tau_n$) and tangential ($\bm{f_{t, \theta}^E}=\boldsymbol{\hat{e}_\theta}.\tau_{\theta}, \bm{f_{t, \Phi}^E}=\boldsymbol{\hat{e}_\Phi}.\tau_{\Phi}$) electric tractions are
\begin{align}
\ \bm{f_{n}^E}=&-N \left(3\alpha^2 \cos2\Phi \sin^2\theta+\frac{1}{2}(\alpha^2-2\beta^2)(1+3\cos2\theta)\right) \boldsymbol{\hat{e}_r}&\\
\ \bm{f_{t}^E}=&S \left(-\alpha^2 \sin\theta \sin2\Phi  \ \boldsymbol{\hat{e}_{\Phi}}+\frac{1}{2}\left(\alpha^2 (\cos2\Phi-1)+2\beta^2 \right)\sin2\theta \ \boldsymbol{\hat{e}_\theta} \right)&
\end{align}
where 
\begin{align}
\ N=&\frac{6 (\sigma_r^2 + (2 C_{mem} \epsilon_r+\epsilon_r^2+\sigma_r^2+ C_{mem}^2 (1 -2 \epsilon_r + \sigma_r^2)) \zeta^2 \omega^2 + (C_{mem}^2 (-1+\epsilon_r)^2+2 C_{mem} \epsilon_r + \epsilon_r^2) \zeta^4 \omega^4)}{16 (4 \sigma_r^2 + (4 (C_{mem} + \epsilon_r)^2 + 4 C_{mem}^2 \sigma_r + (2 + C_{mem})^2 \sigma_r^2) \omega^2 \zeta^2+ (2 \epsilon_r + C_{mem} (2 + \epsilon_r))^2 \omega^4 \zeta^4)}&\\
\ S=&\frac{9 C_{mem} \omega^2 \zeta^2 (C_{mem} (\epsilon_r - \sigma_r) - \sigma_r^2 - \epsilon_r^2 \omega^2 \zeta^2)}{4 (4 \sigma_r^2 + (4 (C_{mem} + \epsilon_r)^2 + 4 C_{mem}^2 \sigma_r + (2 + C_{mem})^2 \sigma_r^2) \omega^2 \zeta^2+ (2 \epsilon_r + C_{mem} (2 + \epsilon_r))^2 \omega^4 \zeta^4)}&      
\end{align}
Therefore the total electric traction is $\bm{f_{tot}^E}=\bm{f_n^E}+\bm{f_t^E}$. \\

In the specific case of $\alpha=1, \beta=0$ (for Y-directional electric field), the total membrane traction can be expressed in terms of vector spherical harmonics by using the identities provided in appendix-C as
\begin{align}
\ \bm{f_{tot}^{E}}=-\frac{N}{2}\left(8 \sqrt{\frac{\pi}{5}} \boldsymbol{y_{202}} + 6 \sqrt{\frac{8\pi}{15}} (\boldsymbol{y_{222}} + \boldsymbol{y_{2-22}})\right)-\frac{S}{2}\left(-\sqrt{\frac{32\pi}{15}} \boldsymbol{y_{200}}\right)+S \sqrt{\frac{4\pi}{5}} (\boldsymbol{y_{220}} + \boldsymbol{y_{2-20}})
\label{tautot}
\end{align}
Eq. \eqref{tautot} can be written in a more compact way as
\begin{align}
\ \bm{f^{E}_{tot}}=\bm{f_n^{E}}+\bm{f_t^{E}}=&\sum_{m=-2}^{m=2}\tau_{2m2}^E\bm{y_{2m2}}+\sum_{m=-2}^{m=2}\tau_{2m0}^E\bm{y_{2m0}}& \nonumber\\
\ \bm{f^{E}_{tot}}=&(\tau_{2-22}^E\bm{y_{2-22}}+\tau_{202}^E\bm{y_{202}}+\tau_{222}^E\bm{y_{222}})+(\tau_{2-20}^E\bm{y_{2-20}}+\tau_{200}^E\bm{y_{200}}+\tau_{220}^E\bm{y_{220}})&
\end{align}
Such that the normal and tangential electric stresses are
\begin{align}
\tau_{2-22}^E=&-2 N \sqrt{\frac{6\pi}{5}}&
\tau_{202}^E=&-4 N \sqrt{\frac{\pi}{5}}&
\tau_{222}^E=&-2 N \sqrt{\frac{6\pi}{5}}&\\
\tau_{2-20}^E=&S \sqrt{\frac{4\pi}{5}}&
\tau_{200}^E=&S \sqrt{\frac{8\pi}{15}}&
\tau_{220}^E=&S \sqrt{\frac{4\pi}{5}}&
\end{align}

The net Maxwell stress is $\tau^{E}=\sum_{l=2}\sum_{m=-l}^{l} (\tau_{lm0}^{E} \boldsymbol{y_{lm0}}+\tau_{lm2}^{E} \boldsymbol{y_{lm2}})$, where $\tau_{lm0}^{E}$ and $\tau_{lm2}^{E}$ are tangential and normal electric stresses, respectively. Here another component of tangential stress $\sum_{l=2}\sum_{m=-l}^{l} \tau_{lm1}^{E} \boldsymbol{y_{lm1}}$ is not taken into account since it  turns out to be zero.\\

\textbf{Membrane mechanics with higher order corrections}\\

The surface of a slightly deformed vesicle is described by 
\begin{align}
\ r_s=\alpha+\sum_{l=0}^{\infty}\sum_{m=-l}^{l}f_{lm}Y_{lm}
\end{align}
where $\alpha$ is obtained by volume conservation constraint $\int r_s^3 \sin\theta d\theta d\phi=4\pi$ which gives $\alpha=1-\sum_{l=0}^{\infty}\sum_{m=-l}^{l}f_{lm}Y_{lm}/(4\pi)$ while the constraint of area conservation  is $\int r_s^2/(\boldsymbol{\hat{e}_r.n}) \sin\theta d\theta d\phi=4\pi+\triangle$ and leads to an excess area stored in the vesicle in the deformed state $\triangle= \sum_{l=0}^{\infty}\sum_{m=-l}^{l}\frac{(l+2)(l-1)}{2}f_{lm}Y_{lm}$\\

The membrane stress on the vesicle can be written as
\begin{align}
\tau_{mem}=(-2\kappa_b (2H^3-2KH+\nabla^2H)+2\sigma H)\boldsymbol{\hat{e}_r}-\nabla_s \sigma
\end{align}
where $\sigma=\sigma_0+\sum_{l=0}^{\infty}\sum_{m=-l}^{l} \sigma_{lm} Y_{lm}$ with $\sigma_0$ is uniform membrane tension and $\sigma_{lm}$ is nonuniform membrane tension which varies along the vesicle surface. On taking curvature terms up to second order in spherical harmonics, the mean curvature is given by
\begin{align}
\ H=&1+\frac{1}{2}(l(l+1)-2)F(\theta,\phi)-(l(l+1)-1)F(\theta,\phi)^2&
\label{curvH} 
\end{align}
and the Gaussian curvature by,
\begin{align}
\ K= &1+(l(l+1)-2)F(\theta,\phi)-3(l(l+1)-1)F(\theta,\phi)^2 -l(l+1)F(\theta,\phi)\left(-l(l+1)F(\theta,\phi)-\frac{d^2F(\theta,\phi)}{d\theta^2}\right)& \nonumber \\
&-\cot^2\theta \csc^2\theta \left(\frac{dF(\theta,\phi)}{d\phi}\right)^2+2\cot\theta \csc^2\theta
 \frac{dF(\theta,\phi)}{d\phi}\frac{d^2F(\theta,\phi)}{d\theta d\phi}-\left(-l(l+1)F(\theta,\phi)-\frac{d^2F(\theta,\phi)}{d\theta^2}\right)^2& \nonumber \\
  &-\csc^2\theta \left(\frac{d^2F(\theta,\phi)}{d\theta d\phi}\right)^2 &
  \label{curvK}
\end{align}
where $F(\theta,\phi)=\sum_{l=2}\sum_{m=-l}^{l} f_{lm} Y_{lm}(\theta, \phi)$ \\

Using these two curvatures ($H$ and $K$) all other linear/non-linear terms in membrane stress can be obtained such that the resulting tangential and normal membrane stresses respectively with higher order correction terms are respectively 
\begin{align}
\tau_{lm0}^{mem}=&\nabla_s \sigma=-Ca^{-1}\sqrt{l(l+1)}\sum_{l=2}\sum_{m=-l}^{l} \sigma_{lm} Y_{lm} &
\label{tmem}
\end{align}
and 
\begin{align}
\tau_{lm2}^{mem}=&-2\kappa_b (2H^3-2KH+\nabla^2H)+2\sigma H&
\label{nmem}
\end{align}
Substitution of the curvature terms (eq. \ref{curvH} and eq. \ref{curvK}) in the eq. \ref{nmem} gives normal membrane stress for $m=-2, 0, 2$ modes as
\begin{align}
\tau_{2-22}^{mem}=&Ca^{-1} \left(\frac{1}{14 \pi} (144 \sqrt{5\pi} f_{20} f_{2-2} + 40 \sqrt{5\pi} f_{20} f_{2-2} \sigma_0) + 4 f_{2-2} (6 + \sigma_0) + 2 \sigma_{2-2}\right) \boldsymbol{y_{2-22}}& \label{nmem1}\\
\tau_{202}^{mem}=&Ca^{-1} \left(\frac{1}{14 \pi} ((144 \sqrt{5\pi} + 
40 \sqrt{5\pi} \sigma_0) f_{2-2} f_{22} - (4 \sqrt{5\pi} (18 + 5 \sigma_0)) f_{20}^2) + 4 (6 + \sigma_0) f_{20} + 2 \sigma_{20}\right) \boldsymbol{y_{202}}& \label{nmem2}\\
\tau_{222}^{mem}=&Ca^{-1} \left(\frac{1}{14 \pi} (144 \sqrt{5\pi} f_{20} f_{22} + 40 \sqrt{5\pi} f_{20} f_{22} \sigma_0) + 4 f_{22} (6 + \sigma_0) + 2 \sigma_{22}\right) \boldsymbol{y_{222}}& \label{nmem3}
\end{align}

\textbf{Overall stress balance}\\

The overall tangential stress balance across the vesicle is given by
\begin{align}
&\ (\tau_{lm0}^{h, ex}-\eta \tau_{lm0}^{h, in})+Mn \tau_{lm0}^E=\tau_{lm0}^{mem} & \label{tstrbal}
\end{align}
Similarly the overall normal stress balance is
\begin{align}
&\ (\tau_{lm2}^{h, ex}-\eta \tau_{lm2}^{h, in})+Mn \tau_{lm2}^E=\tau_{lm2}^{mem} & \label{nstrbal}
\end{align}
The tangential stress balance is used to obtain the non-uniform tension terms ($\sigma_{2-2}, \sigma_{20}, \sigma_{22}$) whereas the normal stress balance gives normal velocity component ($C_{2-22}, C_{202}, C_{222}$), provided in appendix-D. The expressions are derived using the higher order theory for membrane forces.\\

\textbf{Dynamic evolution equation with AC-Shear coupling}\\

With higher order corrections in the membrane stress the evolution equations for different deformation modes when a vesicle is subjected to pure shear flow under electric field is given by \cite{Vla2011ShInt}
\begin{align}
\ \frac{df_{2m}}{dt}= i \frac{m}{2}f_{2m}+C_{2m2}
\label{evolAC}
\end{align}
where $m=-2, 0, 2$ and $C_{2m2}$ is the normal component of membrane velocity due to contribution from both pure shear flow and AC electric field induced stresses (detail provided in appendix-E)
\begin{align}
\ C_{2-22}=&C_{2-2}^{Sh}-24 \left(\frac{7\pi (6+\sigma_0)+\sqrt{5\pi}(18+5\sigma_0)f_{20}}{7\pi (32+23\eta)Ca}\right)f_{2-2}+Mn C_{2-2}^{el}&\\
\ C_{202}=&C_{20}^{Sh}+\left(\frac{-168\pi (6+\sigma_0)f_{20}+12\sqrt{5\pi}(18+5\sigma_0)(f_{20}^2-2f_{22}f_{2-2})}{7\pi (32+23\eta)Ca}\right)+Mn C_{20}^{el}&\\
\ C_{222}=&C_{22}^{Sh}-24 \left(\frac{7\pi (6+\sigma_0)+\sqrt{5\pi}(18+5\sigma_0)f_{20}}{7\pi (32+23\eta)Ca}\right)f_{22}+Mn C_{22}^{el}&
\end{align}
here $C_{2m}^{sh}=-i m \left(\frac{2\sqrt{30\pi}}{32+23\eta}\right)$ is the contribution from pure shear part, and $C_{2m}^{el}$ represents electrostatic contribution. $Mn$ is measure of applied electric field strength relative to the applied shear.\\

In terms of inclination angle ($\psi$), the amplitude of deformation modes can be represented as $f_{22}=R e^{-2i \psi}$ and $f_{2-2}=R e^{2i \psi}$ with $R=\sqrt{A} \cos\theta/2$ which measures the deformation. Separation of real and imaginary parts gives the evolution equations for the inclination angle and the vesicle shape. 
\begin{align}
\label{evoleqn}
\frac{d\psi}{dt}=&\frac{4\sqrt{30\pi}}{(32+23\lambda)}\frac{\cos2\psi}{\cos\theta}-\frac{1}{2}-Mn X_1&\\ \nonumber
\frac{d\theta}{dt}=&-\frac{8\sqrt{30\pi}}{\sqrt{A}(32+23\eta)}\sin\theta \sin2\psi+\frac{720 \sqrt{A}}{7\sqrt{10\pi} (32+23\eta)Ca} \cos3\theta+Mn X_2&
\end{align}
where additional terms due to electrostatic contribution are
\begin{align}
\ X_1=&\frac{\sin2\psi}{2\cos\theta}(14\sqrt{\pi}(3C_{22}^{el}+C_{2-2}^{el}+(-C_{22}^{el}+C_{2-2}^{el})\cos2\theta)+5\sqrt{10}\sqrt{A}(C_{22}^{el}+C_{2-2}^{el} &\nonumber \\&
+(3C_{22}^{el}+C_{2-2}^{el})\cos2\theta)\sin\theta)/(28\sqrt{A}\sqrt{\pi}+5\sqrt{10}A \sin3\theta)&\\
X_2=&\sqrt{\frac{2}{A}}\left(C_{20}^{el}\cos\theta+\frac{1}{2\sqrt{2}}(-3C_{22}^{el}+C_{2-2}^{el}+(C_{22}^{el}+C_{2-2}^{el})\cos2\theta)\cos2\psi/\sin\theta \right)&
\end{align}
here the electrostatic contribution $C_{2m}^{el}=(6 \tau_{2m2} + 2 \sqrt{6} \tau_{2m0})/(32 + 23 \eta)$.

In equation \ref{evoleqn}, the left hand side term represents the torque due to the rotating vesicle. On the right hand side, the first term is the torque due to the elongational part which tries to align the vesicle in the principal direction of the strain rate tensor, the second term is the clockwise torque due to the rotational part of the mean flow, while the third part is the electrical torque. \\

 Only the $l=2$ modes are considered in the analysis since they are excited by the electric and hydrodynamic stresses, and lead to deformation. The symmetry in the problem implies that only  $m=-2,0, 2$ are admitted, where $m=\pm 2$ represents the components in the $X-Y$ plane and $m=0$ mode corresponds to the axisymmetric mode along the Z-axis. The evolution equations so obtained are highly non-linear, transcendental differential equations due to the coupling of higher order terms, and are therefore solved numerically using Mathematica 10. \\
 
\subsection{Results and discussion}
The physics of the tank-treading, trembling and tumbling regimes can be understood as follows. Consider a rigid spherical object in shear flow, the applied flow can be decomposed into a rotational part which tries to rotate the particle with an angular velocity equal to half the shear rate (for pure shear flow the extensional strain rate and the rotational vorticity is identical and equal to half the shear rate), unless it is at rest, and exerts a torque accordingly. The extensional flow part then generates a stresslet, which in a spherical particle can never  exert a net force or a torque due to symmetry. As a result, a torque free rigid sphere means the sphere rotates with the same angular velocity (half the applied shear rate), without offering any resistance, thereby leading to a torque balance. \\

When a rigid ellipsoidal particle is placed in shear flow, it continues to rotate with an angular velocity half the shear rate. However the extensional flow can exert a torque on an ellipsoidal particle, trying to align it in the direction of the extensional strain rate, that is  $45^{o}$ to the direction of flow. The dynamics of the orientation of an ellipsoidal particle can be given by 
\begin{equation} \frac{d \psi}{d t}= -A +B \cos(2 \psi),\end{equation} 
as suggested by ~\cite{Jeffery1922}. Turns out that for rigid ellipsoids, for reasons to be discussed later, the magnitude of $A$ is always greater than or equal to $B$, thereby a steady state solution is never obtained. Thus an ellipsoidal particle rotates (called tumbling in the vesicles/cells literature) when placed in a shearing flow, since it can not simultaneously admit a tangential velocity in the rotating reference frame of the particle.\\
 \begin{figure}[tp]
 \centering
  \includegraphics[width=0.42\linewidth]{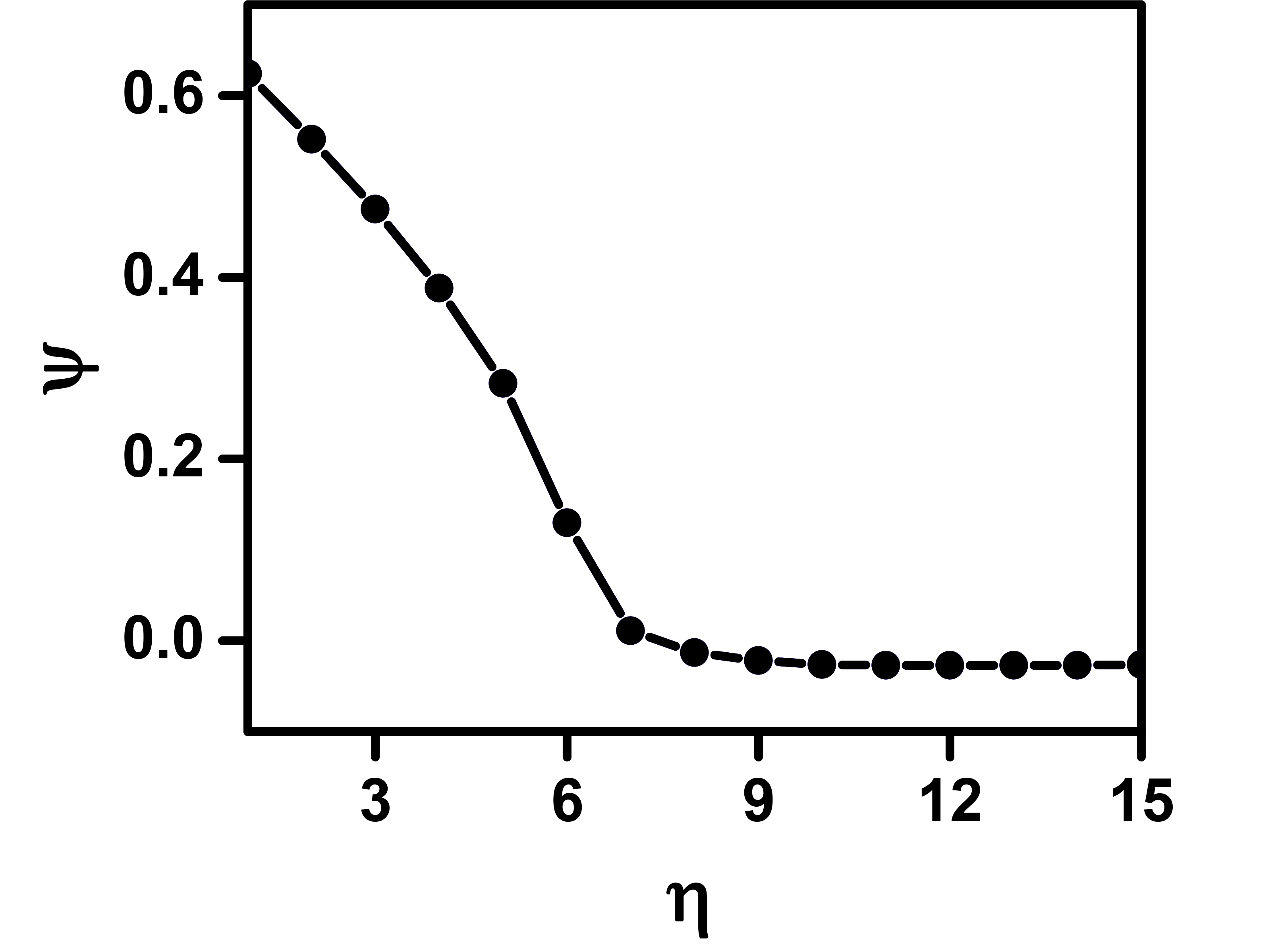}  \includegraphics[width=0.4\linewidth]{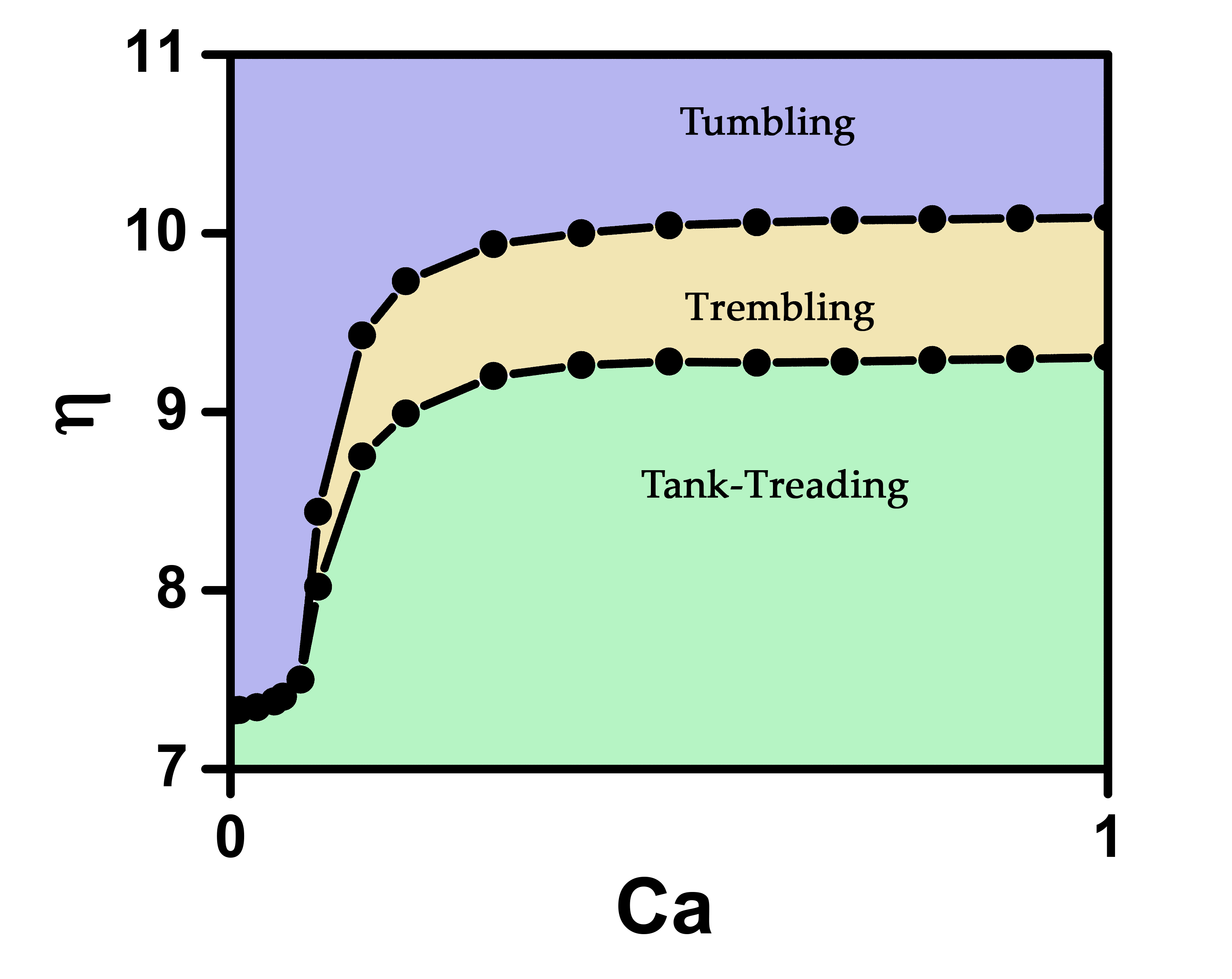}
	\caption{Pure shear: (a) Tank treading angle vs viscosity ratio and (b) the phase diagram showing different regimes for viscosity vs capillary number, in the absence of electric field ($Mn=0, Ca=1, \triangle=0.2$)}
	\label{Mneq0}
\end{figure} 

\begin{figure} [tp] 
\centering
    \hspace{0.12cm}
 \begin{subfigure}[b]{0.485\linewidth}\includegraphics[width=\linewidth]{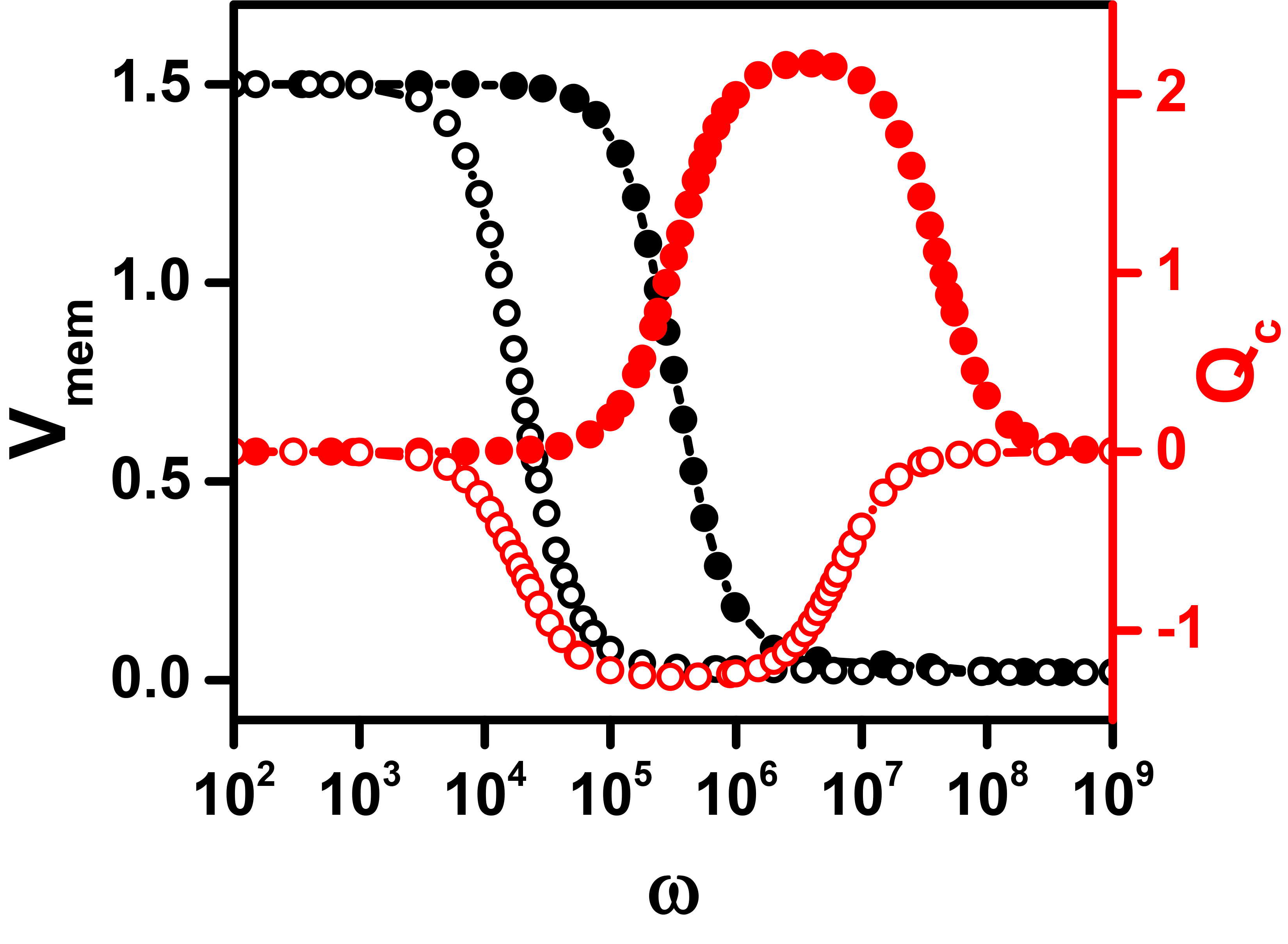}
  \caption{}
 \end{subfigure}
     \hspace{0.12cm}
 \begin{subfigure}[b]{0.482\linewidth}
    \includegraphics[width=\linewidth]{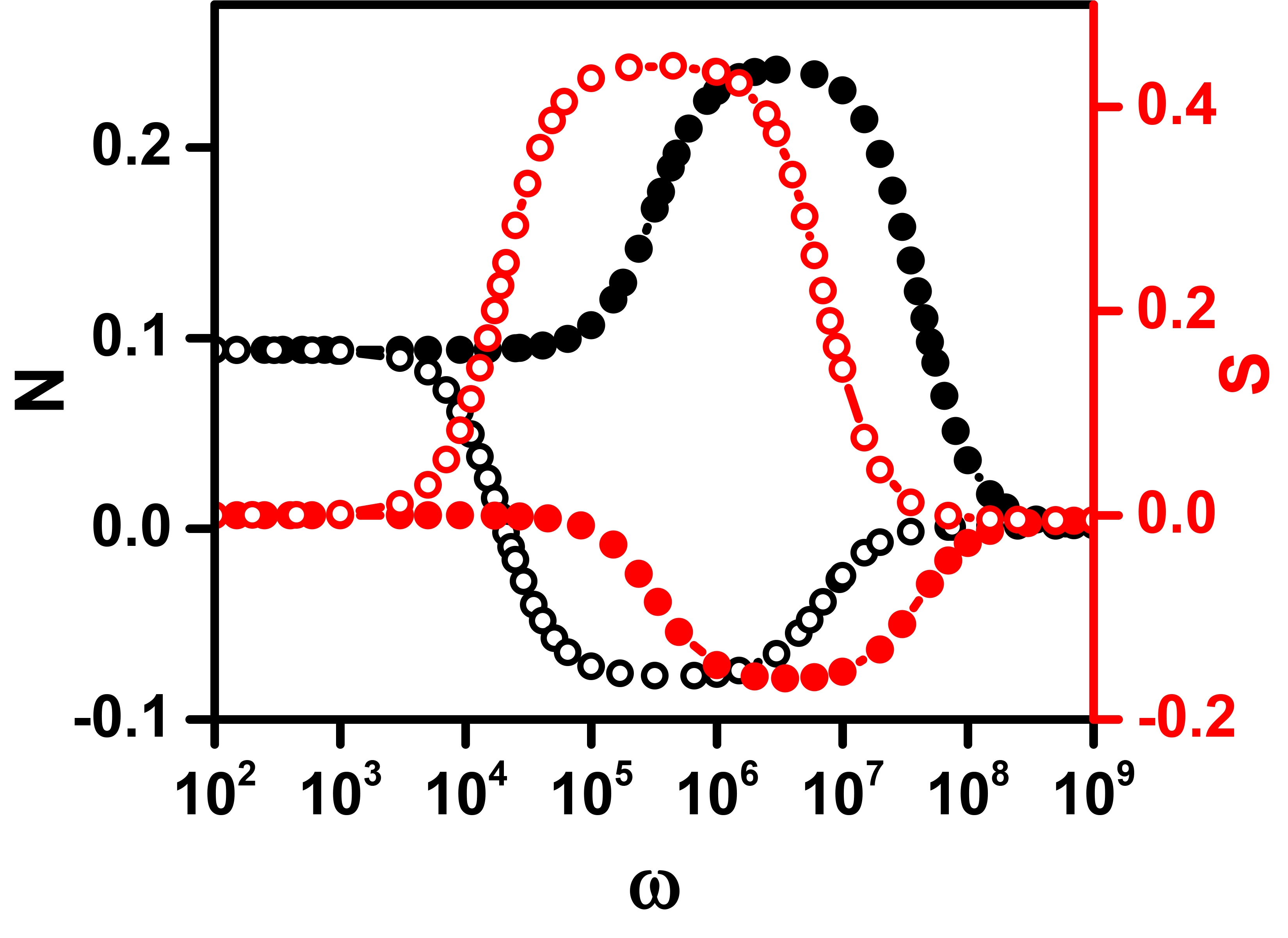}
      \caption{}
  \end{subfigure} 
 \caption{(a) Transmembrane potential ($V_{mem}$, black lines) and interface charge ($Q_c$, red lines) variation with frequency for $\sigma_r=10$ (solid) and $\sigma_r=0.1$ (hollow), (b) Normal stress ($N$, black lines) and tangential stress ($S$, red lines) variation with frequency for $\sigma_r=10$ (solid) and $\sigma_r=0.1$(hollow). ($\zeta=10^{-7}, C_{mem}=50, \epsilon_r=1$)}
    \label{VmStr}
\end{figure}

\begin{figure} [tp] 
\centering
	\hspace{0.12cm}
	\begin{subfigure}[b]{0.48\linewidth}
\includegraphics[width=\linewidth]{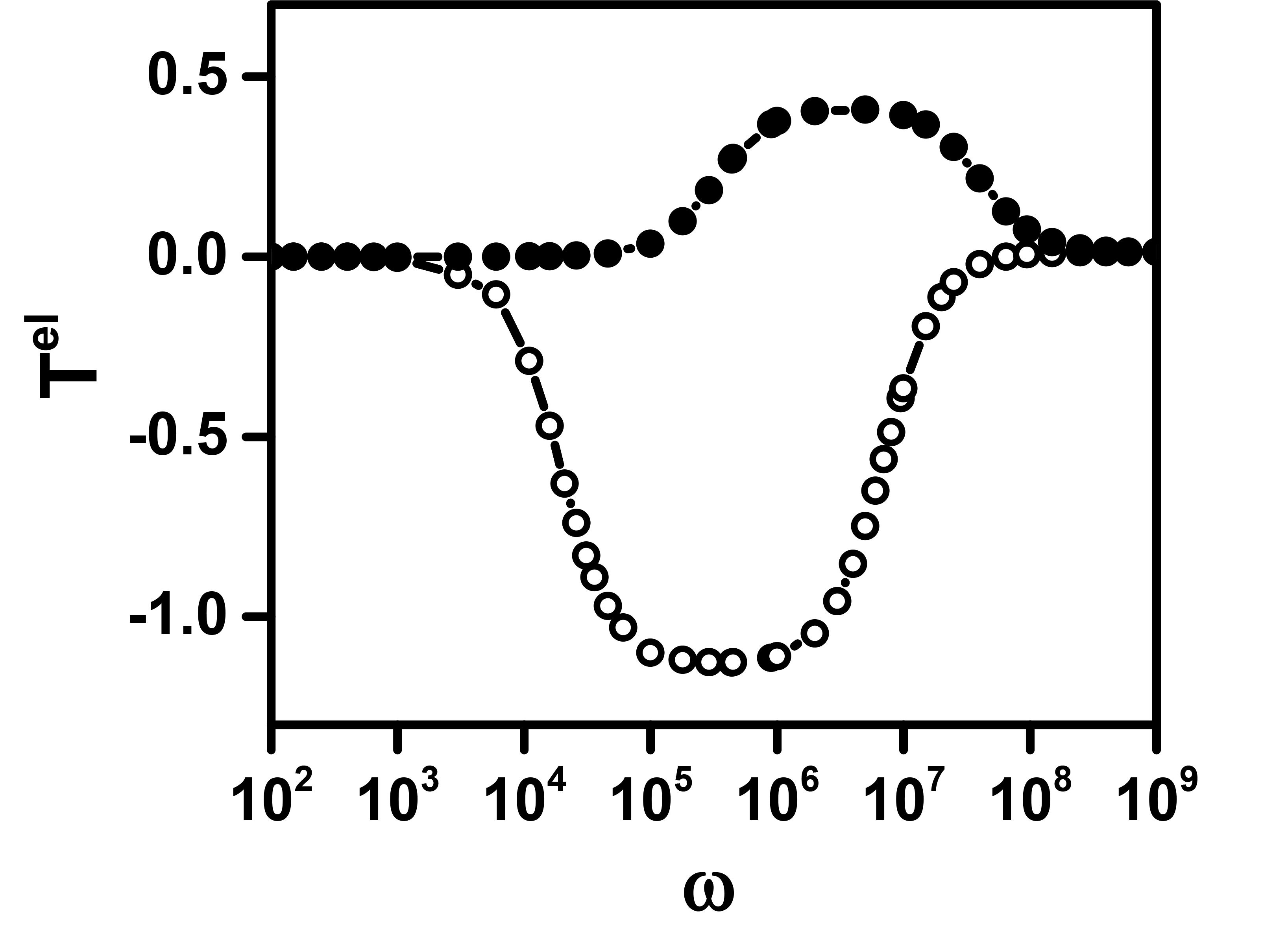}
		\caption{}
	\end{subfigure} 
	\hspace{0.12cm}
	\begin{subfigure}[b]{0.45\linewidth}
	\includegraphics[width=\linewidth]{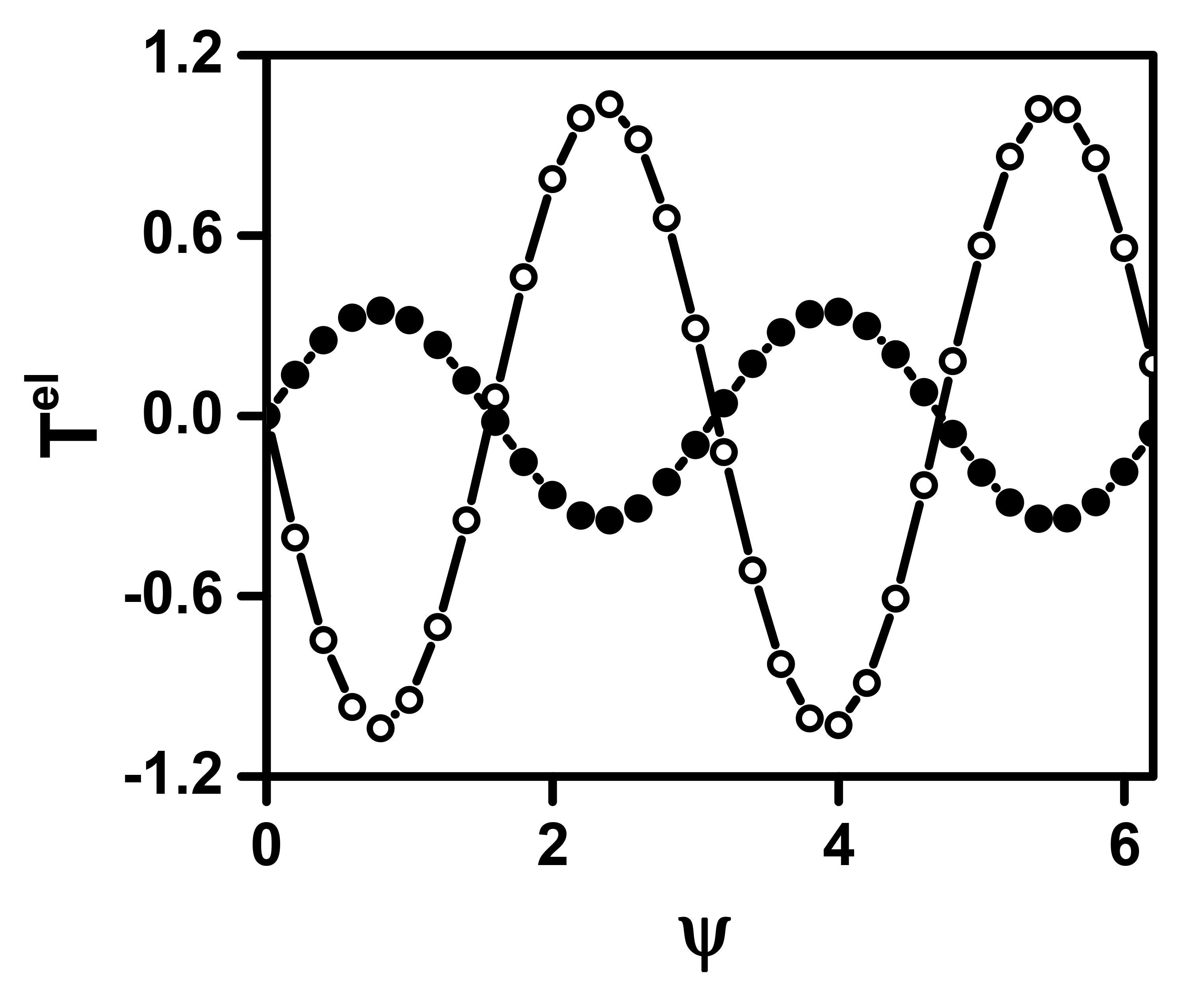}
		\caption{}
	\end{subfigure}    
	\caption{Electric torque ($T^{el}$) on a deformed vesicle in TT state, (a) variation of electric torque with frequency for $\sigma_r>1$, $\sigma_r<1$ (b) electric torque at different inclination angles at $Mn=10$,  $\omega=10^6$ for $\sigma_r>1$ and $\sigma_r<1$; ($\sigma_r=10$ (solid spheres), $\sigma_r=0.1$ (hollow spheres), $\eta=3, Ca=1, \zeta=10^{-7},\epsilon_r=1, \triangle=0.2, C_{mem}=50$)}
	\label{torque}
\end{figure}

Unlike a rigid ellipsoid, a deformable drop placed in a shear flow, can allow a finite tangential velocity of the interface (surface), called the tank-treading velocity, even in the  reference frame of the rotating drop \cite{rumscheidt1961particle}. This leads to $A$ being less than $B$ and a steady value of $\psi$ the orientation angle is observed. In this case, the torque exerted by the shear is partially balanced by the restoring torque of the elongational flow that aligns it at an angle of $45^0$ to the direction of flow, and partially by the interface velocity in the rotating reference frame. \\

The case of a vesicle in shear flow is similar to that of a drop, the tangential velocity of the membrane (tank-treading velocity), helps in reducing the torque due to the applied shear, which is then balanced by the torque due to the elongational part of the shear flow. The torque exerted by the applied shear on the deformed ellipsoidal vesicle, increases with an increase in the inner viscosity of the vesicle, such that the angle of inclination decreases from $\pi/4$ to $0$, with an increase in the inner viscosity. The case of a rigid particle can be envisaged as a limiting case of inner viscosity tending to infinity (Figure \ref{Mneq0}). With an increase in the inner viscosity, the torque due to elongational flow  weakens, and the rotational torque increases such that beyond a critical viscosity ratio, the vesicle no longer tank treads, but undergoes a tumbling transition. The tank treading velocity in the tumbling regime was shown by Skalak and Keller ~\cite{skalak1982ShInt} to vary as $\cos{2 \psi}$.  In the TT regime, the frequency of the tank treading velocity is constant such that the tangential velocity varies in the $\Phi$ direction ~\cite{skalak1982ShInt} for a given orientation angle $\psi$ . Thus a vesicle undergoes a tank treading to trembling to tumbling transition as the viscosity of the inner fluid is increased. \\

It is therefore of interest to understand the effect of electric field on the three dynamical modes of tank-treading, trembling  and tumbling. \\

The non-dimensional numbers used in this work were obtained using dimensional parameter values in a range in which most of the pure shear experiments are typically conducted ~\cite{Steinberg2005ShInt, Steinberg2006ShInt, Steinberg2009PRLShInt, Steinberg2009PNASShInt}. A vesicle of size $R_0=10\mu m$ is assumed to be suspended in an another leaky-dielectric fluid, the ratio of their inside/outside fluid properties are $\epsilon_r=1$, $\sigma_r=0.1, 10$, and $\eta=1-20$. These can be considered to represent electrical conductivities of the order of $1-100 \times 10^{-5}$ $S/m$, and viscosities varying from $1-100 \times 10^{-3}$ $Pa-sec$. Typical shear rates of $\dot{\gamma}=0.1 sec^{-1}$ and frequency of the applied electric field could vary from $\omega=1 kHz-10 MHz$, with electric field strength ($E_0$) varied in the range of  $0.01-1kV/cm$. It is assumed that the membrane is insulating and its non-dimensional capacitance is $C_{mem}=(\epsilon_{mem}/h) (R_0/\epsilon_{ex})=50$, the charging of the membrane then takes place on a time scale of $t_{mem}=(R_{0}/\sigma_{ex}) C_{mem} (1/2 +1/\sigma_r)\sim 10^{-4}-10^{-5}$ sec while the Maxwell Wagner charge relaxation time is of the order of $t_{MW}=\epsilon_{ex}/\sigma_{ex} (2+\epsilon_r)/(2+\sigma_r)\sim 10^{-6}- 10^{-7}$ sec. The  flow capillary number $Ca=\dot{\gamma} \eta_{ex} R_0^3/\kappa_b$, and the electric Mason number $Mn=\epsilon_{ex} E_0^2/(\dot{\gamma}\eta_{ex})$ take values in the range of $0.1-5.0$ and $0.1-100$, respectively. The analysis is presented for two sets of conductivity ratios $\sigma_r=0.1,10$, representing relative conductivities of the inner fluid to be lower or higher than the outer, respectively. \\   
\begin{figure} [tp] 
\centering
    \hspace{0.12cm}
    \begin{subfigure}[b]{0.48\linewidth}       \includegraphics[width=\linewidth]{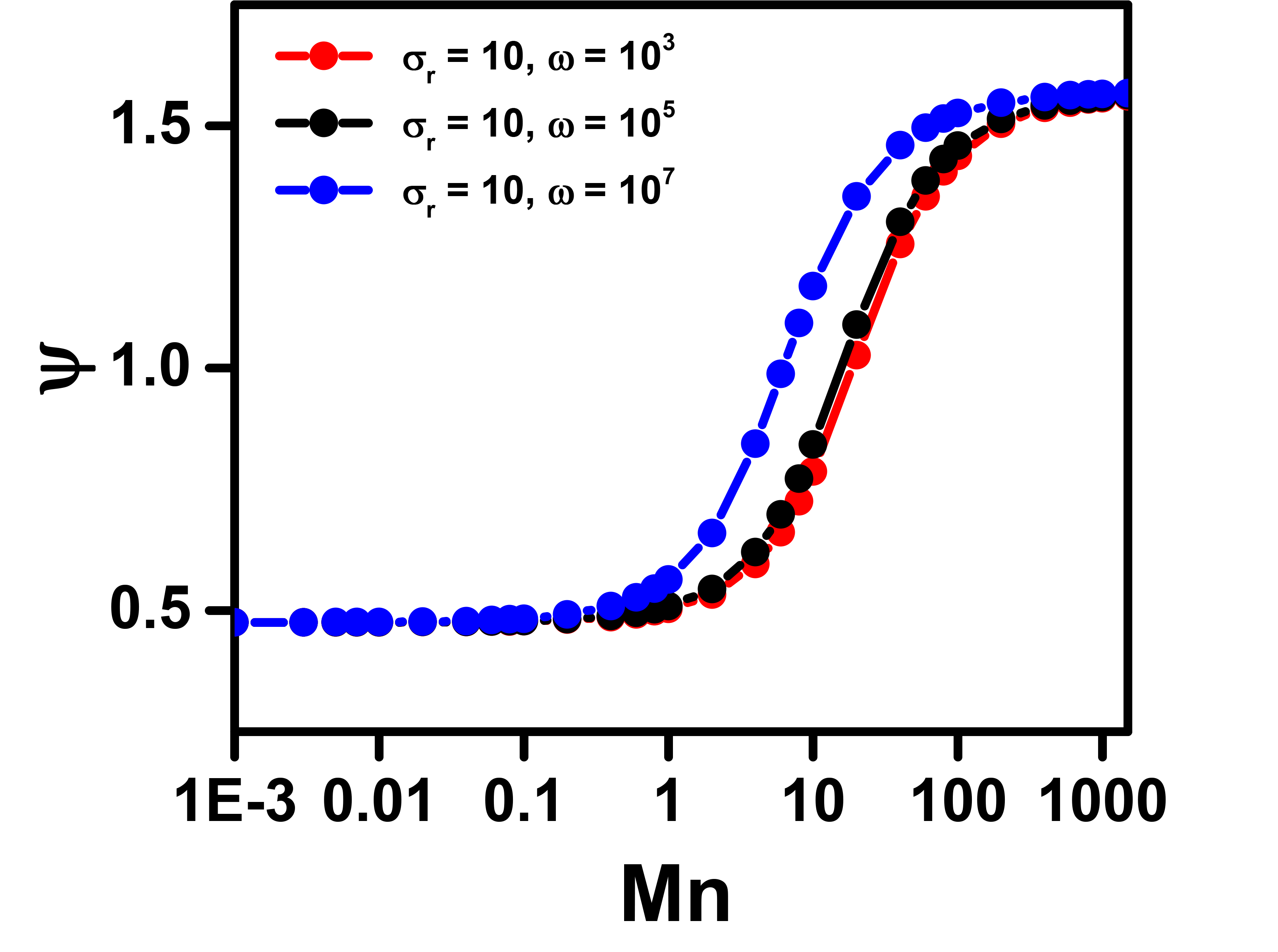}
     \caption{}
     \end{subfigure}
     \hspace{0.12cm}
      \begin{subfigure}[b]{0.48\linewidth}
      \includegraphics[width=\linewidth]{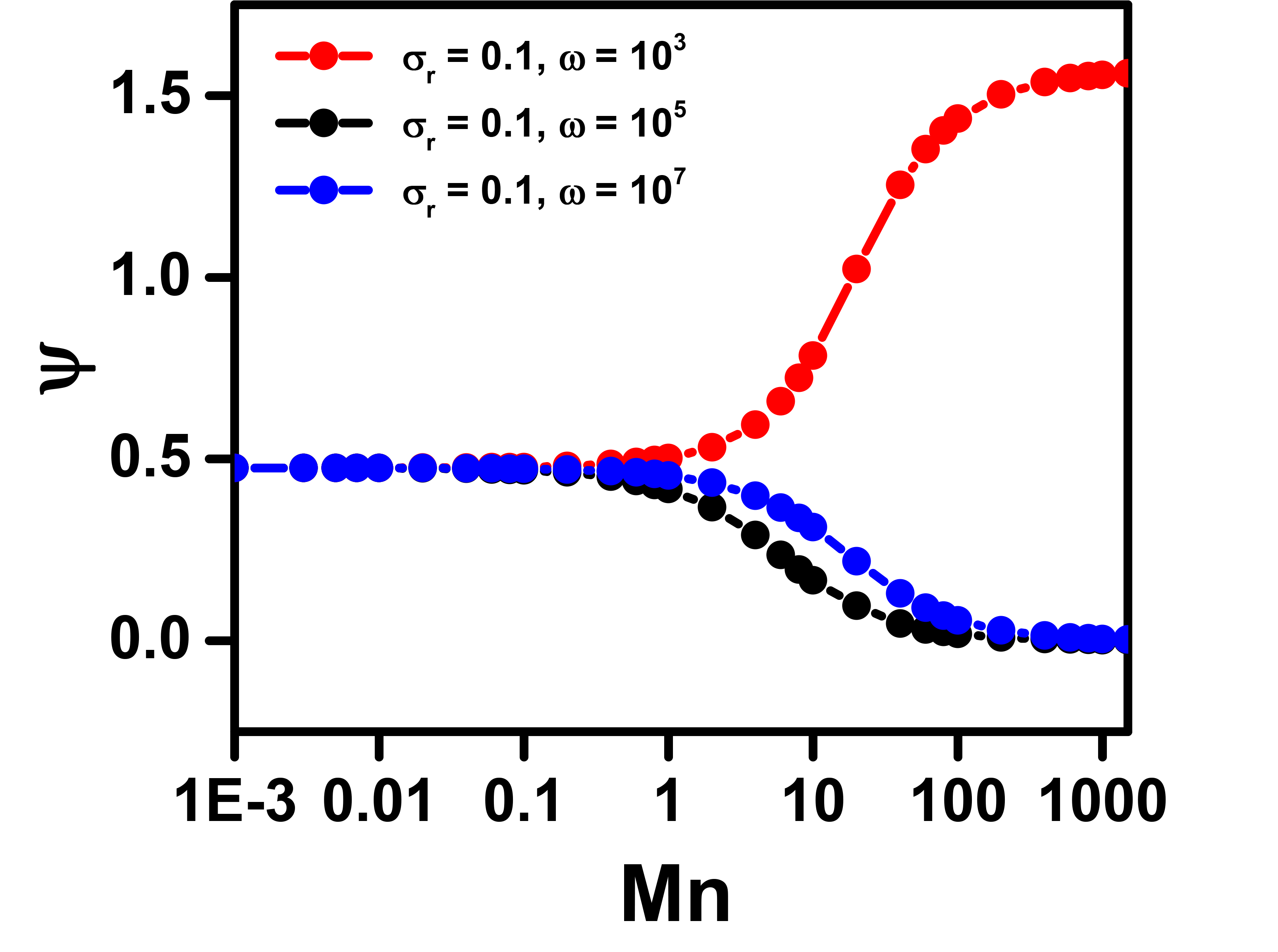}
      \caption{}
     \end{subfigure}
\hspace{0.12cm}
\begin{subfigure}[b]{0.5\linewidth}       \includegraphics[width=\linewidth]{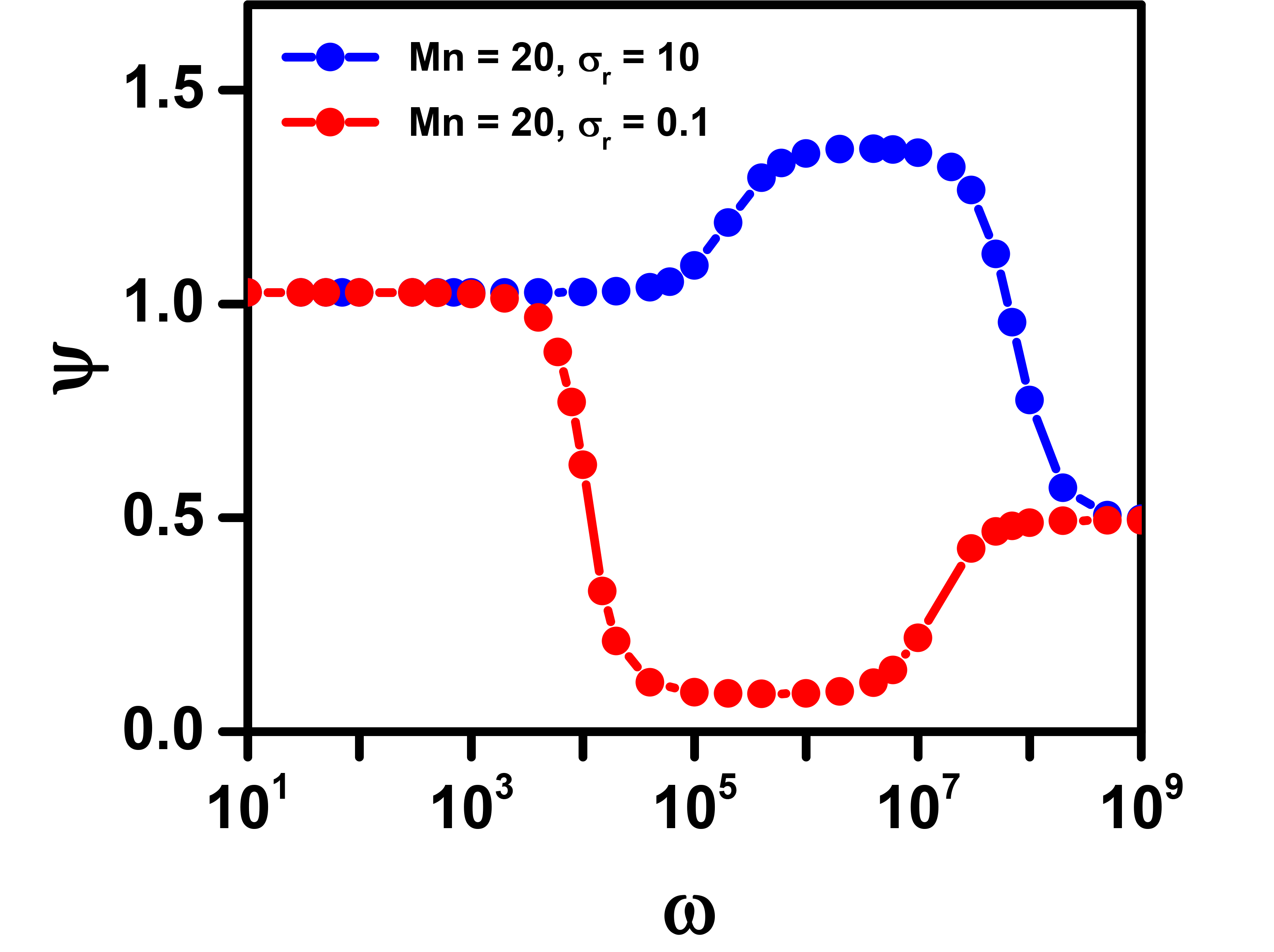}
	\caption{}
\end{subfigure}                      
 \caption{(a), (b): Variation of inclination angle with strength of applied electric field in TT regime for $\sigma_r=10$, and $\sigma_r=0.1$, respectively; (c) Variation of inclination angle with frequency of applied electric field in TT regime for $\sigma_r=10$, $\sigma_r=0.1$ at electric field strength $Mn=20$. ($Ca=1, \eta=3$).
}
\label{PsiVsMn}
\end{figure} 	
The variation of transmembrane potential ($V_{mem}$) with frequency is presented in figure \ref{VmStr}a. The figure shows that at very low-frequencies the transmembrane potential is maximum because of the high impedance of the capacitive membrane that prevents penetration of the field inside. The membrane in this case is fully charged and the net charge is zero on account of equal positive and negative charges on either side of the membrane. The variation of potential is $\sin{2 \Phi}$, and is thereby maximum, although of opposite signs at $\Phi=\pi/2$(positive) and $\Phi=-\pi/2$ (negative).   As the frequency increases beyond the $t_{mem}^{-1}$, the field penetrates the membrane, a fall in $V_{mem}$, and build up of a net positive ($\sigma_r>1$) or negative charge ( $\sigma_r<1$ ) at $\Phi=\pi/2$ at the membrane interface, akin to the case of a liquid drop in electric field is observed. At very high frequencies ($>t_{MW}^{-1}$) a perfect dielectric response of fluid as well as membrane is observed and $V_{mem}$ as well as net charge $Q_c$ (on account of zero absolute charge on each side of the bilayer) tend to zero. \\

The tangential stresses at low and high frequencies are independent of the conductivity ratio, and are zero for completely different reasons. At very high frequencies, the membrane is uncharged, while at very low frequencies, the normal electric field in the outer region, at the membrane interface vanishes, since the membrane acts like a perfect insulator with very high impedance. The normal stresses at very high  frequencies are independent of $\sigma_r$ and identically equal zero due to diminishing contrast of the electrical parameters. At very low frequencies, the normal electric stress is compressive at $\Phi=0$, diminishing towards $\Phi=\pi/2,-\pi/2$. This can be decomposed into an isotropic compressive pressure and a tensile force which is maximum at $\Phi=\pi/2$ and $-\pi/2$. At intermediate frequencies ($>t_{mem}^{-1}$ and  $< t_{MW}^{-1}$), depending upon the value of $\sigma_r$, the electric field acting upon the net charge accumulated at the membrane of a vesicle can result in tangential stresses that can act from $\Phi=0$ to $\Phi=\pi/2$ ($\sigma_r>1$) or from $\Phi=\pi/2$ to $\Phi=0$ ( $\sigma_r<1$). Thus, when $\sigma_r>1$ the net tensile force on the vesicle deforms it into a prolate shape while compressive stress acts cause  oblate deformation (figure \ref{VmStr}b) when $\sigma_r<1$.\\

\begin{figure} [tp] 
\centering
    \hspace{0.12cm}
    \begin{subfigure}[b]{0.25\linewidth}       \includegraphics[width=\linewidth]{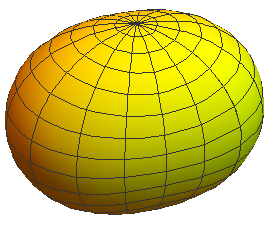}
     \caption{}
     \end{subfigure}
     \hspace{0.12cm}
      \begin{subfigure}[b]{0.25\linewidth}
      \includegraphics[width=\linewidth]{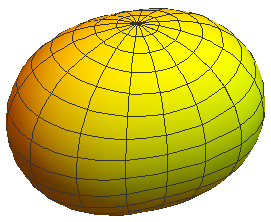}
      \caption{}
     \end{subfigure} 
 \hspace{0.12cm}
      \begin{subfigure}[b]{0.25\linewidth}
      \includegraphics[width=\linewidth]{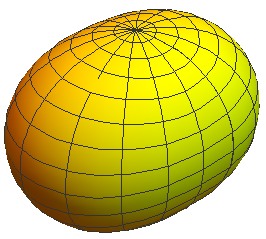}
      \caption{}
     \end{subfigure} 
    \hspace{0.12cm}
    \begin{subfigure}[b]{0.25\linewidth}       \includegraphics[width=\linewidth]{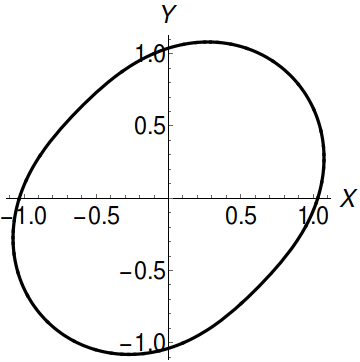}
     \caption{}
     \end{subfigure}
     \hspace{0.12cm}
      \begin{subfigure}[b]{0.25\linewidth}
      \includegraphics[width=\linewidth]{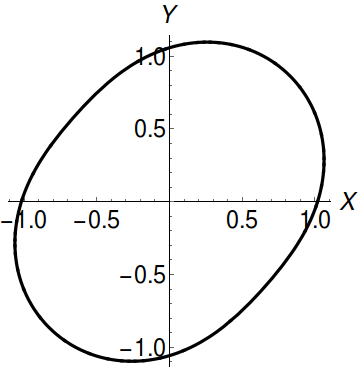}
      \caption{}
     \end{subfigure} 
 \hspace{0.12cm}
      \begin{subfigure}[b]{0.25\linewidth}
      \includegraphics[width=\linewidth]{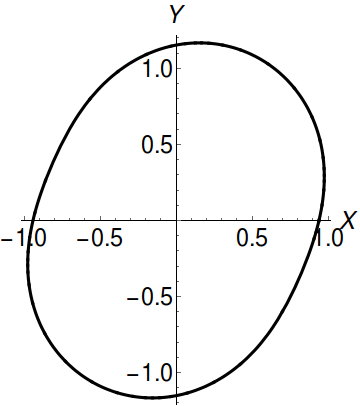}
      \caption{}
     \end{subfigure} 
     \hspace{0.12cm}
     \begin{subfigure}[b]{0.25\linewidth}       \includegraphics[width=\linewidth]{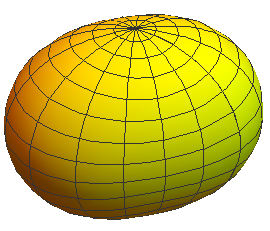}
      \caption{}
      \end{subfigure}
      \hspace{0.12cm}
       \begin{subfigure}[b]{0.25\linewidth}
       \includegraphics[width=\linewidth]{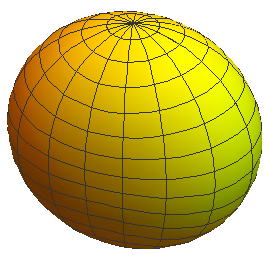}
       \caption{}
      \end{subfigure} 
  \hspace{0.12cm}
       \begin{subfigure}[b]{0.25\linewidth}
       \includegraphics[width=\linewidth]{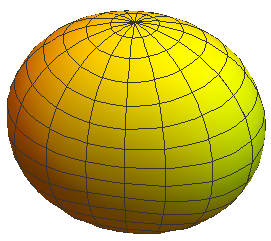}
       \caption{}
      \end{subfigure} 
    \hspace{0.12cm}
      \begin{subfigure}[b]{0.25\linewidth}       \includegraphics[width=\linewidth]{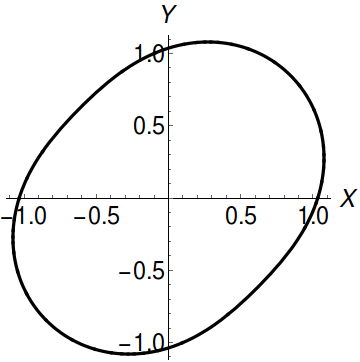}
       \caption{}
       \end{subfigure}
       \hspace{0.12cm}
        \begin{subfigure}[b]{0.25\linewidth}
        \includegraphics[width=\linewidth]{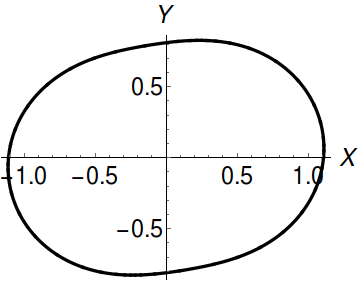}
        \caption{}
       \end{subfigure} 
   \hspace{0.12cm}
        \begin{subfigure}[b]{0.25\linewidth}
        \includegraphics[width=\linewidth]{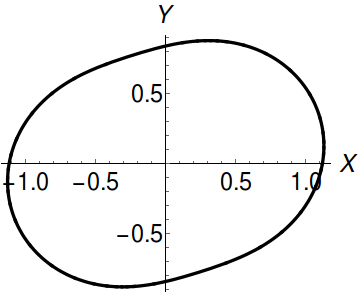}
        \caption{}
       \end{subfigure}                   
 \caption{Vesicle orientation under the effect of AC field in shear flow at fixed $Mn$. (a)-(f): $\sigma_r=10$, (g)-(l): $\sigma_r=0.1$. In each set $\omega$ increases as $10^3, 10^5, 10^7$ from left to right, $\eta=3, Mn=10, Ca=1, \epsilon_r=1, \triangle=0.2, C_{mem}=50$}
    \label{shapes1}
\end{figure}

\begin{figure} [tp] 
\centering
    \hspace{0.12cm}
    \begin{subfigure}[b]{0.25\linewidth}       \includegraphics[width=\linewidth]{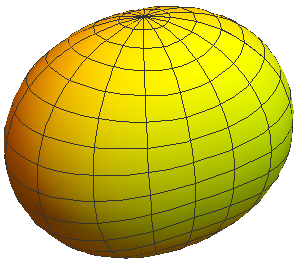}
     \caption{}
     \end{subfigure}
      \hspace{0.12cm}
       \begin{subfigure}[b]{0.25\linewidth}
    \includegraphics[width=\linewidth]{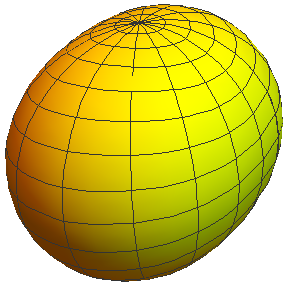}
           \caption{}
          \end{subfigure} 
      \hspace{0.12cm}
       \begin{subfigure}[b]{0.25\linewidth}
       \includegraphics[width=\linewidth]{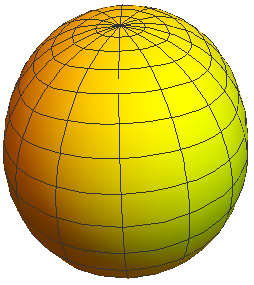}
       \caption{}
      \end{subfigure}          
     \hspace{0.12cm}
      \begin{subfigure}[b]{0.25\linewidth}
      \includegraphics[width=\linewidth]{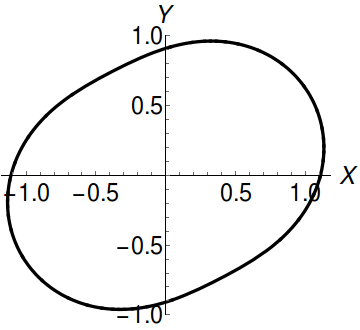}
      \caption{}
     \end{subfigure} 
    \hspace{0.12cm}
    \begin{subfigure}[b]{0.25\linewidth}       \includegraphics[width=\linewidth]{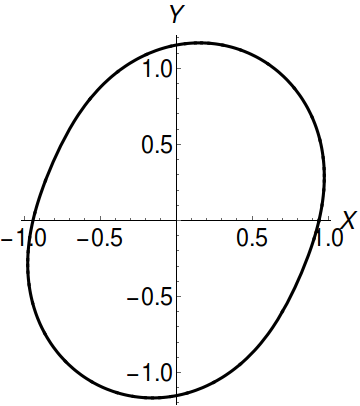}
     \caption{}
     \end{subfigure}
 \hspace{0.12cm}
      \begin{subfigure}[b]{0.25\linewidth}
      \includegraphics[width=\linewidth]{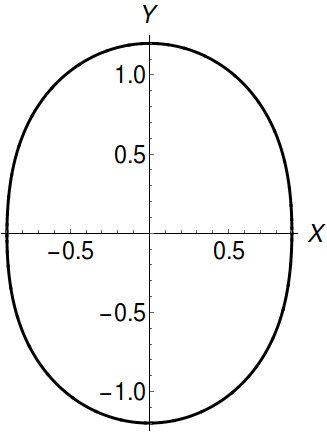}
      \caption{}
     \end{subfigure} 
     \hspace{0.12cm}
     \begin{subfigure}[b]{0.25\linewidth}       \includegraphics[width=\linewidth]{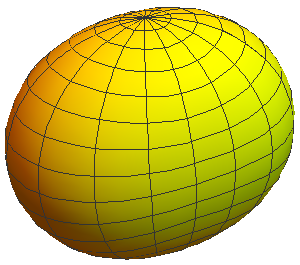}
      \caption{}
      \end{subfigure}
    \hspace{0.12cm}
         \begin{subfigure}[b]{0.25\linewidth}
         \includegraphics[width=\linewidth]{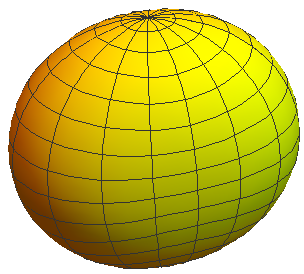}
         \caption{}
        \end{subfigure} 
        \hspace{0.12cm}
         \begin{subfigure}[b]{0.25\linewidth}
         \includegraphics[width=\linewidth]{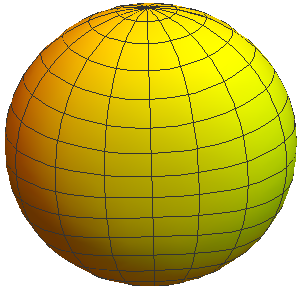}
         \caption{}
        \end{subfigure}            
      \hspace{0.12cm}
       \begin{subfigure}[b]{0.25\linewidth}
       \includegraphics[width=\linewidth]{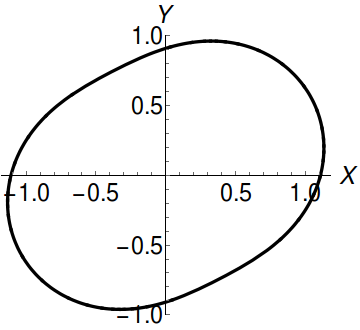}
       \caption{}
      \end{subfigure} 
    \hspace{0.12cm}
      \begin{subfigure}[b]{0.25\linewidth}       \includegraphics[width=\linewidth]{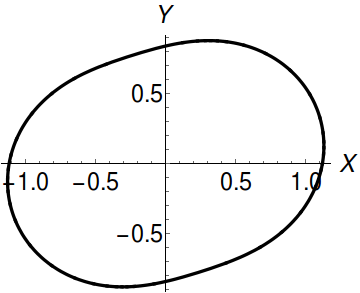}
       \caption{}
       \end{subfigure}
   \hspace{0.12cm}
        \begin{subfigure}[b]{0.25\linewidth}
        \includegraphics[width=\linewidth]{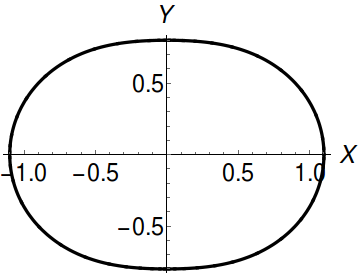}
        \caption{}
       \end{subfigure}                   
 \caption{Vesicle orientation under the effect of AC field in shear flow at fixed $\omega$. (a)-(f): $\sigma_r=10$, (g)-(l): $\sigma_r=0.1$. In each set $Mn$ increases as $0.001, 10, 1000$ from left to right, $\eta=3, \omega=10^{7}, Ca=1, \zeta=10^{-7}, \epsilon_r=1, \triangle=0.2, C_{mem}=50$}
    \label{shapes2}
\end{figure}
Thus in general, one can observe conductivity ratio $\sigma_r$ dependent behavior at intermediate frequencies whereas the behavior should be independent of $\sigma_r$ for very high and very low frequencies.\\

\subsubsection{Tank treading regime}
A vesicle, under linear shear flow, with viscosity ratio less than a critical value, gets deformed into an ellipsoidal shape and orients itself by making an angle with the direction of shear flow. The vesicle shape remains unchanged at that inclination angle while the membrane undergoes a continuous rotation around its fixed shape, called tank-treading.  The inclination angle is modified by the application of an AC electric field, when a vesicle is in the TT regime. The inclination angle $\psi$ increases with respect to the direction of flow and reaches its maximum value of $\psi=\pi/2$ when $\sigma_r>1$ or can decrease to zero  when $\sigma_r<1$, especially in the intermediate frequency range. The dynamics is decided by the relative magnitudes of the hydrodynamic, electric, and membrane torques. The electric torque can depend upon the frequency, electric field and the conductivity ratio.\\

Figure \ref{torque}(a) shows the variation of the total electric torque with frequency at $\psi=\pi/4$ in the first quadrant (in X-Y plane, $0<\psi<\pi/2$)   for the two conductivity ratios, where the torque is calculated on a deformed sphere oriented such that $0<\psi<\pi/2$), and the electrostatics is determined on an undeformed sphere. The torque is always anticlockwise (positive) in the first quadrant at all frequencies for $\sigma_r>1$, whereas it is clockwise (negative) at intermediate frequencies for $\sigma_r<1$. Figure \ref{torque}(b) shows that at an intermediate frequency {($10^3<\omega<10^5$)}, the torque in the first quadrant ($0<\psi<\pi/2$) is clockwise (negative) for $\sigma_r<1$, indicating a tendency to rotate the vesicle along the $X$ direction. On the other hand for $\sigma_r>1$, the torque in the first quadrant is anticlockwise (positive), thereby rotating the vesicle towards the $Y$ axis. The stability of the stationary point $\psi=0$ (X axis) for $\sigma_r<1$ can be seen from the anticlockwise torque (positive) in the fourth quadrant ($3 \pi/2<\psi<2 \pi$). Similarly the $\psi=\pi/2$ is a stable point for $\sigma_r>1$ and can be seen from the clockwise torque (negative) in the 2nd quadrant ($\pi/2<\psi<\pi$). The torque in the very low and very high frequencies for both the conductivity ratio is anticlockwise in the first quadrant thereby favoring $\psi=\pi/2$.\\

The effect of the electric torque is clearly seen in figure \ref{PsiVsMn}(a,b) in the TT regime which shows the variation of inclination angle ($\psi$) with the applied field strength for three different frequency values. Selection of these frequencies are based on the non-dimensional (by the shear rate) membrane charging time $t_{mem}^{-1}\sim10^{5}$. Thus the three values of $\omega$ of interest are $\omega< t_{mem}^{-1}$, $\omega \sim t_{mem}^{-1}$, and $\omega>t_{mem}^{-1}$. Figure \ref{shapes1} shows the shape of the deformed vesicle as a function of frequency for a given $Mn$ for $\sigma_r=0.1,10$. Similarly, figure \ref{shapes2} shows the shape of a deformed vesicle as a function of $Mn$ at intermediate frequencies for the two conductivity ratios. A clear dependence of both the shape (prolate or oblate) and the orientation (near $\psi=\pi/2$ or $\psi=0$) can be clearly seen.
For $\sigma_r>1$, (figure \ref{PsiVsMn}a), a vesicle acquires a prolate ellipsoidal shape (figure \ref{shapes2}a,b,c) and the inclination angle increases for a given frequency from a value corresponding to $Mn=0$ to its maximum value of $\pi/2$, with the major axis parallel to the applied electric field. A stronger effect of electric field is seen at intermediate frequencies when the Maxwell stresses and thereby the electric torques are higher.  When $\sigma_r<1$ (figure \ref{PsiVsMn}b), in the low frequency regime($\omega<t_{mem}^{-1}$) a vesicle shape remains prolate ellipsoidal (figure \ref{shapes2}g) and shows an increase in inclination angle upto $\psi=\pi/2$  with an increase in $Mn$.  When the frequency $\omega>t_{mem}^{-1}$ the vesicle acquires an oblate ellipsoidal shape (figure \ref{shapes2}h, i) and shows a decrement in inclination angle with $Mn$ for a given frequency, finally attaining a zero inclination angle. Thus the shape- Maxwell stress coupling leads to interesting dependence of inclination angle on the applied frequency. \\

The variation of the inclination ange with frequency for a particular value of $Mn$ is shown in figure \ref{PsiVsMn}(c) for the $\sigma_r=10, 0.1$ cases and is exactly similar to that seen for the variation of electric torque with frequency (figure \ref{torque}). This confirms that the electric torque determines the inclination angle of a vesicle in simultaneous shear and electric fields. When $\sigma_r=10$, the inclination angle remains constant at low frequency and then start increasing with $\omega$ to attain its maximum value at $\omega=10^7$. Further increase in frequency decreases vesicle orientation angle. When $\sigma_r=0.1$ in the low frequency a prolate spheroidal shape is seen at low frequency, similar to the case $\sigma_r=10$. Remarkably when $\omega$ is increased, the inclination angle decreases almost to zero and remains at that orientation for a range of frequency value $10^5-10^7$, beyond which it again increases to a value similar to that seen at $\sigma_r=10$ (but not shown in figure \ref{PsiVsMn}b). 

\subsubsection{Trembling-tumbling transition}

\begin{figure} [tp] 
    \hspace{0.cm}
    \begin{subfigure}[b]{0.45\linewidth}       \includegraphics[width=\linewidth]{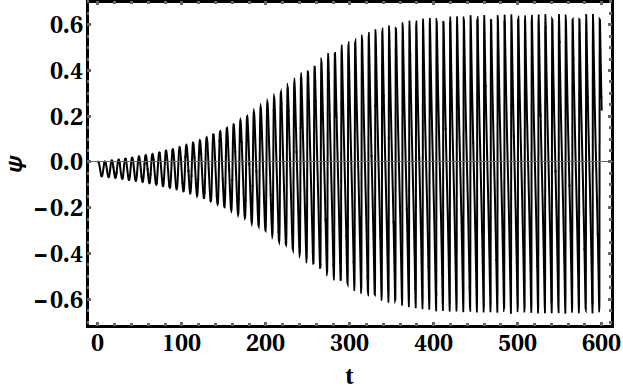}
     \caption{}
     \end{subfigure}
     \hspace{1.05cm}
      \begin{subfigure}[b]{0.45\linewidth}
      \includegraphics[width=\linewidth]{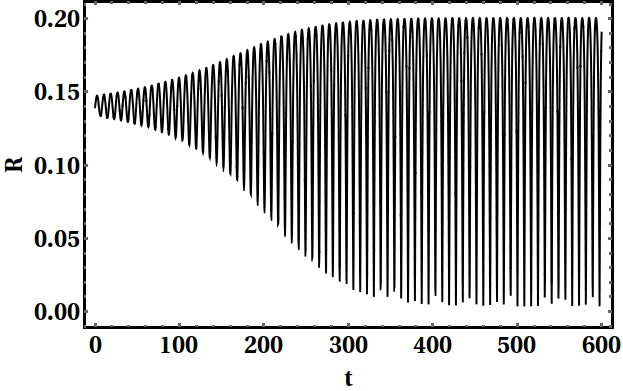}
      \caption{}
     \end{subfigure} 
     \hspace{0.12cm}
      \begin{subfigure}[b]{0.45\linewidth}
      \includegraphics[width=\linewidth]{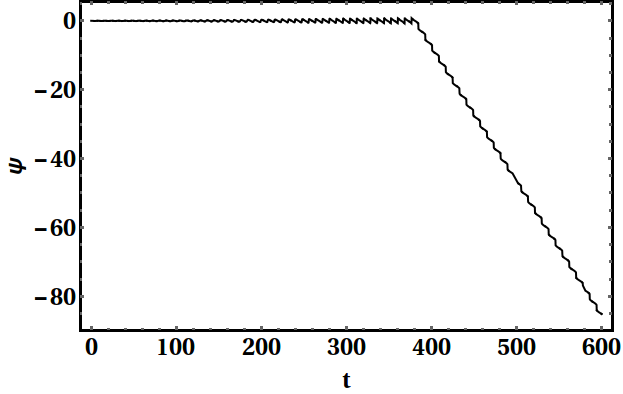}
      \caption{}
     \end{subfigure} 
       \hspace{1.05cm}
        \begin{subfigure}[b]{0.45\linewidth}
        \includegraphics[width=\linewidth]{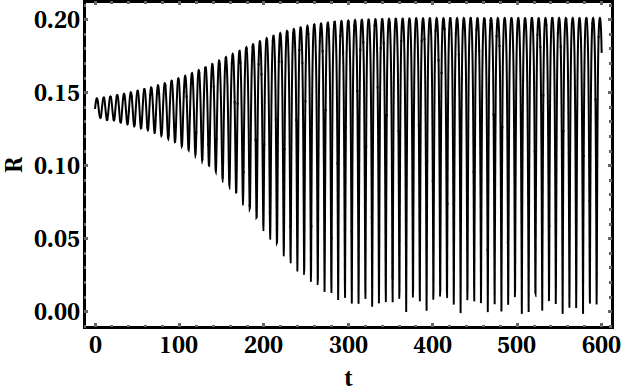}
        \caption{}
       \end{subfigure}             
 \caption{
 TR to TU transition as a function of time for $\sigma_r=0.1$: (a), (b) - TR motion and corresponding shape deformation at $\eta=10.6$; (c), (d) - transition from TR to TU motion and corresponding shape deformation at $\eta=10.7$. ($Mn=0.1, \omega=10^6, Ca=1, \zeta=10^{-7},\epsilon_r=1, \triangle=0.2, C_{mem}=50$)}
    \label{theta}
\end{figure}

A vesicle subjected to linear shear flow shows TR as an intermediate regime between TT and TU (Figures \ref{PD1} and \ref{PD2} and discussed in detail in the appendix).
Figure \ref{TRTU} presents the transition from TR to TU regime when $\sigma_r>1$.  A vesicle in the TR regime exhibits small oscillations about an average orientation (figure \ref{TRTU}a and \ref{TRTU}b). The TU motion of a vesicle is characterized by a continuous periodic flipping (figure \ref{TRTU}c and \ref{TRTU}d). Both the TR and TU regimes show correspoding shape oscillations apart from orientation oscillations and rotations respectively \cite{Steinberg2005ShInt}. When $\sigma_r<1$ (figure \ref{theta}),  a similar TR-TU transition is observed. However, the transition for $\sigma_r<1$ occurs at much higher viscosity ratio than $\sigma_r>1$.  This is despite the fact that the electric torque is anticlockwise for $\sigma_r>1$ and vice-versa for $\sigma_r<1$. This could  be attributed to the higher hydrodynamic torque and a rigid ellipsoid like behavior for $\sigma_r>1$ which promotes elongation in the $X-Y$ plane in the first quadrant. The transition viscosity ratio not only depends upon the conductivity ratio, but is also significantly different than the transition viscosity ratio in the absence of electric field ($Mn = 0$). Thus electric field alters the TT-TR-TU transition of a vesicle in shear flow. 

\begin{figure} [tp] 
    \hspace{0.cm}
    \begin{subfigure}[b]{0.49\linewidth}       \includegraphics[width=\linewidth]{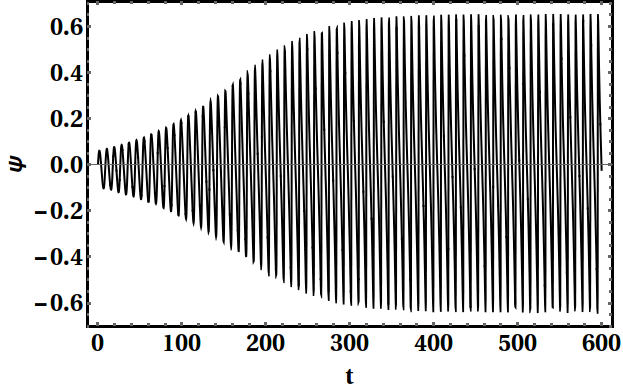}
     \caption{}
     \end{subfigure}
     \hspace{0.9cm}
      \begin{subfigure}[b]{0.48\linewidth}
      \includegraphics[width=\linewidth]{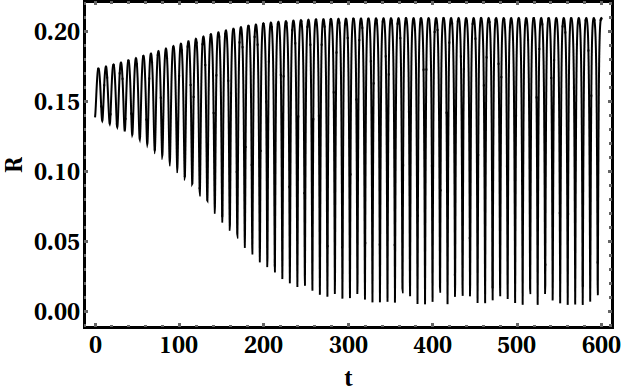}
      \caption{}
     \end{subfigure} 
     \hspace{0.12cm}
      \begin{subfigure}[b]{0.475\linewidth}
      \includegraphics[width=\linewidth]{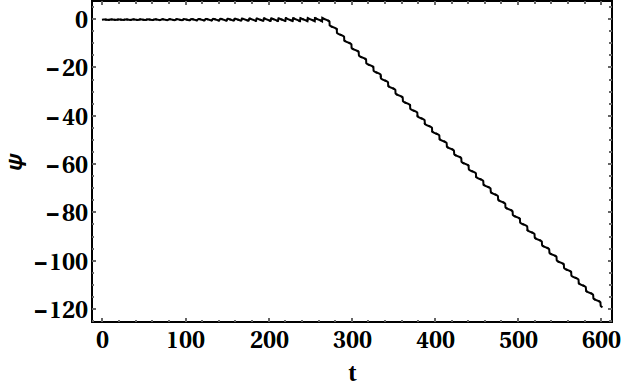}
      \caption{}
     \end{subfigure}
   \hspace{0.8cm}
    \begin{subfigure}[b]{0.47\linewidth}       \includegraphics[width=\linewidth]{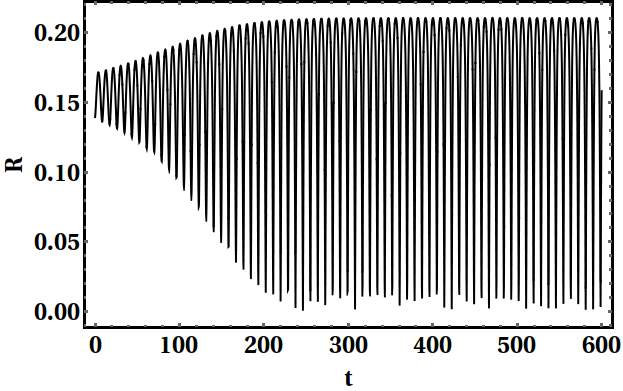}
     \caption{}
     \end{subfigure}               
 \caption{ TR to TU transition as a function of time for $\sigma_r=10$: (a), (b) - TR motion and corresponding shape defprmation at $\eta=9.3$; (c), (d) - transition from TT to TU motion and corresponding shape deformation at $\eta=9.4$. ($Mn=0.1, \omega=10^6, Ca=1, \zeta=10^{-7},\epsilon_r=1, \triangle=0.2, C_{mem}=50$)}
    \label{TRTU}
\end{figure}

\subsubsection{Phase diagram}
\begin{figure} [tp] 
\centering
    \hspace{0.0cm}
    \begin{subfigure}[b]{0.476\linewidth}\includegraphics[width=\linewidth]{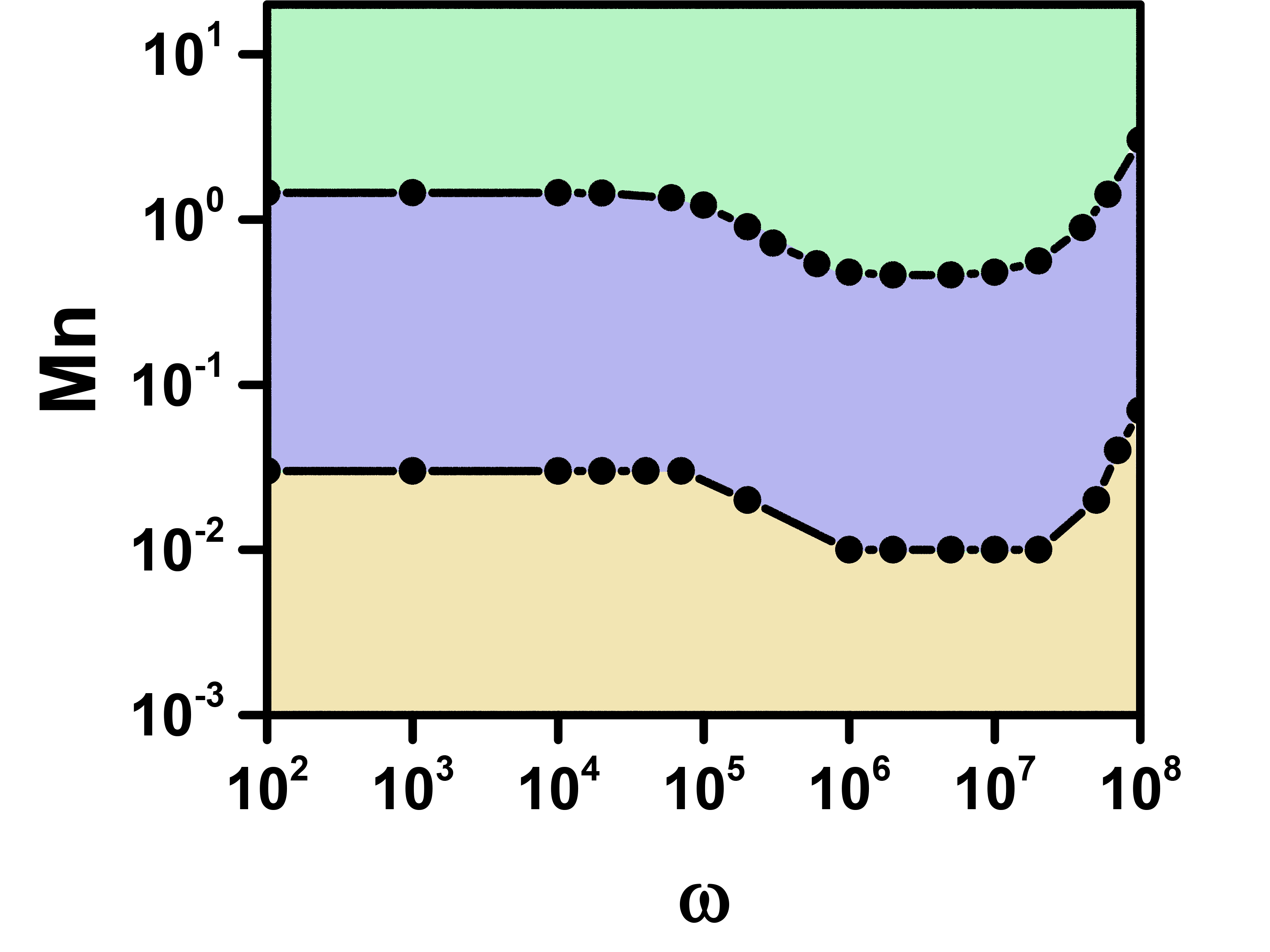}
     \caption{}
     \end{subfigure}
    \hspace{0.5cm}
      \begin{subfigure}[b]{0.47\linewidth}
      \includegraphics[width=\linewidth]{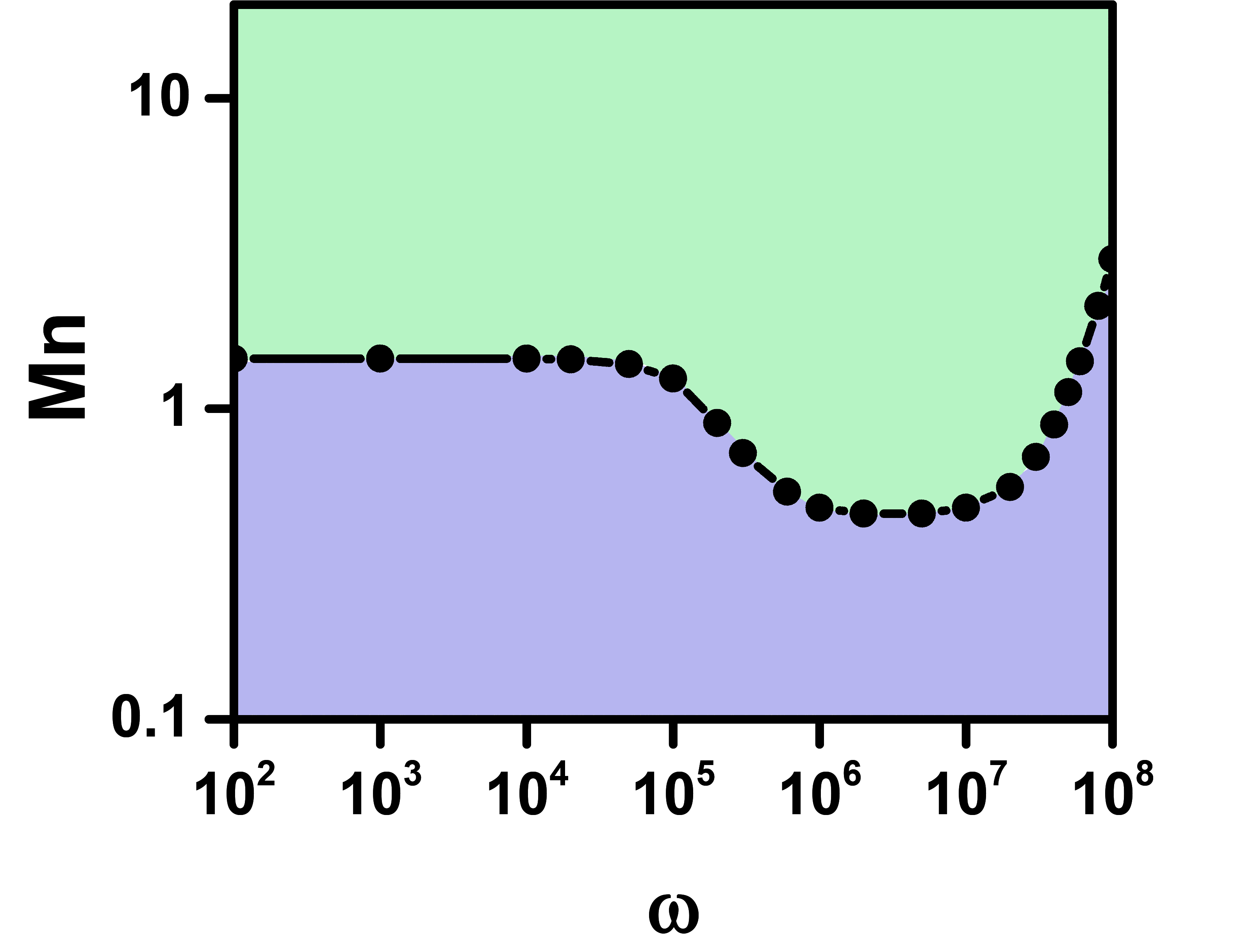}
      \caption{}
     \end{subfigure} 
  \hspace{0.9cm}
      \begin{subfigure}[b]{0.46\linewidth}
      \includegraphics[width=\linewidth]{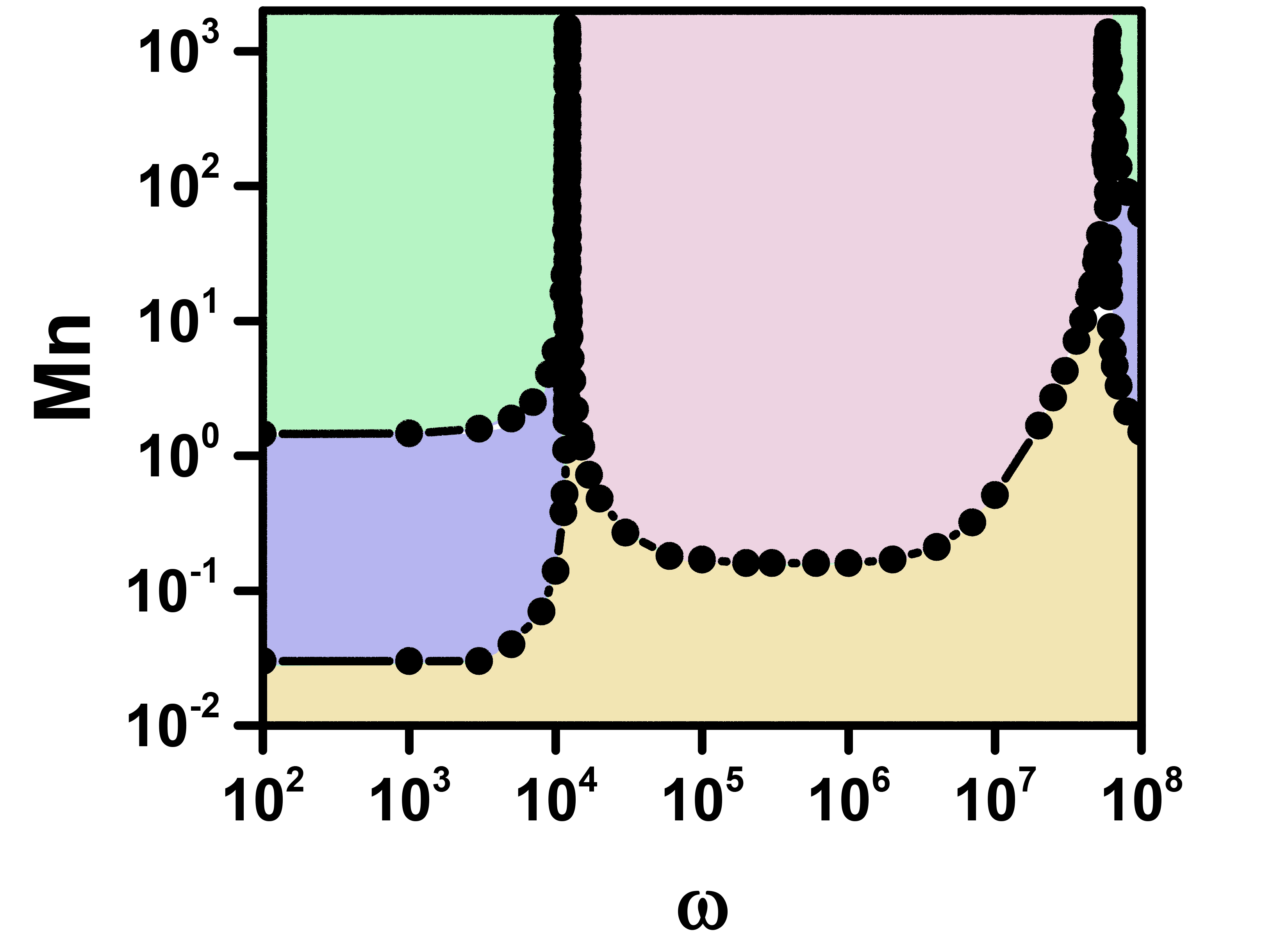}
      \caption{}
     \end{subfigure} 
 \hspace{0.9cm}
      \begin{subfigure}[b]{0.47\linewidth}
      \includegraphics[width=\linewidth]{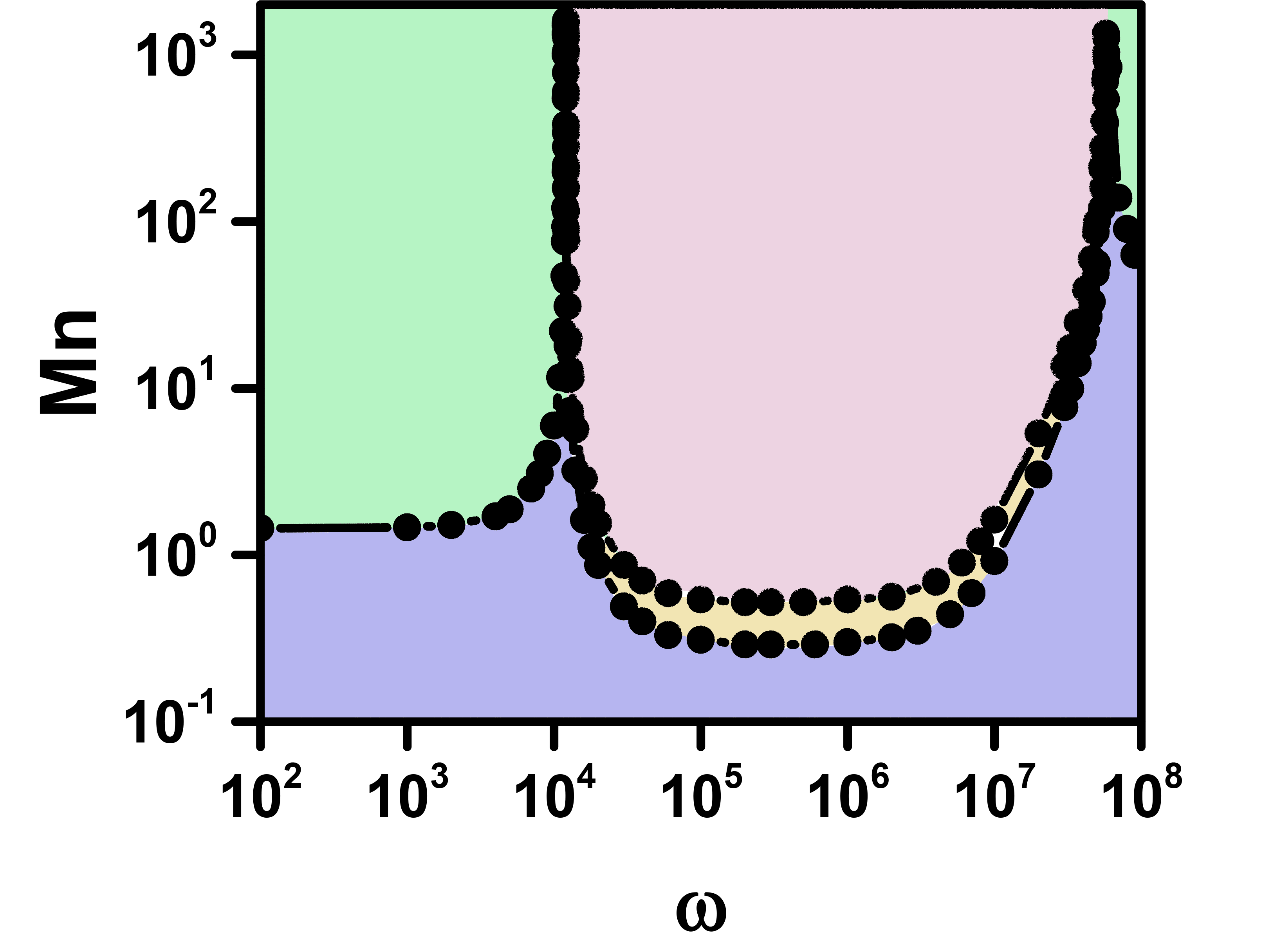}
      \caption{}
     \end{subfigure}           
 \caption{Phase diagram for transition between dynamic states (a) $\eta=10, \sigma_r=10$, (b) $\eta=12, \sigma_r=10$, (c) $\eta=10, \sigma_r=0.1$, and (d) $\eta=12, \sigma_r=0.1$. TR (yellow region), TT (green region), TT in intermediate frequency for $\sigma_r<1$ (purple), TU (blue region) ($\epsilon_r=1, \triangle=0.2, C_{mem}=50, Ca=1, \zeta=10^{-7}$)}
    \label{PD3}
\end{figure}

There is no significant variation in the $\eta-Ca$ phase diagram  (Figures \ref{PD1},\ref{PD2} presented in the appendix) with respect to the capillary number, owing to which the dynamics of a vesicle in simultaneous shear flow and electric fields is best presented in the $Mn-\omega$ coordinates. Two viscosity values are selected such that when $Mn=0$ the vesicle either shows TR ($\eta=10$) or TU ($\eta=12$) modes. In each of these dynamic modes, the effect of $Mn$ and $\omega$ is investigated for conductivity ratio $\sigma_r>1$ and $\sigma_r<1$. \\
\\
\textbf{Case I: $\sigma_r>1$}\\
In this case the electric torque is always anticlockwise, and is maximum when the $t_{mem}^{-1}<\omega<t_{MW}^{-1}$. Correspondingly, in the low $\eta$ regime, the anticlockwise torque suppresses the low $Mn$ TR modes into high $Mn$ TT modes. The transition $Mn$ for TR to TU to TT is lower when $t_{memb}^{-1}<\omega<t_{MW}^{-1}$ (figure \ref{PD3}a). Similar arguments can be given to explain the TU to TT transition for the higher viscosity contrast (high $\eta$) case (figure \ref{PD3}b). \\ 
\\
\textbf{Case II: $\sigma_r<1$}\\
 
 When $\sigma_r<1$ the transition is frequency dependent. In the low viscosity case, at low and high-frequencies, the transition from TR to TT occurs via the TU mode, similar to the $\sigma_r>1$ case (figure \ref{PD3}c,d). However at an intermediate frequency, a  TR-TT transition is seen and the TU mode is suppressed. Figure  \ref{PD3}d for the high viscosity case, shows dynamic transitions such that a direct TR-TU-TT transition is observed at  low and high frequencies. In the intermediate frequency range, as $Mn$ is increased,  there is an appearance of TR regime before entering into the TT regime. In all the cases, an oscillatory relaxation to TT state (long time relaxation) takes place near the TR-TT phase boundary. An important feature of the $\sigma_r<1$ case is the transition from the TT($\psi=\pi/2$) at the low and high frequencies to the $\psi=0$ at intermediate frequencies, especially at high $Mn$. In the low $Mn$ regime, the two TT modes differ with the low frequency showing a long tumbling mode is observed in the dynamics before switching to TT, whereas in the intermediate frequency case, the TT state is attained instantaneously.   The electric torque at these transition frequencies $\omega=t_{mem}^{-1}$ and $\omega=t_{MW}^{-1}$ changes sign, and goes through zero, thereby enabling the TR and TU modes in figures \ref{PD3}b and d respectively, corresponding to the Mn=0 case.  \\

\section{Conclusions}
A systematic study on the effect of electric field on the different dynamical modes such as TT, TR and TU that are commonly observed in a vesicle in shear flow. The electrical parameters of interest here are the conductivity ratio, the frequency of applied field and the Mason number, apart from the viscosity contrast between inner and outer fluid. Our study shows that apart from the role of the electric torque on deciding the dynamical modes of  a vesicle in simultaneous electric and shear field, a complicated coupling between the elongation caused by high $Mn$ can lead to unexpected appearance of an intermediate TU regime in some cases. The phase diagrams presented here can enable the judicious use of electrical parameters in either promoting or prohibiting specific dynamical modes. For example, TU mode may be more desirable if mixing of the vesicles content as well as agitation in the system is desired. On the other hand, if undisturbed, streamlined flow of vesicles is desired, it might be desirable to be in the TT regime.  We show here that electric field parameters can easily allow enforcing a desired dynamical mode in such systems. The dynamical modes of a suspension of emulsions, as encountered say in dielectrophoretic devices, can determine the effective residence time as well as the total possible suspension density in biotechnological applications involving continuous processing. The present study in that case would form the basis for more detailed calculations. \\

\textbf{\large Acknowledgment}\\ 

Authors would like to acknowledge the Department of Science and Technology, India, for financial support.

\newpage
\bibliographystyle{model1-num-names}
\bibliography{bibshearac}

\newpage
\appendix
\section*{\LARGE Appendix}

\section{Hydrodynamics}
In the low Reynolds number limit, velocity field induced (\cite{Vla2006ShInt}) in the two regions are given by the solution of Stokes equation (\cite{Lamb1932}) 
\begin{align}
&\ \boldsymbol{u_{in}}= C_{lm0}^g \boldsymbol{u_{lm0}^g}+C_{lm1}^g \boldsymbol{u_{lm1}^g}+C_{lm2}^g \boldsymbol{u_{lm2}^g} &\\
&\ \boldsymbol{u_{ex}}=(C_{lm0}^d \boldsymbol{u_{lm0}^d}+C_{lm1}^d \boldsymbol{u_{lm1}^d}+C_{lm2}^d \boldsymbol{u_{lm2}^d})+(C^\infty_{lm0} \boldsymbol{u_{lm0}^g}+C^\infty_{lm1} \boldsymbol{u_{lm1}^g}+C^\infty_{lm2} \boldsymbol{u_{lm2}^g}) &
\end{align}  
where coefficients $C_{lm0}^g, C_{lm1}^g, C_{lm2}^g$ are  coefficients corresponding to the growing harmonics in the interior region and $C_{lm0}^d, C_{lm1}^d, C_{lm2}^d$ are decaying harmonic coefficients in the exterior region. These unknown coefficients are determined by using the velocity continuity, membrane incompressibility conditions and stress balance at the surface of the vesicle.\\ 

$C^\infty_{lm0}, C^\infty_{lm1}, C^\infty_{lm2}$ are coefficients associated with the applied unperturbed external flow (when $\boldsymbol{u_{ex}}=\boldsymbol{u^{\infty}}$ ) and they depend on the elongational ($s$) as well as rotational ($\Omega$) component of applied flow described as (for $l=2, m= \pm 2$ mode)
\begin{align}
&\ C^\infty_{2\pm20}=\mp2 i \left(\sqrt{\frac{\pi}{5}}\right) s & \\
&\ C^\infty_{2\pm21}=0&\\
&\ C^\infty_{2\pm22}=\mp2 i \left(\sqrt{\frac{2 \pi}{15}}\right) s & \\
&\ C^\infty_{1\pm10}=2 i \left(\sqrt{\frac{2\pi}{3}}\right) \Omega &
\end{align}
where $\Omega=s=\dot{\gamma}/2$ for the case of pure shear.\\

$\boldsymbol{u_{lm0}^g}, \boldsymbol{u_{lm1}^g}, \boldsymbol{u_{lm2}^g}$ and $\boldsymbol{u_{lm0}^d}, \boldsymbol{u_{lm1}^d}, \boldsymbol{u_{lm2}^d}$ are growing and decaying velocity eigen functions for inner and outer fluid, respectively, defined as (\cite{Vla2006ShInt})
\begin{align}
&\ \boldsymbol{u_{lm0}^g}=\frac{1}{2} r^{l-1}\left(-(l+1)+(l+3)r^{2}\right)\boldsymbol{y_{lm0}}-\frac{1}{2}r^{l-1}[l(l+1)]^{1/2}\left(1-r^{2}\right)\boldsymbol{y_{lm2}} & \\
&\ \boldsymbol{u_{lm1}^g}=r^{l}\boldsymbol{y_{lm1}} &\\
&\ \boldsymbol{u_{lm2}^g}=\frac{1}{2}r^{l-1}(3+l)\left(\frac{l+1}{l}\right)^{1/2}\left(1-r^{2}\right)\boldsymbol{y_{lm0}}+\frac{1}{2}r^{l-1}\left(l+3-(l+1)r^{2}\right)\boldsymbol{y_{lm2}}& \\
&\ \boldsymbol{u_{lm0}^d}=\frac{1}{2} r^{-l}\left(2-l+lr^{-2}\right)\boldsymbol{y_{lm0}}+\frac{1}{2}r^{-l}[l(l+1)]^{1/2}\left(1-r^{-2}\right)\boldsymbol{y_{lm2}} &\\
&\ \boldsymbol{u_{lm1}^d}=r^{-l-1}\boldsymbol{y_{lm1}} &\\
&\ \boldsymbol{u_{lm2}^d}=\frac{1}{2}r^{-l}(2-l)\left(\frac{l}{l+1}\right)^{1/2}\left(1-r^{-2}\right)\boldsymbol{y_{lm0}}+\frac{1}{2}r^{-l}\left(l+(2-l)r^{-2}\right)\boldsymbol{y_{lm2}}& 
\end{align} 
Here $\boldsymbol{y_{lm0}}, \boldsymbol{y_{lm1}}, \boldsymbol{y_{lm2}}$ are vector spherical harmonics defined as
\begin{align}
&\ \boldsymbol{y_{lm0}}=\frac{1}{\sqrt{l(l+1)}}\frac{\partial Y_{lm}}{\partial \theta} \boldsymbol{\hat{e}_\theta}+\frac{im}{\sqrt{l(l+1)}}\frac{Y_{lm}}{\sin \theta}\boldsymbol{\hat{e}_\Phi} &\\
&\ \boldsymbol{y_{lm1}}=-\frac{m}{\sqrt{l(l+1)}}\frac{Y_{lm}}{\sin\theta} \boldsymbol{\hat{e}_\theta}-\frac{i}{\sqrt{l(l+1)}}\frac{\partial Y_{lm}}{\partial \theta}\boldsymbol{\hat{e}_\Phi} &\\
& \boldsymbol{y_{lm2}}=\boldsymbol{\hat{e}_r}Y_{lm} &
\end{align}
with scalar spherical harmonics
\begin{equation}
\ Y_{lm}(\theta, \Phi)=\sqrt{\frac{2l+1}{4\pi}\frac{(l-m)!}{(l+m)!}}(-1)^m P_{lm}(\cos\theta)e^{im\Phi}
\end{equation} 
\subsection{Solution for hydrodynamic stress}

The hydrodynamic stress in the interior and exterior ($j=in, ex$) region of a vesicle is given by
\begin{equation}
\ \boldsymbol{\tau_{j}^h}=-p_{j}\boldsymbol{I}+\mu_{j} \left[\nabla \boldsymbol{u_{j}}+(\nabla \boldsymbol{u_{j}})^T\right]
\end{equation}
where $\boldsymbol{I}$ is the identity matrix and superscript $T$ represents the transpose of the matrix, $p$ and $\boldsymbol{u}$ are pressure and velocity field, respectively. Normal stress traction exerted on the vesicle surface (at $r=1$) are given by
\begin{align}
&\ \tau_{in}^h.\bm{\hat{e}_r}=-p_{in} \bm{\hat{e}_r}+Z_{in} \qquad \mathrm{at\ r=1}& \label{Hin}\\ 
&\ \tau_{ex}^h.\bm{\hat{e}_r}=-p_{ex} \bm{\hat{e}_r}+Z_{ex} \qquad \mathrm{at\ r=1}  \label{Hex}&
\end{align}
with
\begin{equation}
\  Z_{j}=\bm{\hat{e}_r}.\left(\nabla \boldsymbol{u_{j}}+(\nabla \boldsymbol{u_{j}})^T\right)=r \frac{d}{dr}\left(\frac{\boldsymbol{u_{j}}}{r}\right)+\frac{1}{r}\nabla((\boldsymbol{u_{j}}.\boldsymbol{\hat{e}_r})r)
\end{equation}
where $j=in, ex$. Expression for $Z_{in}|_{r=1}, Z_{ex}|_{r=1}$ are
\begin{align}
\ Z_{in}=&(2 C_{lm1}^g (-1 + l) \boldsymbol{y_{lm1}} + 
  C_{lm2}^g ((-6 \sqrt{1 + 1/l} - 2 \sqrt{1 + 1/l} l + 
 2 \sqrt{l (1 + l)}) \boldsymbol{y_{lm0}} - 6 \boldsymbol{y_{lm2}}) &\nonumber\\
        &+ (2 C_{lm0}^g \sqrt{l (1 + l)} + 
     C_{lm2}^g (-(1 + l) (2 + l) + l (3 + l))) \boldsymbol{y_{lm2}} + 
  C_{lm0}^g (2 (1 - l^2 + l (2 + l)) \boldsymbol{y_{lm0}} &\nonumber\\
         &+ 2 \sqrt{l (1 + l)} \boldsymbol{y_{lm2}}))/2&\\
\ Z_{ex}=&(-3 C^\infty_{lm2} \sqrt{
     1 + 1/l} + (3 \sqrt{l/(1 + l)}) (C_{lm2}^g - C^\infty_{lm2}) - (C_{lm0}^g - 
       2 C^\infty_{lm0}) (1 + 2 l) ) \boldsymbol{y_{lm0}} &\nonumber\\
       &+ (C^\infty_{lm1} l - (2 C_{lm1}^d + 
      C^\infty_{lm1}) - C_{lm1}^d l ) \boldsymbol{y_{lm1}} + (2 C_{lm0}^g \sqrt{l (1 + l)}-4 C_{lm2}^g) \boldsymbol{y_{lm2}}&
\end{align}

The pressure field in each region in terms of growing/decaying harmonics can be expressed as a solution of Laplace equation ($\nabla^2 p=0$)
\begin{align}
&\ p_{in}=A_{in} r^l \boldsymbol{y_{lm2}}&\\
&\ p_{ex}=A^g_{ex} r^l \boldsymbol{y_{lm2}}+A^d_{ex} r^{-l-1} \boldsymbol{y_{lm2}} &
\end{align}
 where $A_{in}, A^g_{ex}, A^d_{ex}$ are pressure coefficients obtained by solving momentum equation for interior and exterior fluid (i.e., $\nabla p_{j}=\mu_{j}\nabla^2 \boldsymbol{u_{j}}$ with $j=in, ex$). Full expressions are \\
\begin{align}
\ A_{in}=&\frac{(1 + l) (C_{lm0}^g (3 + 2 l) - 
   C_{lm2}^g (3 \sqrt{1 + 1/l} + \sqrt{1 + 1/l} l + \sqrt{l (1 + l)}))}{\sqrt{l (1 + l)}}&\\
\ A_{ex}^g=&- \frac{((3 + 2 l) (C^\infty_{lm2} + C^\infty_{lm2} l - 
   C^\infty_{lm0} \sqrt{l} \sqrt{1 + l}))}{l}&\\
\ A_{ex}^d=&\frac{(\sqrt{l} (-1 + 2 l) (C_{lm2}^d \sqrt{l} + C_{lm0}^d \sqrt{1 + l}))}{(1 + l)}&   
\end{align}

The hydrodynamic stress traction acting on the vesicle at inner surface is (from equation \ref{Hin})
\begin{align}
&\  \tau^h_{in}.\bm{\hat{e}_r}|_{r=\lambda}=(-A_{in} r^l \boldsymbol{y_{lm2}}+Z_{in})|_{r=1} &
\end{align}
Similarly hydrodynamic stress traction acting at the outer surface of the vesicle is (from equation \ref{Hex})
\begin{align}
&\  \tau^h_{ex}.\bm{\hat{e}_r}|_{r=1}=(-A^g_{ex} r^l \boldsymbol{y_{lm2}}-A^d_{ex} r^{-l-1} \boldsymbol{y_{lm2}}+Z_{ex})|_{r=1} &
\end{align}
Substitution of pressure coefficients ($A_{in}, A^d_{ex}, A^g_{ex}$) and Z values ($Z_{in}, Z_{ex}$) into the above equations gives stress traction on the inner and outer surface of the vesicle (at $r=1$) due to inner fluid and exterior fluid, respectively in matrix form as \cite{Vla2006ShInt}
\begin{align}
\tau^h_{in}.\bm{\hat{e}_r}|_{r=1}=
  \begin{pmatrix}
    2 l+1 & 0 & -3\sqrt{\frac{l+1}{l}}  \\
    0 & l-1 & 0  \\
    -3 \sqrt{\frac{l+1}{l}} & 0 & 1+\frac{3}{l}+2l 
  \end{pmatrix}
\begin{pmatrix} C^g_{lm0} & C^g_{lm1} & C^g_{lm2} \end{pmatrix}
\begin{pmatrix} \boldsymbol{y_{lm0}} \\ \boldsymbol{y_{lm1}} \\ \boldsymbol{y_{lm2}} \end{pmatrix}
\label{mat1}  
\end{align}

\begin{align}
\tau^h_{ex}.\bm{\hat{e}_r}|_{r=1}=&
  \begin{bmatrix}
    1+2l & 0 & -3 \sqrt{\frac{l+1}{l}}  \\
    0 & -l+1 & 0  \\
    -3 \sqrt{\frac{l+1}{l}} & 0 & 1+\frac{3}{l}+2l 
  \end{bmatrix}
  \begin{pmatrix} C^\infty_{lm0} & C^\infty_{lm1} & C^\infty_{lm2} \end{pmatrix}
  \begin{pmatrix} \boldsymbol{y_{lm0}} \\ \boldsymbol{y_{lm1}} \\ \boldsymbol{y_{lm2}} \end{pmatrix} + \nonumber\\&
\begin{bmatrix}
    -1-2l & 0 & 3\sqrt{\frac{l}{l+1}}  \\
    0 & -2-l & 0 \\
    3\sqrt{\frac{l}{l+1}} & 0 & -1-2l-\frac{3}{l+1}  
  \end{bmatrix}
  \begin{pmatrix} C^d_{lm0} & C^d_{lm1} & C^d_{lm2} \end{pmatrix}
  \begin{pmatrix} \boldsymbol{y_{lm0}} \\ \boldsymbol{y_{lm1}} \\ \boldsymbol{y_{lm2}} \end{pmatrix} 
\label{mat4}   
\end{align}
This can be further expressed as the tangential hydrodynamic stress inside and outside\\
\begin{align}
\tau_{lm0}^{h,in}=&\left((2l+1)C^g_{lm0}+\left(-3\sqrt{\frac{l+1}{l}}\right)C^g_{lm2}\right)\boldsymbol{y_{lm0}}& \label{Hin1}\\
\tau_{lm0}^{h,ex}=&\left((4l+2)C^\infty_{lm0}+\left(-3\sqrt{\frac{l+1}{l}}-3\sqrt{\frac{l}{l+1}}\right)C^\infty_{lm2}\right)\boldsymbol{y_{lm0}}+\left(\left(-2l-1\right) C^d_{lm0}+\left(3\sqrt{\frac{l}{l+1}}\right)C^d_{lm2}\right)\boldsymbol{y_{lm0}}& \label{Hin2}
\end{align} 
while the normal hydrodynamic stress inside and outside are\\
\begin{align}
\tau_{lm2}^{h,in}=&\left(\left(-3\sqrt{\frac{l+1}{l}}\right)C^g_{lm0}+\left(2l+1+\frac{3}{l}\right)C^g_{lm2}\right)\boldsymbol{y_{lm2}}& \label{Hex1}\\
\tau_{lm2}^{h,ex}=&\left(\left(-3\sqrt{\frac{l+1}{l}}-3\sqrt{\frac{l}{l+1}}\right)C^\infty_{lm0}+\left(4l+2+\frac{3}{l}+\frac{3}{l+1}\right)C^\infty_{lm2}\right)\boldsymbol{y_{lm2}} &\nonumber\\
&+\left(\left(3\sqrt{\frac{l}{l+1}}\right) C^d_{lm0}+\left(-2l-1-\frac{3}{l+1}\right)C^d_{lm2}\right)\boldsymbol{y_{lm2}}& \label{Hex2}
\end{align}
where $C_{lm2}=\frac{\sqrt{l(l+1)}}{2}C_{lm0}$

\section{Resultant normal and tangential Maxwell stress}
The net $\boldsymbol{\hat{e}_{r}}$ directional normal electric stress
\begin{align}
\tau_r=&\frac{1}{4}\left(Z_1+\frac{Z_2}{(4 \sigma_r^2 + \zeta^2 (4 (C_{mem} + \epsilon_r)^2 + 4 C_{mem}^2 \sigma_r + (2 + C_{mem})^2 \sigma_r^2) \omega^2 + (2 \epsilon_r + 
     C_{mem} (2 + \epsilon_r))^2 \zeta^4 \Omega^4)}\right)&
\end{align}
where
\begin{align}
\ Z_1=&\frac{9 C_{mem}^2 \epsilon_r \zeta^2 \omega^2 (1 + \zeta^2 \omega^2) ((\alpha^2 - 2 \beta^2) \cos[2 \theta] + \alpha (\alpha + 2 \alpha \cos2\Phi \sin\theta^2 - 4 \beta \sin 2\Theta \sin\Phi))}{8 \sigma_r^2 + 
   2 \zeta^2 (4 (C_{mem} + \epsilon_r)^2 + 
 4 C_{mem}^2 \sigma_r + (2 + C_{mem})^2 \sigma_r^2) \omega^2 + 2 (2 \epsilon_r + C_{mem} (2 + \epsilon_r))^2 \zeta^4 \omega^4}& \nonumber\\
\ Z_2=&9 ((-\alpha^2 \cos\Phi^2 - \beta^2 \sin\theta^2) (1 + 
\zeta^2 \omega^2) (\sigma_r^2 + (C_{mem} + \epsilon_r)^2 \zeta^2 
\omega^2) + \alpha \beta (\sigma_r^2 + \zeta^2 ((C_{mem} + 
\epsilon_r)^2 &\nonumber\\
&+ (1 + C_{mem}^2) \sigma_r^2) \omega^2 + (C_{mem}^2 + 2 C_{mem} \epsilon_r + (1 + C_{mem}^2) \epsilon_r^2) \zeta^4 \omega^4) \sin2\theta \sin\Phi + 
   C_{mem}^2 \alpha^2 \zeta^2 \omega^2 &\nonumber\\
   &(\sigma_r^2 + \epsilon_r^2 \zeta^2 \omega^2) \sin\theta^2 \sin\Phi^2 + \cos\theta^2 (C_{mem}^2 \beta^2 \zeta^2 \omega^2 (\sigma_r^2 + \epsilon_r^2 \zeta^2 \omega^2) - \alpha^2 (1 + \zeta^2 \omega^2) &\nonumber\\
      &(\sigma_r^2 + (C_{mem} + \epsilon_r)^2 \zeta^2 \omega^2)
\sin\Phi^2))& \nonumber
\end{align}
The isotropic normal electric stress
\begin{align}
\tau_0=&(3 \sqrt{\pi} (\alpha^2 + \beta^2) (-2\sigma_r^2 +(-4\epsilon_r C_{mem} + C_{mem}^2 (-2+\epsilon_r+\sigma_r^2)-2 (\epsilon_r^2+\sigma_r^2)) \omega^2 \zeta^2 +(-4 C_{mem} \epsilon_r &\nonumber\\
&- 2 \epsilon_r^2+ C_{mem}^2 (-2 + \epsilon_r +\epsilon_r^2))
\omega^4 \zeta^4))/(2 (4 \sigma_r^2 + (4 (C_{mem} + \epsilon_r)^2 + 
 4 C_{mem}^2 \sigma_r + (2 + C_{mem})^2 &\nonumber\\
   &\sigma_r^2) \omega^2 \zeta^2+ (2 \epsilon_r  + C_{mem} (2 + \epsilon_r))^2 \omega^4 \zeta^4))&
\end{align}
The deformation causing net normal electric stress
\begin{align}
\tau_n=&(3 (\sigma_r^2 + \zeta^2 (2 C_{mem} \epsilon_r + \epsilon_r^2 + \sigma_r^2 + C_{mem}^2 (1 - 2 \epsilon_r + \sigma_r^2)) \omega^2 + (C_{mem}^2 (-1 + \epsilon_r)^2 + 2 C_{mem} \epsilon_r + \epsilon_r^2) \zeta^4 \omega^4)&\nonumber\\
& (-2 (\alpha^2 - 2 \beta^2) (1 + 3 \cos{2 \theta}) - 12 \alpha^2 \cos{2 \Phi} \sin{\theta}^2 + 
24 \alpha \beta \sin{2 \theta} \sin{\Phi}))/(32 (4 \sigma_r^2 + \zeta^2 (4 (C_{mem} + \epsilon_r)^2 &\nonumber\\
&+ 4 C_{mem}^2 \sigma_r + (2 + 
C_{mem})^2 \sigma_r^2) \omega^2 + (2 \epsilon_r + 
C_{mem} (2 + \epsilon_r))^2 \zeta^4 \omega^4))&
\end{align}
The net $\boldsymbol{\hat{e}_{\theta}}$ and $\boldsymbol{\hat{e}_{\Phi}}$ directional tangential electric stress
\begin{align}
\tau_\theta=&\frac{9 C_{mem} \zeta^2 \omega^2 (C_{mem} (\epsilon_r - \sigma_r) - \sigma_r^2 - \epsilon_r^2 \zeta^2 \omega^2) (-\alpha^2 + 
   2 \beta^2 + \alpha^2 \cos{2 \Phi}) \sin{2 \theta}}{8 (4 \sigma_r^2 + \zeta^2 (4 (C_{mem} + \epsilon_r)^2 + 4 C_{mem}^2 \sigma_r + (2 + C_{mem})^2 \sigma_r^2) \omega^2 + (2 \epsilon_r + C_{mem} (2 + \epsilon_r))^2 \zeta^4 \omega^4)}&\\          
\tau_\Phi=&-\frac{9 C_{mem} \alpha \zeta^2 \omega^2 (C_{mem} (\epsilon_r - \sigma_r) - \sigma_r^2 - \epsilon_r^2 \zeta^2 \omega^2) \cos{\Phi} (\beta \cos{\theta} + \alpha \sin{\theta} \sin{\Phi})}{8 \sigma_r^2 + 2 \zeta^2 (4 (C_{mem} + \epsilon_r)^2 + 4 C_{mem}^2 \sigma_r + (2 + C_{mem})^2 \sigma_r^2) \omega^2 + 2 (2 \epsilon_r + C_{mem} (2 + \epsilon_r))^2 \zeta^4 \omega^4}&
\end{align}

\section{Vector spherical harmonics}
\begin{align}
\ \boldsymbol{y_{200}}=&-\sqrt{\frac{15}{32 \pi}} \sin2\theta \  \boldsymbol{\hat{e}_{\theta}} &\\
\ \boldsymbol{y_{202}}=&\frac{1}{8}\sqrt{\frac{5}{\pi}} (1+3\cos2\theta) \ \boldsymbol{\hat{e}_r} &\\
\ \boldsymbol{y_{222}}+\boldsymbol{y_{2-22}}=&\sqrt{\frac{15}{8 \pi}} \cos2\Phi \sin^2\theta \ \boldsymbol{\hat{e}_{r}} &\\
\ \boldsymbol{y_{220}}+\boldsymbol{y_{2-20}}=&\sqrt{\frac{5}{4\pi}} (\cos2\Phi \sin2\theta \ \boldsymbol{\hat{e}_{\theta}}-\sin2\Phi \sin\theta \ \boldsymbol{\hat{e}_{\Phi}}) &
\end{align}

\section{Overall stress balance}
At the vesicle surface membrane stresses up to second order approximation are balanced by hydrodynamic stress and electric stress. Thus overall tangential stress balance across the vesicle is given by
\begin{align}
&\ (\tau_{lm0}^{h, ex}-\eta \tau_{lm0}^{h, in})+Mn \tau_{lm0}^E=\tau_{lm0}^{mem} & \label{tstrbal}
\end{align}

Similarly overall normal stress balance
\begin{align}
&\ (\tau_{lm2}^{h, ex}-\eta \tau_{lm2}^{h, in})+Mn \tau_{lm2}^E=\tau_{lm2}^{mem} & \label{nstrbal}
\end{align}

Substitution of tangential hydrodynamic stress (\ref{Hin1}, \ref{Hex1}), electric stress, and membrane stress (\ref{tmem}) into equation (\ref{tstrbal}) gives nonuniform tension acting on the vesicle  
\begin{align}
\sigma_{lm}=Ca\left(-C^\infty_{lm0}\frac{2(2l+1)}{\sqrt{l(l+1)}}+C^\infty_{lm2} \frac{3(2l+1)}{l(l+1)}+C_{lm0} \frac{l+2+\eta(l-1)}{2\sqrt{l(l+1)}}-Mn \frac{\tau_{lm0}^E}{\sqrt{l(l+1)}} \right)
\end{align}
which for $m=-2, 0, 2$ gives 
\begin{align}
\sigma_{2-20}=&\frac{Ca (-20 C_{2-20}^{\infty}+5\sqrt{6}C_{2-22}^\infty+(4+\eta) C_{2-20}))}{2\sqrt{6}}-\frac{Mn}{\sqrt{6}} \tau_{2-20}^E&\\
\sigma_{200}=&\frac{Ca (-20 C_{200}^{\infty}+5\sqrt{6}C_{202}^\infty+(4+\eta) C_{200}))}{2\sqrt{6}}-\frac{Mn}{\sqrt{6}} \tau_{200}^E&\\
\sigma_{220}=&\frac{Ca (-20 C_{220}^{\infty}+5\sqrt{6}C_{222}^\infty+(4+\eta) C_{220}))}{2\sqrt{6}}-\frac{Mn}{\sqrt{6}} \tau_{220}^E&
\end{align}
Solving eq. (\ref{nstrbal}) using normal hydrodynamic stresses (\ref{Hin2}, \ref{Hex2}), electric stress, and membrane stresses (\ref{nmem1}, \ref{nmem2}, \ref{nmem3}) by using membrane incompressibility ($C_{lm2}=\sqrt{l(l+1)} C_{lm0}/2$) gives normal velocity component in the form
\begin{align}
\ C_{lm2}=C_{lm}^{Sh}+\frac{\Gamma(\sigma_0)}{Ca} f_{lm}+Mn C_{lm}^{el} \label{normvel}
\end{align}
For $m=-2, 0, 2$ modes this equation become\\
\begin{align}
\ C_{2-22}=&\frac{-24 f_{2-2} (7 \pi (6 + \sigma_0) + 
    \sqrt{5 \pi} f_{20} (18 + 5 \sigma_0)) + 
 35 (\sqrt{6} C^\infty_{2-20} + 9 C^\infty_{2-22}) \pi Ca}{7 \pi (32 + 23 \eta) Ca}+ Mn C_{2-2}^{el}&\\
\ C_{202}=&\frac{12 \sqrt{5 \pi} (f_{20}^2 - 2 f_{2-2} f_{22}) (18 + 5 \sigma_0) - 
 7 \pi (24 f_{20} (6 + \sigma_0) - 5 (\sqrt{6} C^\infty_{200} + 9 C^\infty_{202}) Ca)}{7 \pi (32 + 23 \eta) Ca}+ Mn C_{20}^{el}&\\
\ C_{222}=&\frac{-24 f_{22} (7 \pi (6 + \sigma_0) + 
    \sqrt{5 \pi} f_{20} (18 + 5 \sigma_0)) + 
 35 (\sqrt{6} C^\infty_{220} + 9 C^\infty_{222}) \pi Ca}{7 \pi (32 + 23 \eta) Ca}+ Mn C_{22}^{el}&
\end{align}
where $C_{220}^\infty = -i \sqrt{\pi/5},
C_{2-20}^\infty = i \sqrt{\pi/5},
C_{222}^\infty = -i \sqrt{2 \pi/15},
C_{2-22}^\infty = i \sqrt{2 \pi/15},
C_{200}^\infty = 0,
 C_{202}^\infty = 0$ and
\begin{align}
\ C_{2-2}^{el}=&\frac{(6\tau_{2-22}^E+2\sqrt{6}\tau_{2-20}^E)}{32+23\lambda}&\\
\ C_{20}^{el}=&\frac{(6\tau_{202}^E+2\sqrt{6}\tau_{200}^E)}{32+23\lambda}&\\
\ C_{22}^{el}=&\frac{(6\tau_{222}^E+2\sqrt{6}\tau_{220}^E)}{32+23\lambda}&
\end{align}

From the above equations uniform tension ($\sigma_0$) is estimated by the constraint of area conservation ($\dot{\triangle}$=0 where over dot represents derivative with respect to time), that is $\dot{f}_{22} f_{2-2}+\dot{f}_{20} f_{20}+\dot{f}_{2-2} f_{22}=0$; here $\dot{f_{2-2}}, \dot{f_{20}}, \dot{f_{22}}$ expression is provided in appendix-E (eq. E1). This gives
\begin{align}
\ \sigma_0=&-\frac{18}{5}+((1008 \sqrt{\pi} \triangle) + 
  28 \sqrt{5} \pi (6 f_{20} Mn P - 2 f_{20} Mn S + 
  \sqrt{6} f_{22} (-5 i + 3 Mn P - Mn S) + & \nonumber \\ &
  \sqrt{6} f_{2-2} (5 i  + 3 Mn P- Mn S)) Ca - 
  35 \sqrt{\pi} (5 (\sqrt{6} C^\infty_{200} + 
9 C^\infty_{202}) f_{20} Ca))/(300 \sqrt{5} f_{20} (f_{20}^2 & \nonumber \\ &- 6 f_{2-2} f_{22}) - 420 \triangle \sqrt{\pi})&
\label{isotens}
\end{align}

\section{Evolution equation}

By using equation \ref{normvel}, the final evolution equation is of the form (with $\omega=1$)
\begin{align}
\ \frac{df_{lm}}{dt}=i\frac{m}{2}\omega+C_{lm2} \label{evoleq1}
\end{align} 

Substituting $\sigma_0$ from eq. \ref{isotens} into eq. \ref{evoleq1} gives final evolution with higher order membrane correction as \\
\begin{align}
\frac{df_{2-2}}{dt}=&-i f_{2-2} + C_{2-2}^{sh} + C_{2-2}^{el} Mn+2 f_{2-2} ((-7 \pi (32 + 23 \eta) - 
115 f_{20} \sqrt{5\pi} \eta - 10) (C_{2-2}^{sh} f_{22} + C_{22}^{sh} f_{2-2}& \nonumber\\
& + (C_{20}^{el} f_{20} + C_{2-2}^{el} f_{22} + C_{22}^{el} f_{2-2}) Mn) Ca - 
16 f_{20} \sqrt{5\pi} (-9 \triangle - 9 f_{20}^2 + 54 f_{22} f_{2-2}))/(Ca (7 \triangle \pi & \nonumber\\
 &- 5 f20 (f_{20}^2 - 6 f_{22} f_{2-2}) \sqrt{5\pi}) (32 + 23 \eta))&\\
\frac{df_{20}}{dt}=&2 f20 \pi ((-224 - 161 \eta) (C_{20}^{el} f_{20} + C_{2-2}^{el} f_{22} + C_{22}^{el} f_{2-2}) Ca Mn + 144 (f_{20}^3 - 6 f_{20} f_{22} f_{2-2}) \sqrt{5/\pi} & \nonumber\\
& - (C_{2-2}^{sh} f_{22} + C_{22}^{sh} f_{2-2} ) (32 + 23 \eta) Ca) + (f_{20}^2 - 2 f_{22} f_{2-2}) \sqrt{5\pi} (-144 \triangle + 5 Ca (C_{2-2}^{sh} f_{22} + & \nonumber\\
&C_{22}^{sh} f_{2-2}  + (C_{20}^{el} f_{20} + C_{2-2}^{el} f_{22} + C_{22}^{el} f_{2-2}) Mn) (32 + 23 \eta))&\\
\frac{df_{22}}{dt}=&i f_{22} + C_{22}^{sh} + C_{22}^{el} Mn+2 f_{22} ((-7 \pi (32 + 23 \eta) - 
115 f_{20} \sqrt{5\pi} \eta - 10) (C_{2-2}^{sh} f_{22} + C_{22}^{sh} f_{2-2}& \nonumber\\
& + (C_{20}^{el} f_{20} + C_{2-2}^{el} f_{22} + C_{22}^{el} f_{2-2}) Mn) Ca - 
16 f_{20} \sqrt{5\pi} (-9 \triangle - 9 f_{20}^2 + 54 f_{22} f_{2-2}))/(Ca (7 \triangle \pi & \nonumber\\
 &- 5 f20 (f_{20}^2 - 6 f_{22} f_{2-2}) \sqrt{5\pi}) (32 + 23 \eta))&
\end{align}

\section{Effect of $Mn$ and $\omega$ on the $\eta-Ca$ phase diagram}

\begin{figure} [ht] 
	\hspace{0.12cm}
	\begin{subfigure}[b]{0.31\linewidth}       \includegraphics[width=\linewidth]{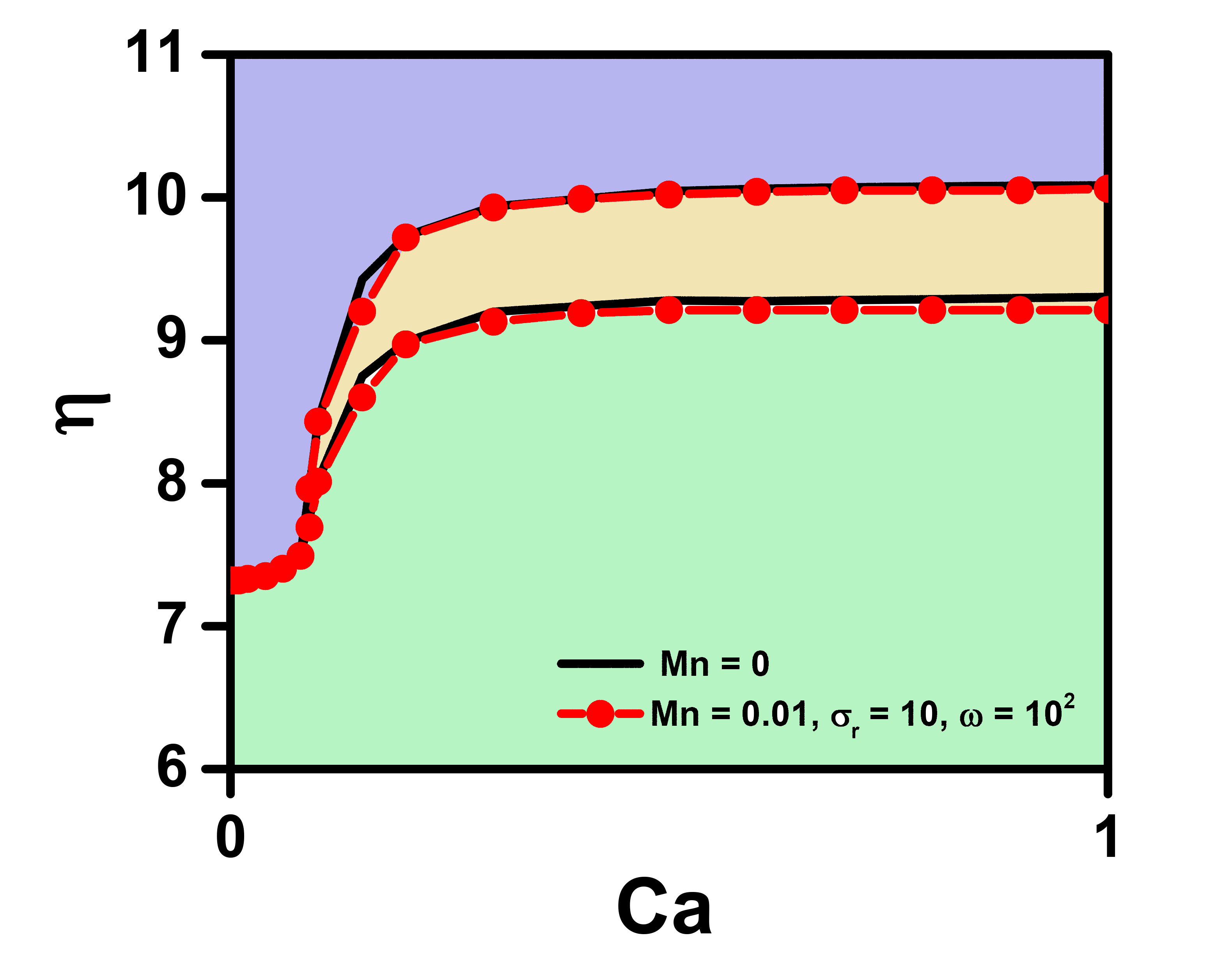}
		\caption{}
	\end{subfigure}
	\hspace{0.12cm}
	\begin{subfigure}[b]{0.31\linewidth}
		\includegraphics[width=\linewidth]{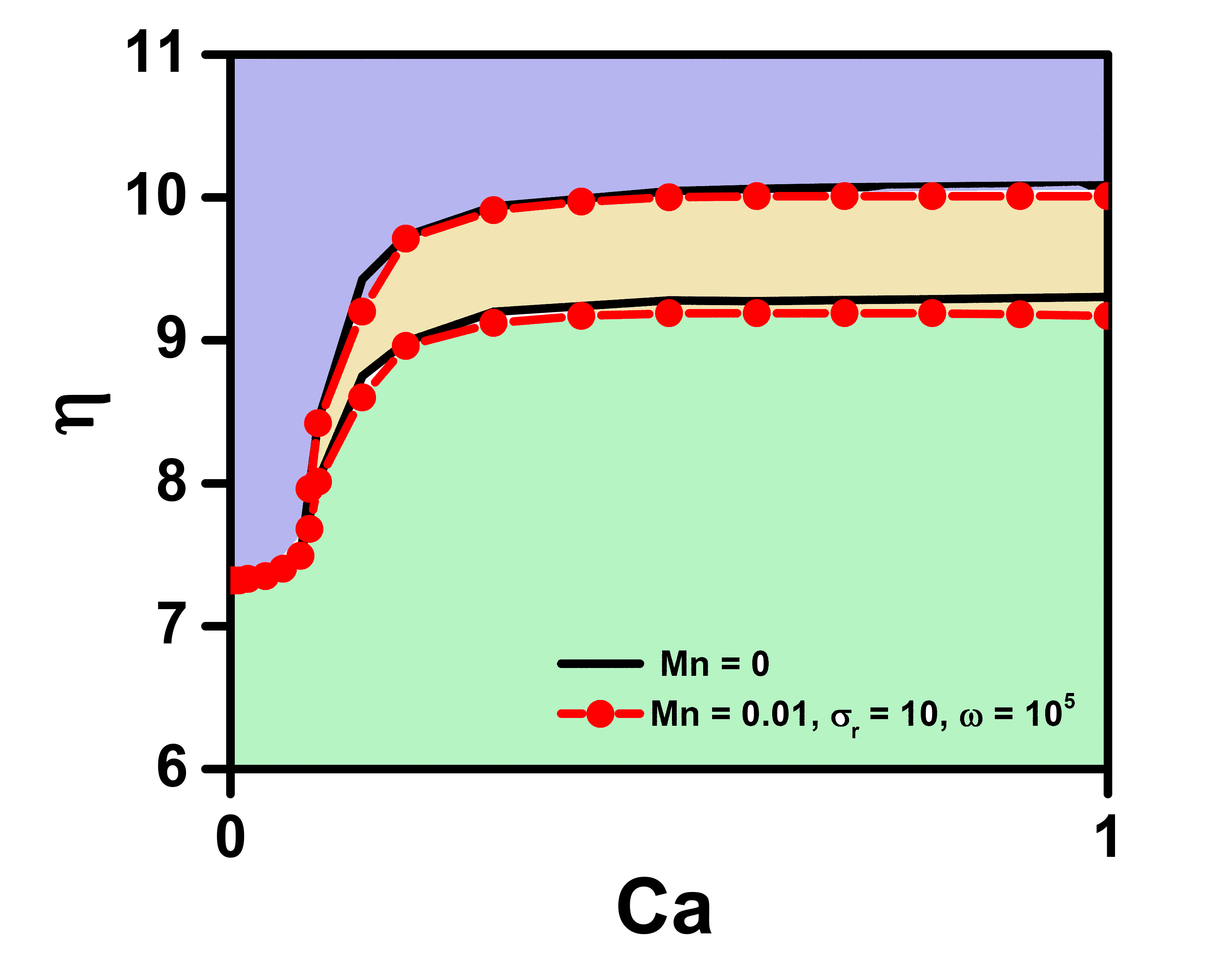}
		\caption{}
	\end{subfigure} 
	\hspace{0.12cm}
	\begin{subfigure}[b]{0.31\linewidth}
		\includegraphics[width=\linewidth]{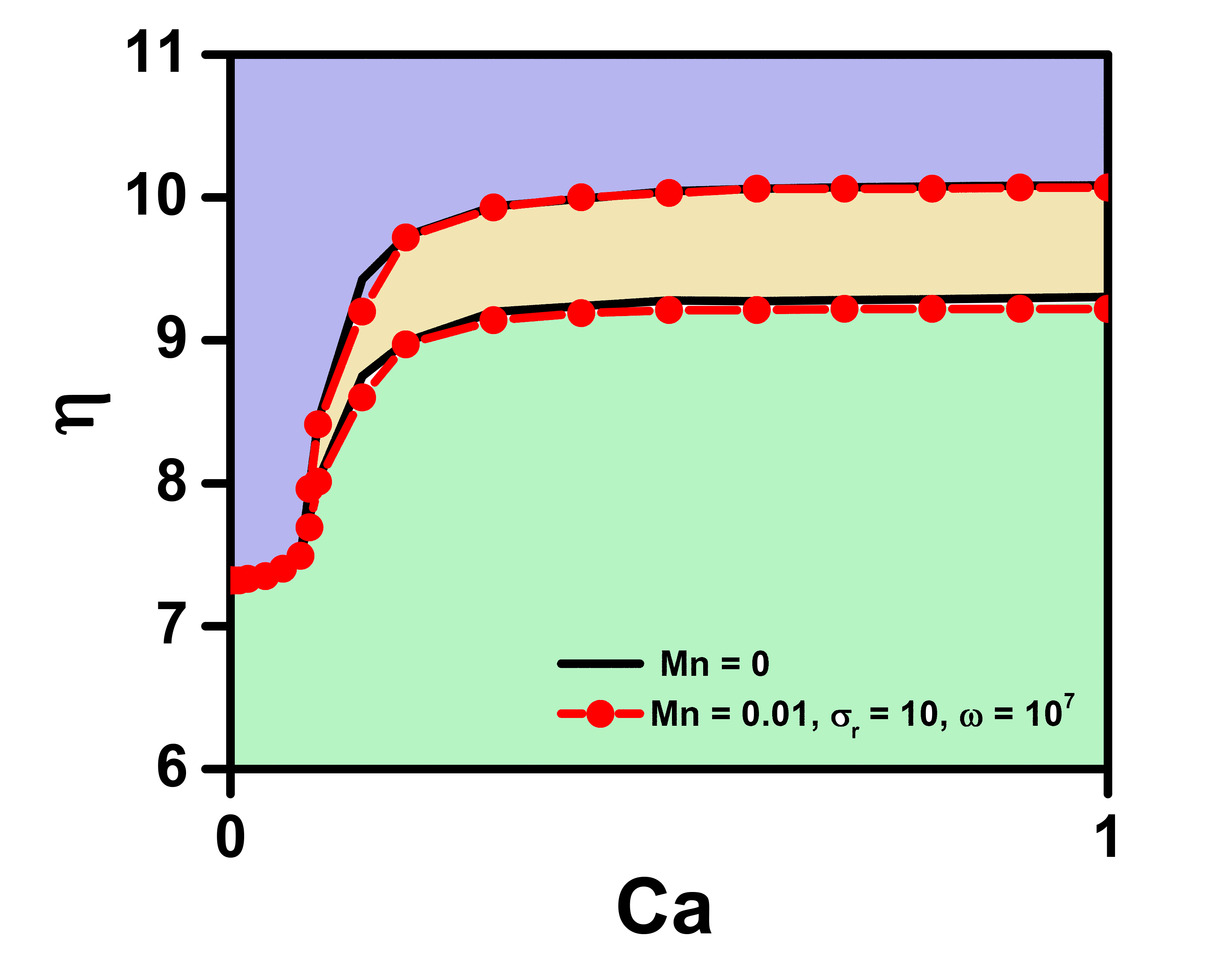}
		\caption{}
	\end{subfigure} 
	\hspace{0.12cm}
	\begin{subfigure}[b]{0.31\linewidth}       \includegraphics[width=\linewidth]{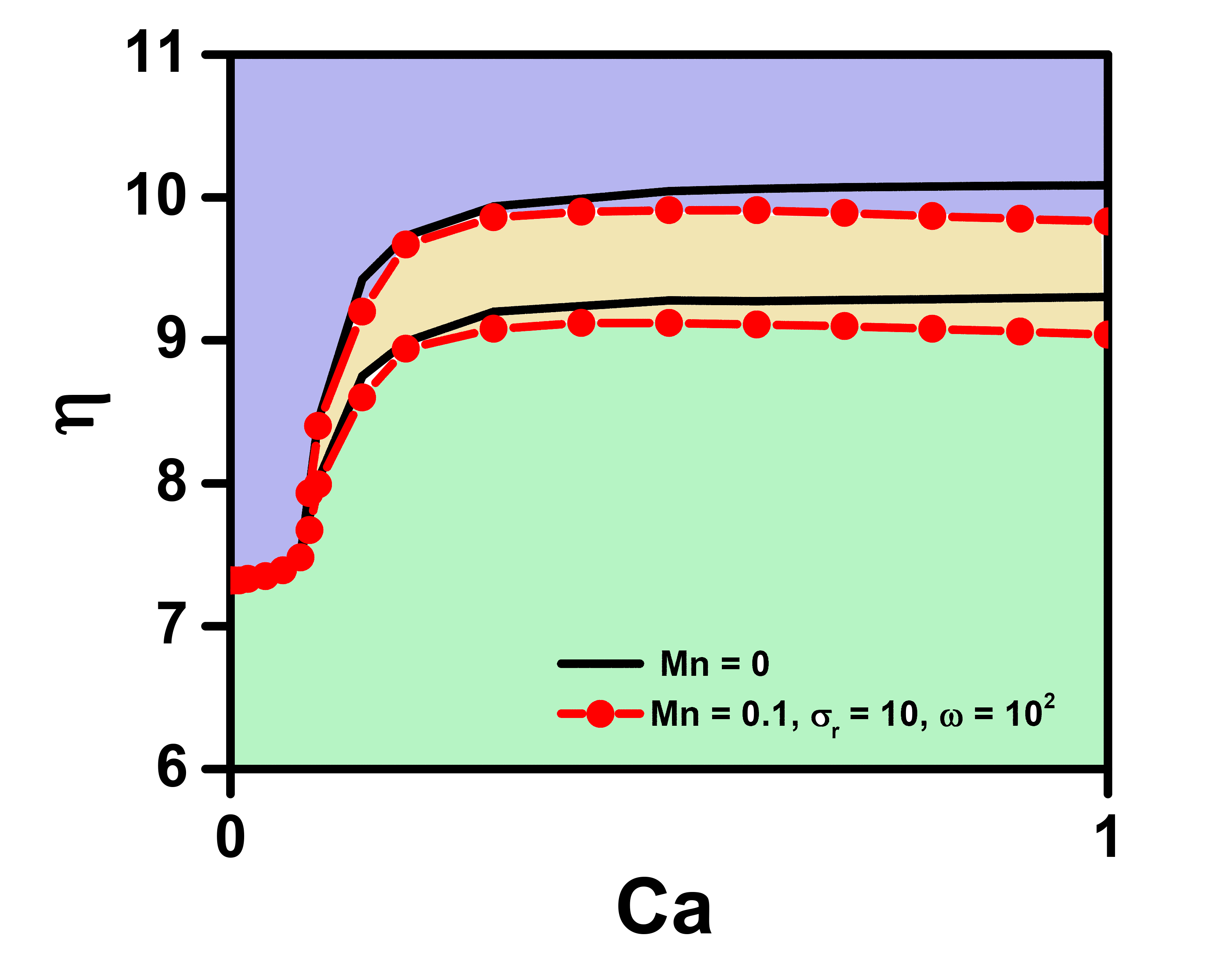}
		\caption{}
	\end{subfigure}
	\hspace{0.12cm}
	\begin{subfigure}[b]{0.31\linewidth}
		\includegraphics[width=\linewidth]{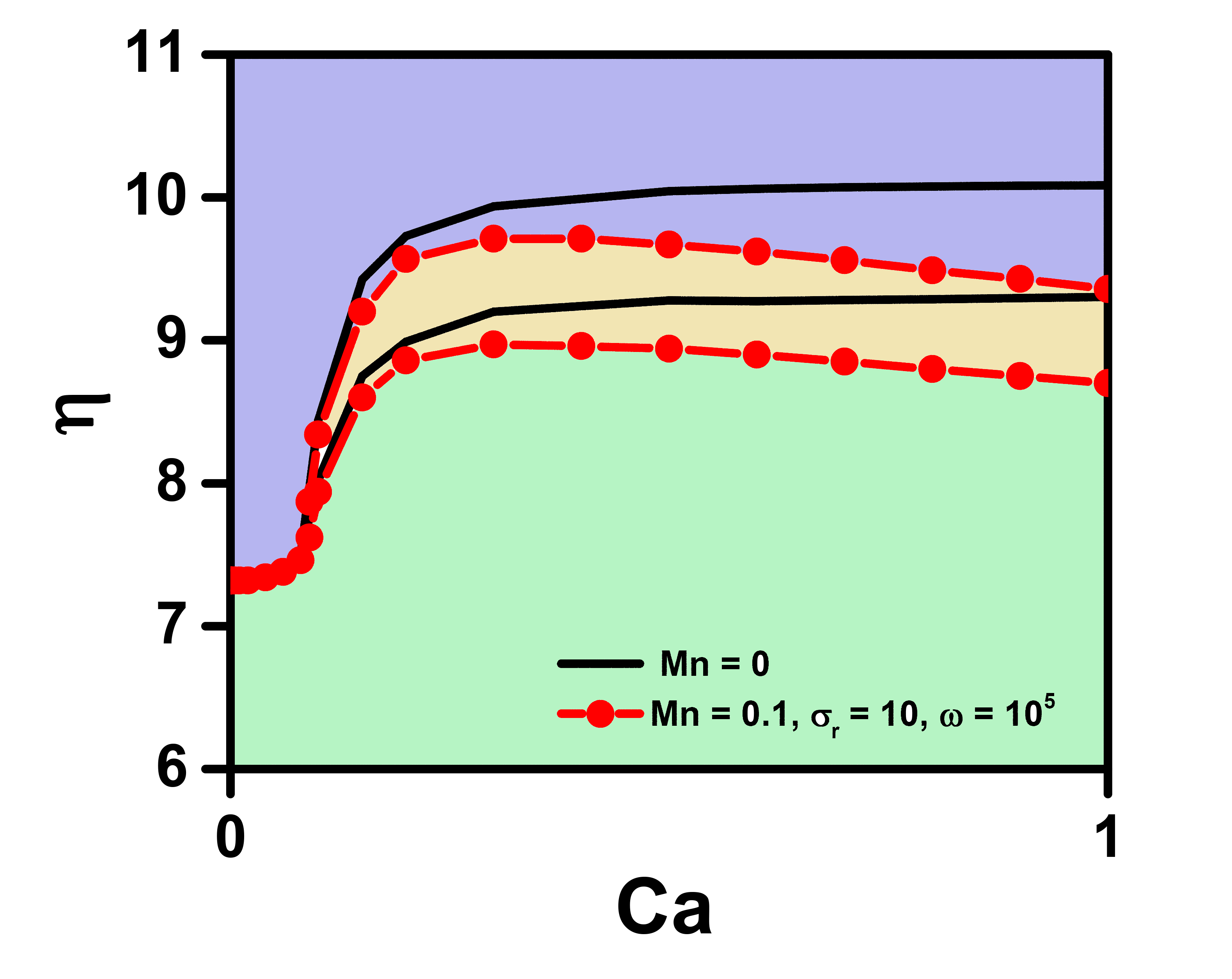}
		\caption{}
	\end{subfigure} 
	\hspace{0.12cm}
	\begin{subfigure}[b]{0.31\linewidth}
		\includegraphics[width=\linewidth]{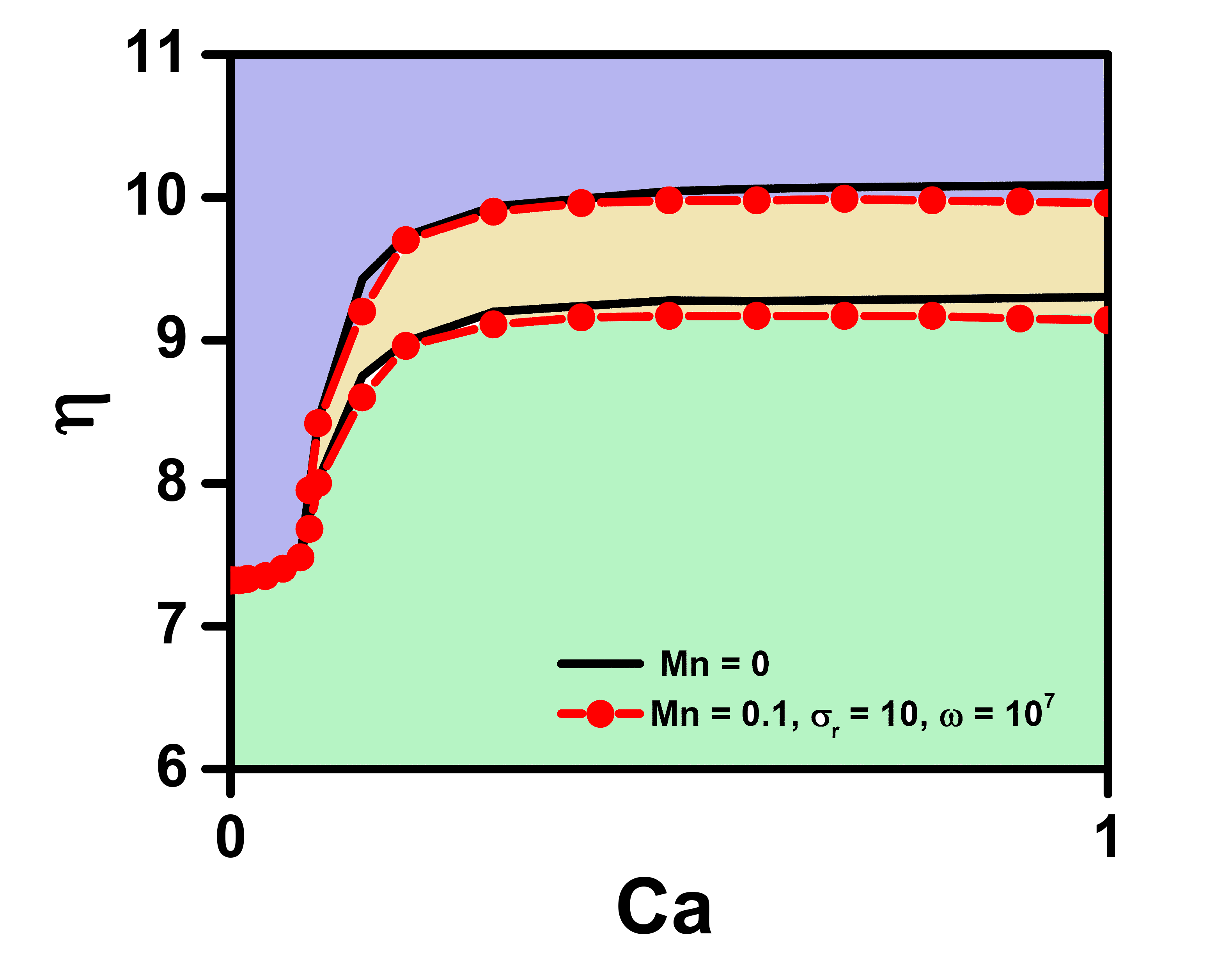}
		\caption{}
	\end{subfigure} 
	\hspace{0.12cm}
	\begin{subfigure}[b]{0.31\linewidth}       \includegraphics[width=\linewidth]{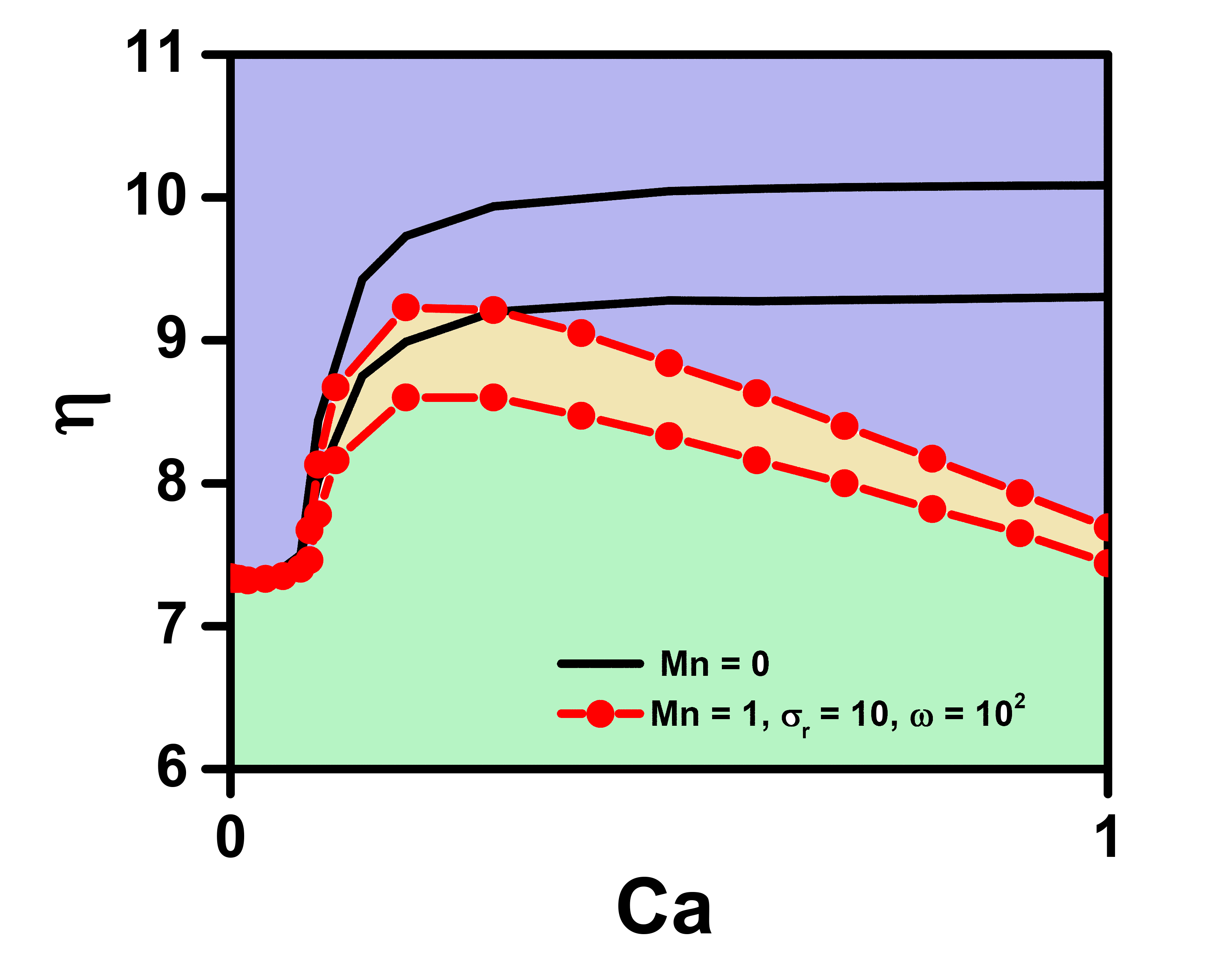}
		\caption{}
	\end{subfigure}
	\hspace{0.12cm}
	\begin{subfigure}[b]{0.31\linewidth}
		\includegraphics[width=\linewidth]{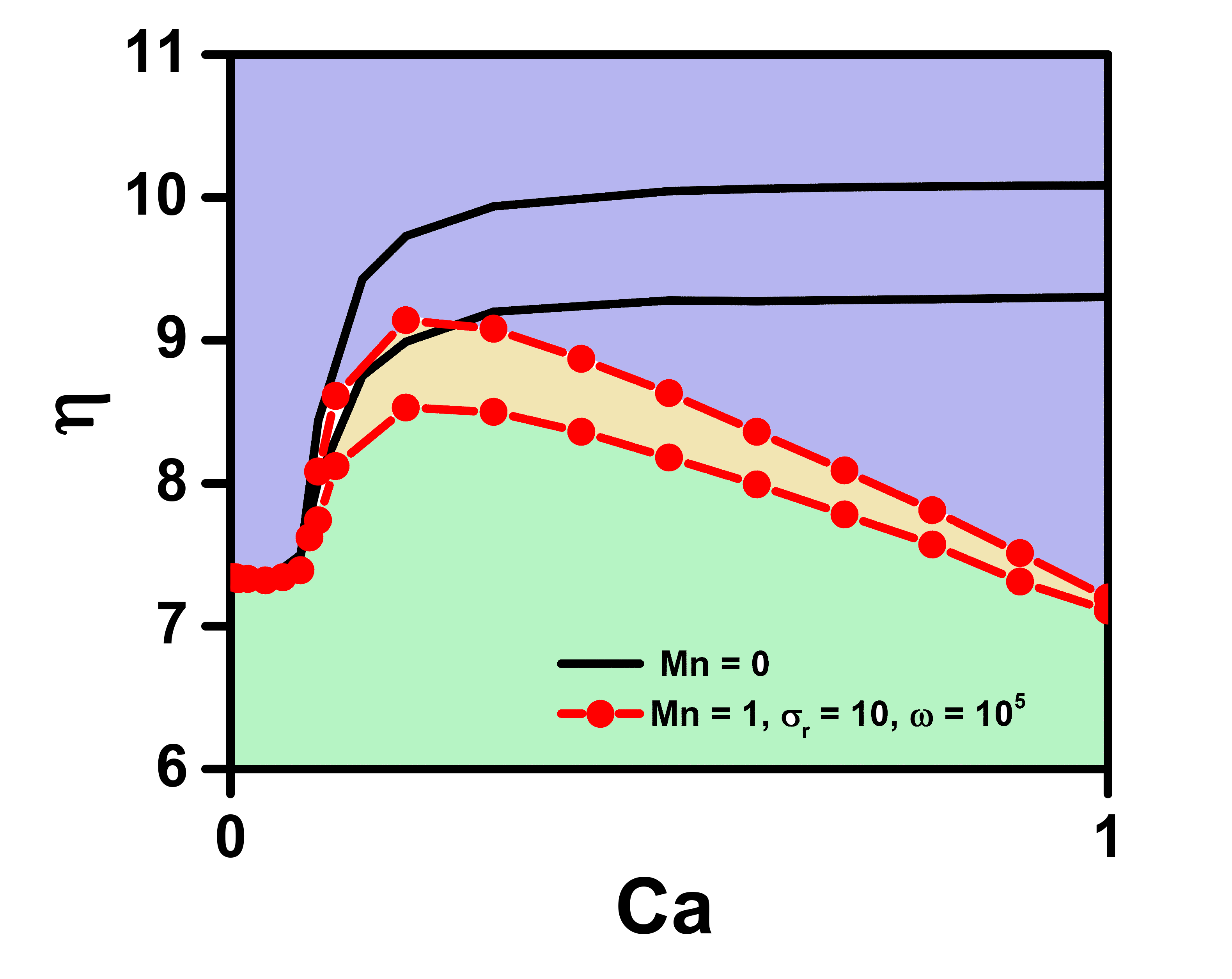}
		\caption{}
	\end{subfigure} 
	\hspace{0.12cm}
	\begin{subfigure}[b]{0.31\linewidth}
		\includegraphics[width=\linewidth]{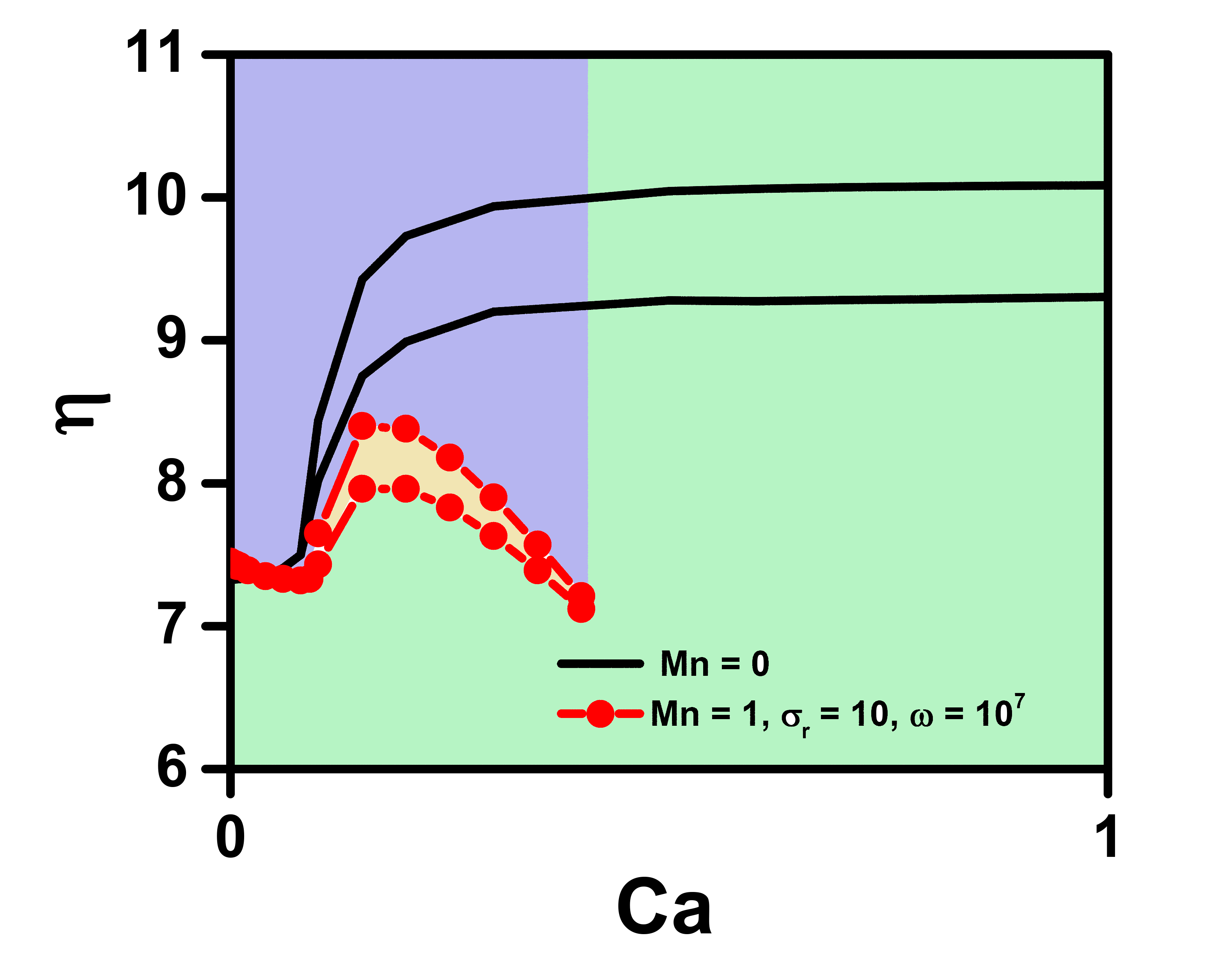}
		\caption{}
	\end{subfigure}                
	\caption{Phase diagram of transition in dynamic states (TT-green region, TR-yellow region, TU-blue region) for $\sigma_r=10$. (a)-(c): $Mn=0.01$, (d)-(f): $Mn=0.1$, and (d)-(f): $Mn=1$. In each set $\omega$ varies as $10^2, 10^5, 10^7$. ($\triangle=0.2, C_{mem}=50, \zeta=10^{-7}$)}
	\label{PD1}  
\end{figure}

\begin{figure} [ht] 
	\hspace{0.12cm}
	\begin{subfigure}[b]{0.31\linewidth}       \includegraphics[width=\linewidth]{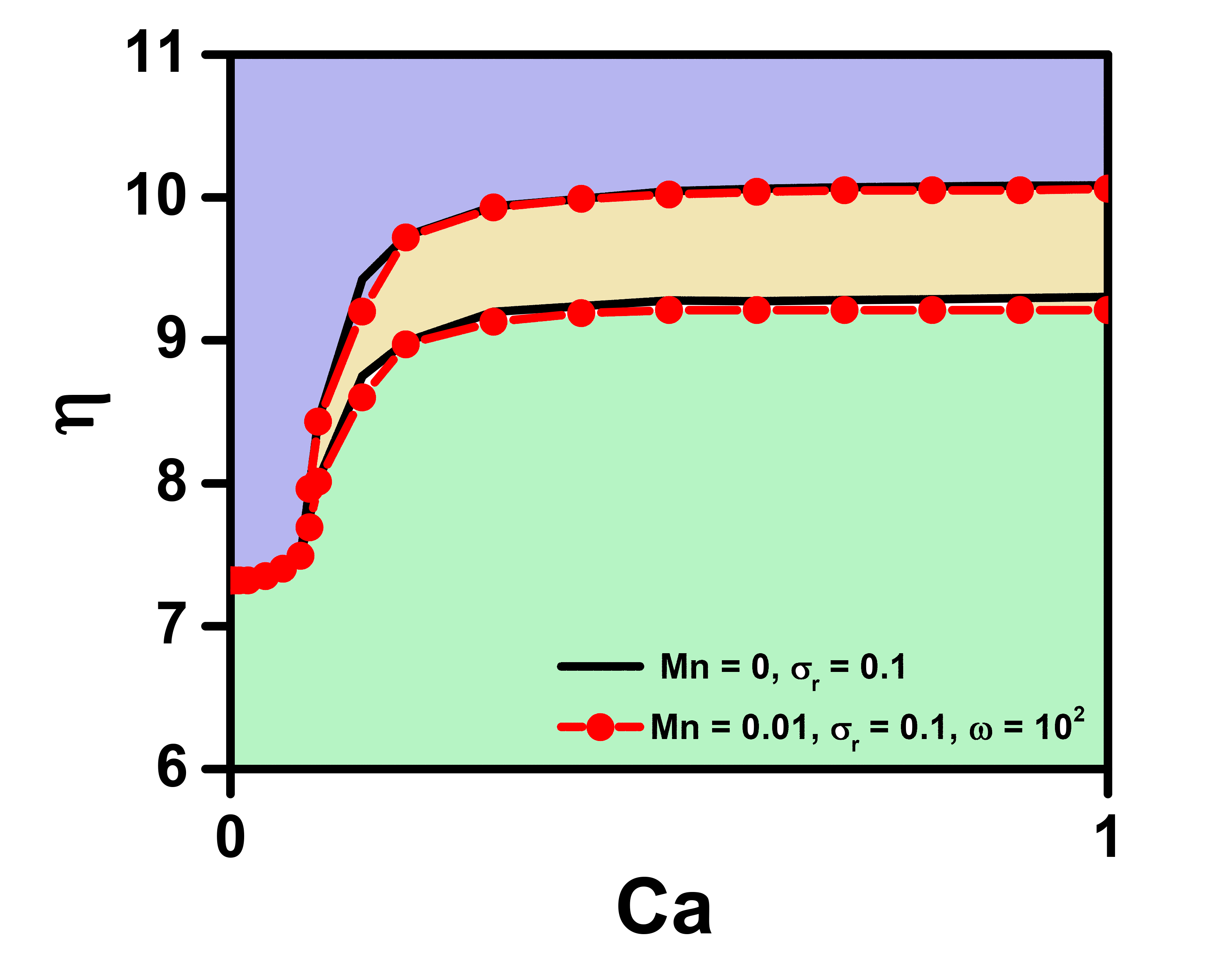}
		\caption{}
	\end{subfigure}
	\hspace{0.12cm}
	\begin{subfigure}[b]{0.31\linewidth}
		\includegraphics[width=\linewidth]{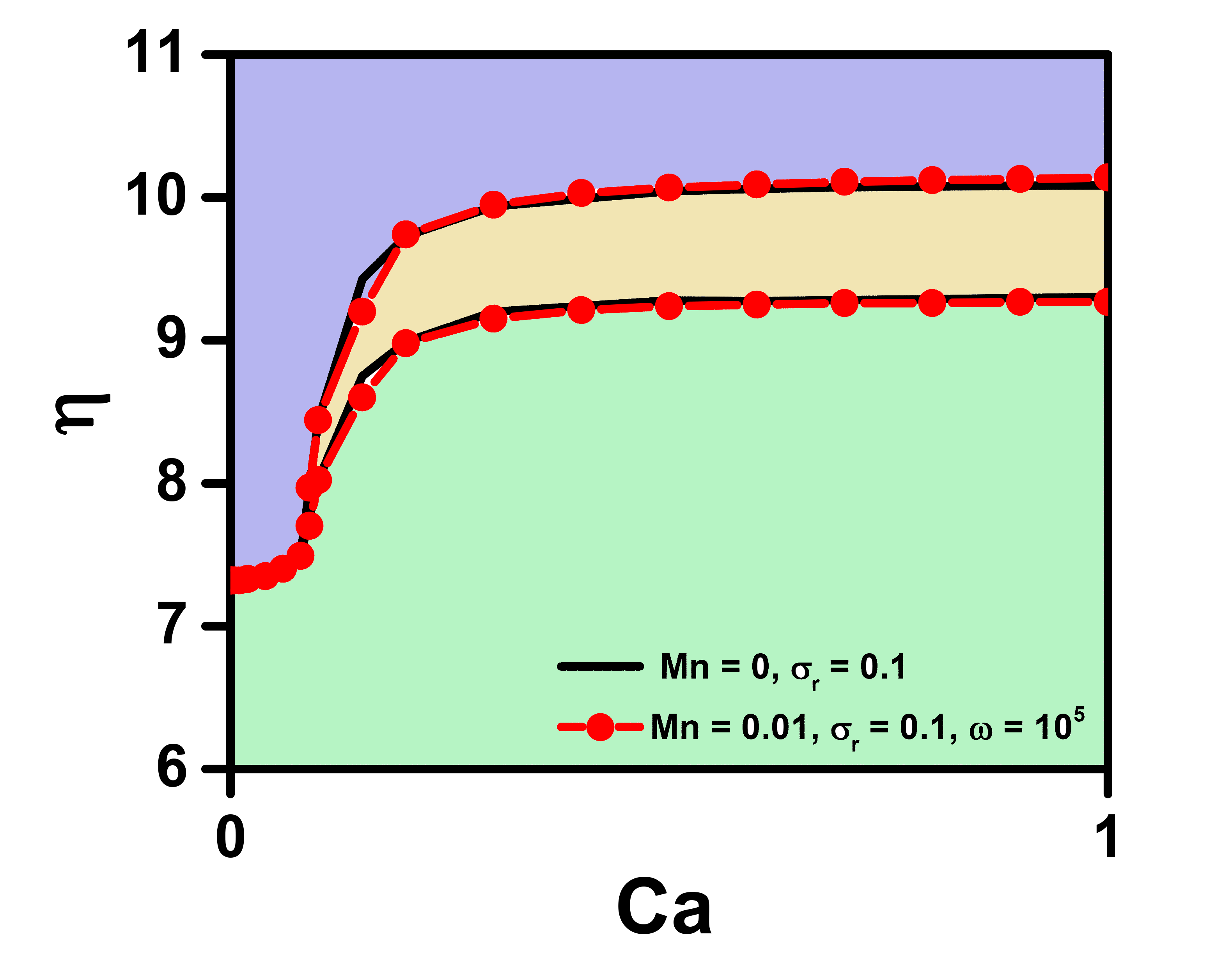}
		\caption{}
	\end{subfigure} 
	\hspace{0.12cm}
	\begin{subfigure}[b]{0.31\linewidth}
		\includegraphics[width=\linewidth]{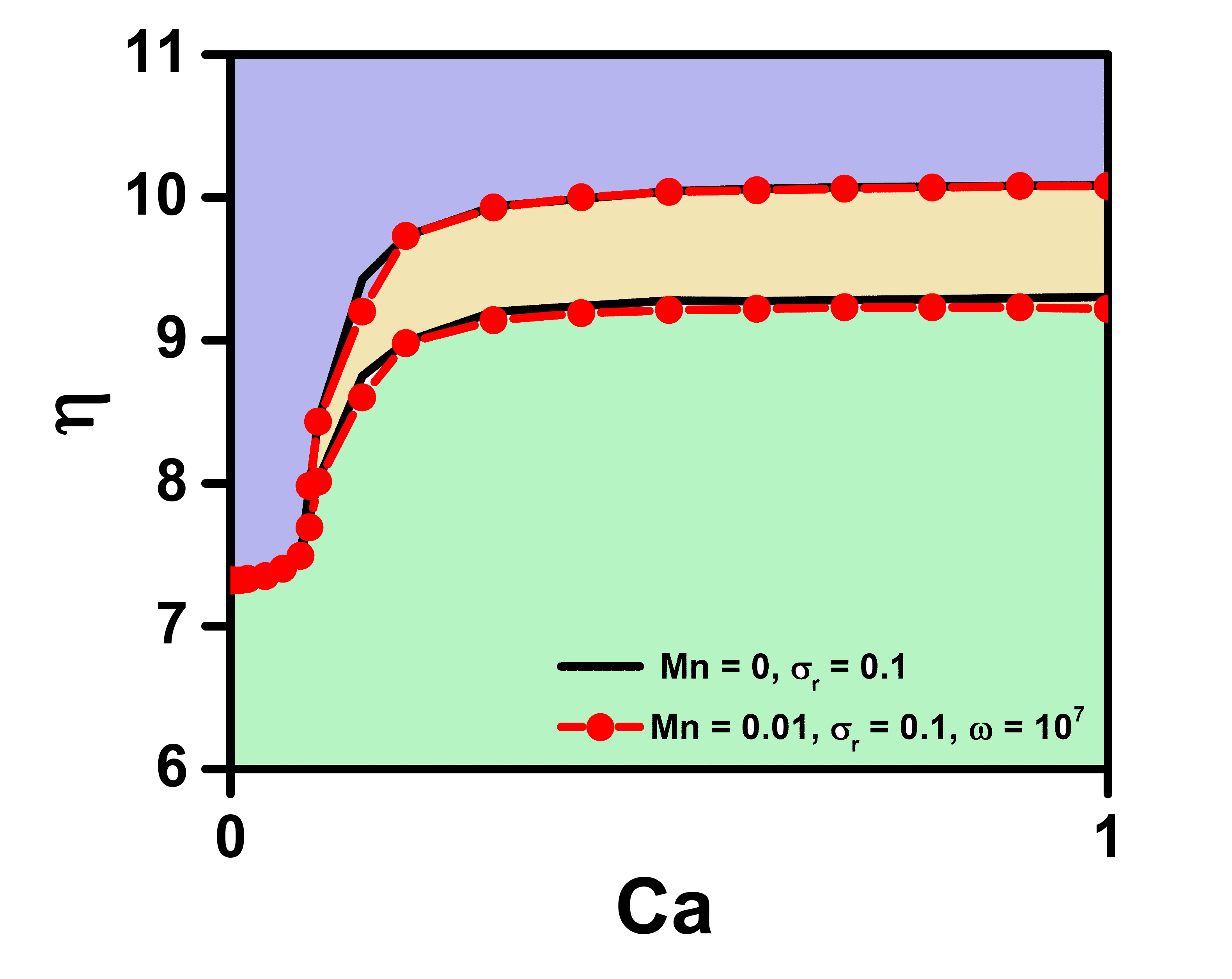}
		\caption{}
	\end{subfigure} 
	\hspace{0.12cm}
	\begin{subfigure}[b]{0.31\linewidth}       \includegraphics[width=\linewidth]{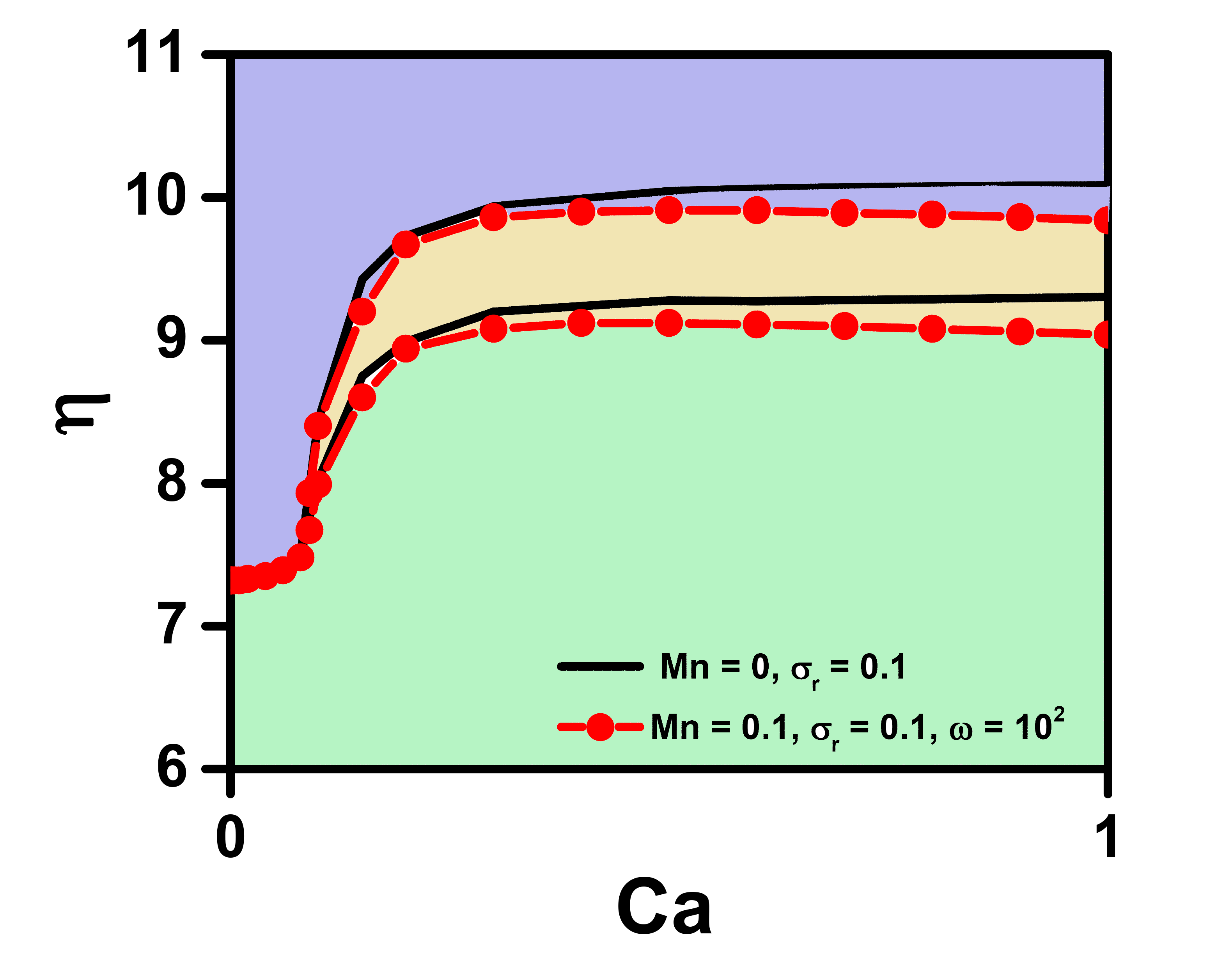}
		\caption{}
	\end{subfigure}
	\hspace{0.12cm}
	\begin{subfigure}[b]{0.31\linewidth}
		\includegraphics[width=\linewidth]{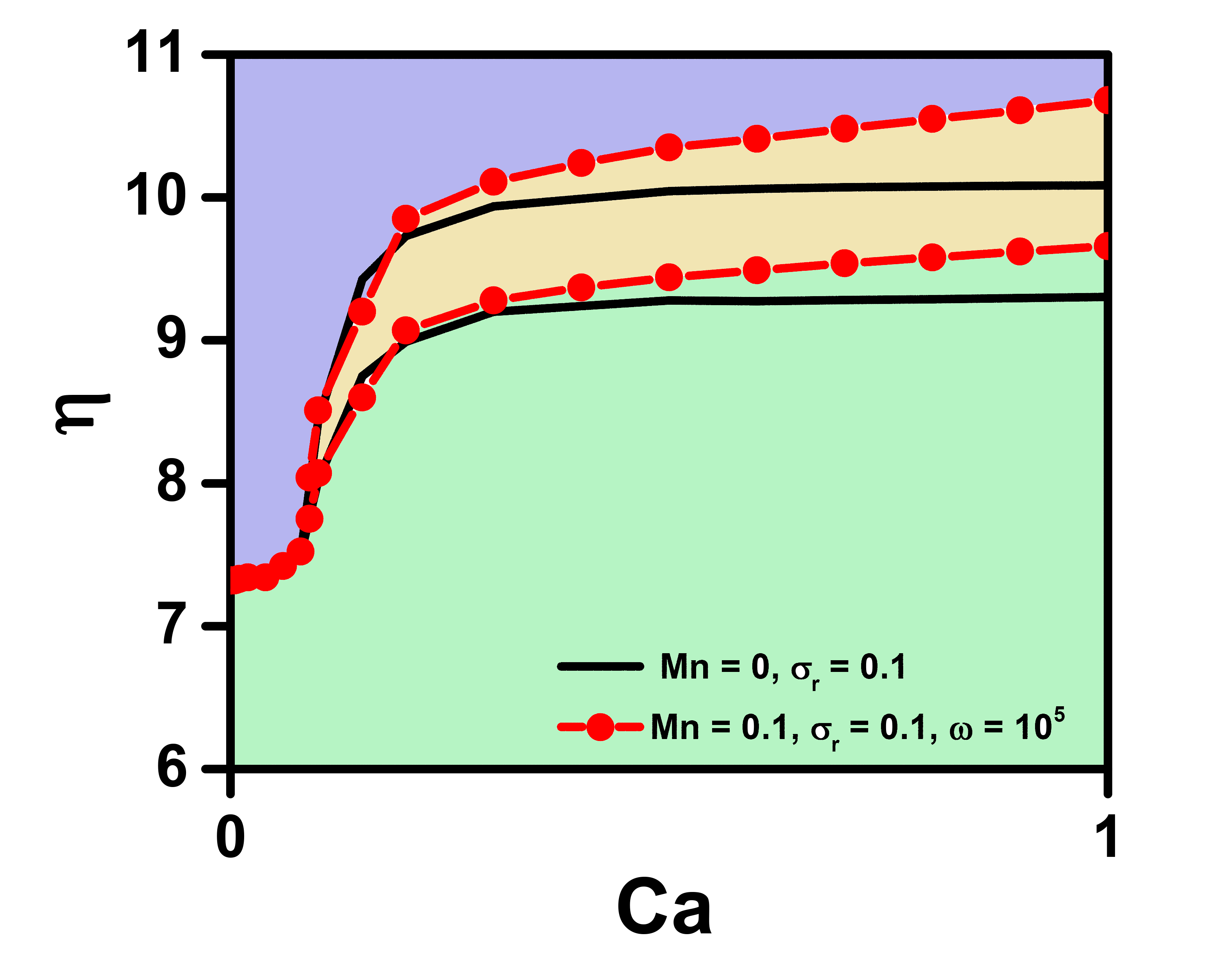}
		\caption{}
	\end{subfigure} 
	\hspace{0.12cm}
	\begin{subfigure}[b]{0.31\linewidth}
		\includegraphics[width=\linewidth]{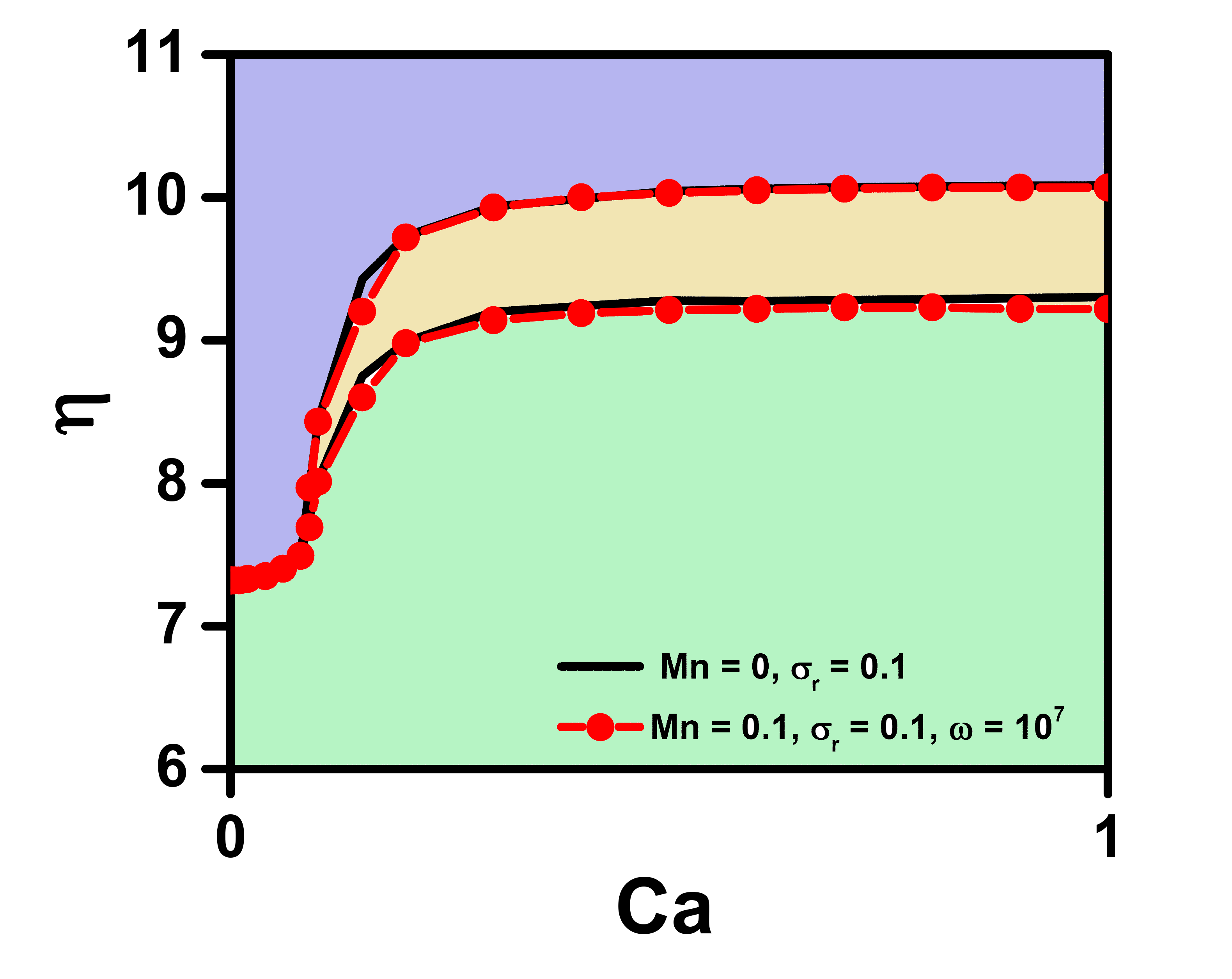}
		\caption{}
	\end{subfigure} 
	\hspace{0.12cm}
	\begin{subfigure}[b]{0.31\linewidth}       \includegraphics[width=\linewidth]{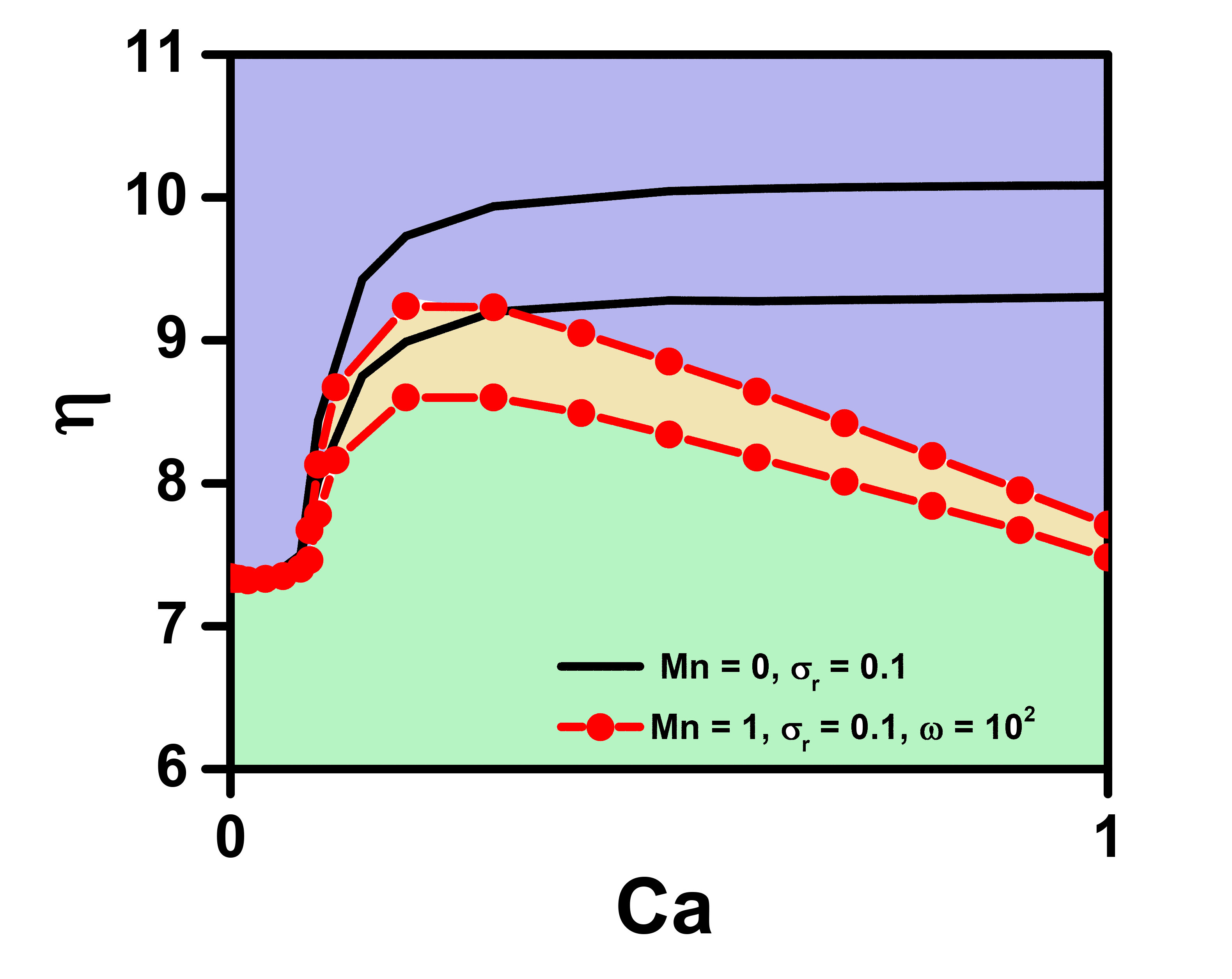}
		\caption{}
	\end{subfigure}
	\hspace{0.12cm}
	\begin{subfigure}[b]{0.31\linewidth}
		\includegraphics[width=\linewidth]{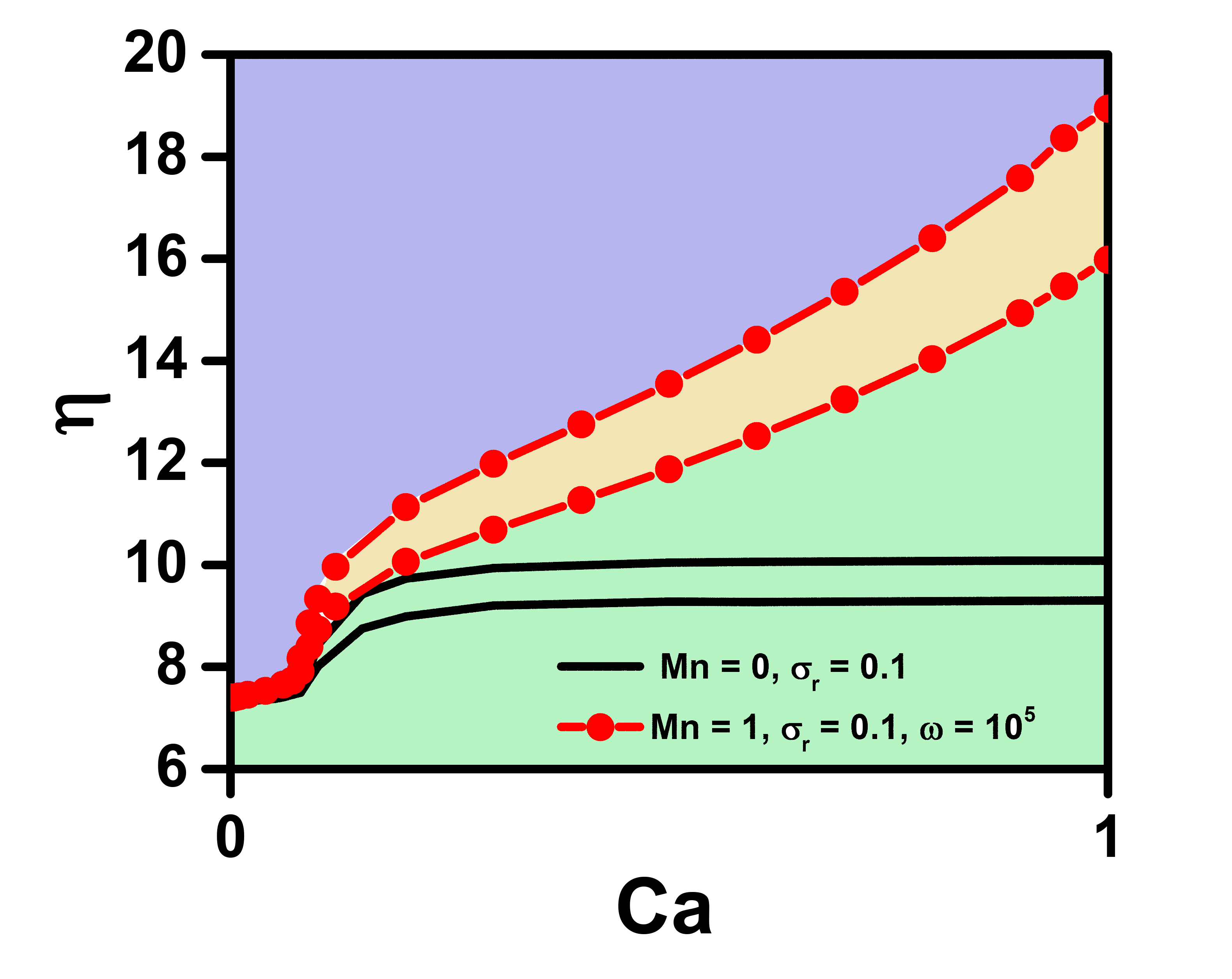}
		\caption{}
	\end{subfigure} 
	\hspace{0.12cm}
	\begin{subfigure}[b]{0.31\linewidth}
		\includegraphics[width=\linewidth]{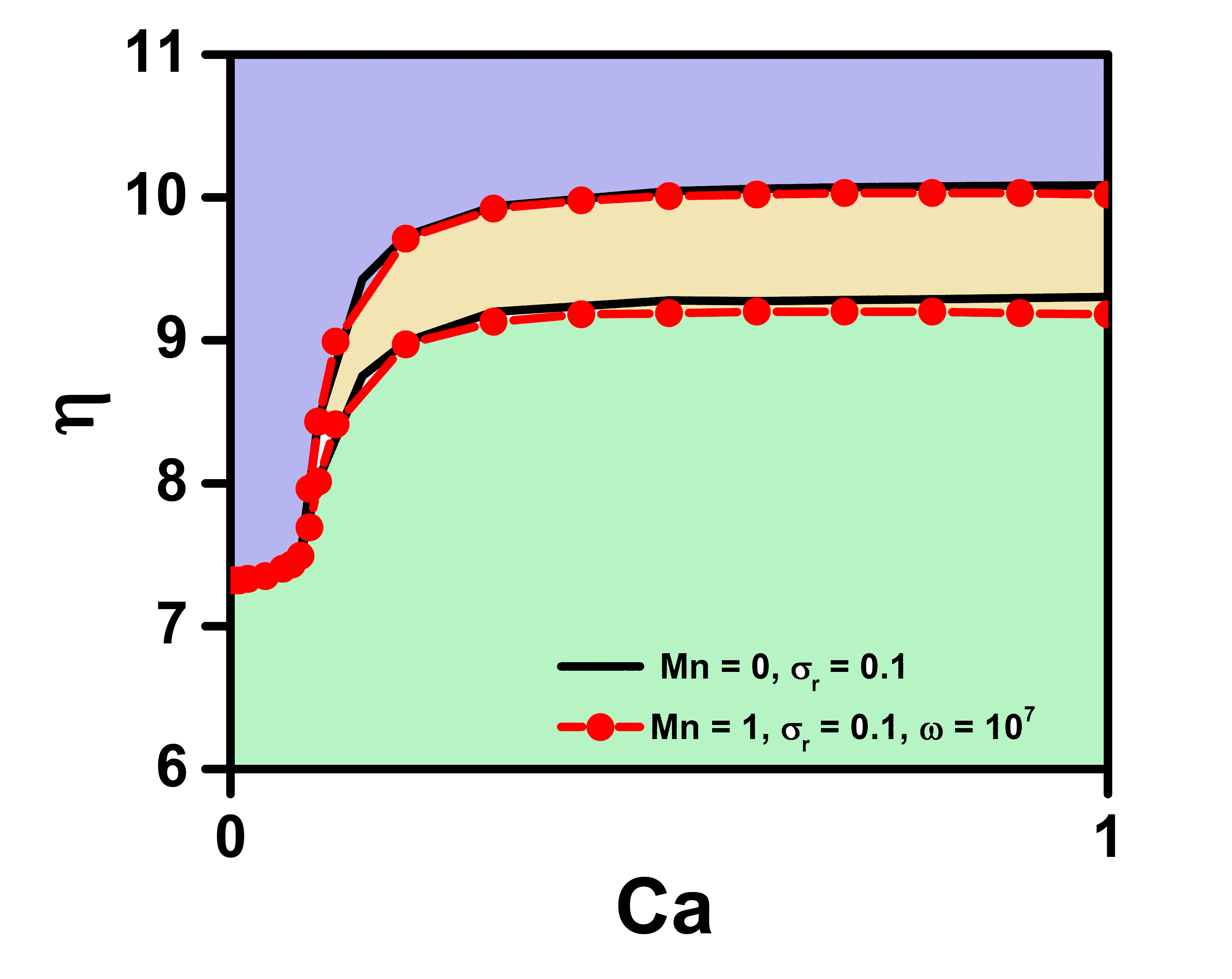}
		\caption{}
	\end{subfigure}            
	\caption{Phase diagram of transition in dynamic states (TT-green region, TR-yellow region, TU-blue region) for $\sigma_r=0.1$. (a)-(c): $Mn=0.01$, (d)-(f): $Mn=0.1$, and (d)-(f): $Mn=1$. In each set $\omega$ varies as $10^2, 10^5, 10^7$. ($\triangle=0.2, C_{mem}=50, \zeta=10^{-7}$)}
	\label{PD2}
\end{figure}

Phase diagrams in figure \ref{PD1} and \ref{PD2} show the transition between TT-TU, TT-TR, and TR-TU modes for $\sigma_r>1$ and $\sigma_r<1$, respectively. In both cases the study is limited to capillary numbers  up to $Ca=1$ only. Beyond that nonlinear hydrodynamic corrections are important which are not considered in this work. Also for both the conductivity ratio cases,  results show deviation from pure shear results and the system loses its character of $Ca$ number independent TT-TR and TR-TU transition. A clear shift is seen in the TT-TR, TR-TU phase transition boundaries such that the $\eta$ value at which TT-TR, TR-TU takes place depends upon flow capillary number. The analysis was conducted for three values of $Mn$, 0.01, 0.1, and 1.  With an increase in $Mn$ the boundary separating the two regimes especially TT-TR and TR-TU shows a narrowing of TR region, and lower transition viscosity which is frquency dependent.  In all these transitions (a) to (i) TT-TU transition value is fixed at around $\eta=7.4$ for a given excess area of $\triangle=0.2$. \\

Figure \ref{PD2} shows a similar study for $\sigma_r<1$ case. Low-frequency behavior is same as for $\sigma_r>1$ case and TR regime gets suppressed with an increase of $Ca$. At intermediate frequency TT-TR and TR-TU transition boundaries are pushed to higher $\eta$ with an increase in the capillary number,  as the $Mn$ is increased from 0.01 to 1. This  also shows widening of TR regime with $Ca$. Moreover for all the three $Mn$ values, high-frequency phase diagrams coincide with pure shear phase diagram, unlike $\sigma_r>1$ case.  \\

The results are a bit counterintutive. One would have expected that a high anticlockwise torque (at high $Mn$) for $\sigma_r>1$ would have delayed the TT-TU transition. Similarly a clockwise torque would have hastened the TT-TU transition for $\sigma_r<1$. We postulate that for $\sigma_r>1$, a prolate shape is favored by a high $Mn$. This in turn increases the anti-clockwise torque due to the $Ca$ thereby admitting lower transition viscosity as the capillary number is increased. On the other hand for $\sigma_r<1$ at intermediate frequencies, the $\psi\sim0$, whereby an increase in capillary number decreases the deformation of the vesicle, leading to an increase in the transition viscosity (equation \ref{evoleqn}).

\end{document}